\documentclass[fleqn,usenatbib]{mnras}

\usepackage{newtxtext,newtxmath}
\usepackage[T1]{fontenc}

\DeclareRobustCommand{\VAN}[3]{#2}
\let\VANthebibliography\thebibliography
\def\thebibliography{\DeclareRobustCommand{\VAN}[3]{##3}\VANthebibliography}

\usepackage{rotating}
\usepackage{amsmath}	% Advanced maths commands
\usepackage{siunitx}
\usepackage[dvipsnames]{xcolor}

\usepackage{totcount} % bibliography counting
\newtotcounter{citnum} %From the package documentation
\def\oldbibitem{} \let\oldbibitem=\bibitem
\def\bibitem{\stepcounter{citnum}\oldbibitem}

\newcommand{\teff}{$T_{\rm eff}$}

\title[Optical Data-Driven Cool Dwarf Parameters]{Cool and Data-Driven: An Exploration of Optical Cool Dwarf Chemistry with Both Data-Driven and Physical Models}

\author[Adam D. Rains et al.]{Adam D. Rains,$^{1,2}$\thanks{E-mail: adam.rains@physics.uu.se (ADR)}
Thomas Nordlander,$^{2,3}$
Stephanie Monty,$^{4}$ 
Andrew R. Casey,$^{3,5}$
B\'arbara Rojas-Ayala,$^{6}$ \newauthor
Maru\v{s}a \v{Z}erjal,$^{2,7,8}$
Michael J. Ireland,$^{2}$
Luca Casagrande,$^{2,3}$
and Madeleine McKenzie$^{2,3}$
\\
$^{1}$Department of Physics and Astronomy, Uppsala University, Box 516, 75120 Uppsala, Sweden\\
$^{2}$Research School of Astronomy and Astrophysics, Australian National University, Canberra, ACT 2611, Australia\\
$^{3}$ARC Centre of Excellence for Astrophysics in Three Dimensions (ASTRO-3D), Australia\\
$^{4}$Institute of Astronomy, University of Cambridge, Madingley Rd, Cambridge, CB3 0HA, UK\\
$^{5}$School of Physics \& Astronomy, Monash University, Wellington Road, Clayton 3800, Victoria, Australia\\
$^{6}$Instituto de Alta Investigaci\'on, Universidad de Tarapac\'a, Casilla 7D, Arica, Chile \\
$^{7}$Instituto de Astrof{\'{\i}}sica de Canarias, E-38205 La Laguna, Tenerife, Spain \\
$^{8}$Universidad de La Laguna, Dpto. Astrof{\'{\i}}sica, E-38206 La Laguna, Tenerife, Spain \\
}

\date{Accepted XXX. Received YYY; in original form ZZZ}

\pubyear{2015}

\begin{document}
\label{firstpage}
\pagerange{\pageref{firstpage}--\pageref{lastpage}}
\maketitle

\begin{abstract}

Detailed chemical studies of F/G/K---or Solar-type---stars have long been routine in stellar astrophysics, enabling studies in both Galactic chemodynamics, and exoplanet demographics. However, similar understanding of the chemistry of M and late-K dwarfs---the most common stars in the Galaxy---has been greatly hampered both observationally and theoretically by the complex molecular chemistry of their atmospheres. Here we present a new implementation of the data-driven \textit{Cannon} model, modelling $T_{\rm eff}$, $\log g$, [Fe/H], and [Ti/Fe] trained on low--medium resolution optical spectra ($4\,000-7\,000\,$\SI{}{\angstrom}) from 103 cool dwarf benchmarks. Alongside this, we also investigate the sensitivity of optical wavelengths to various atomic and molecular species using both data-driven and theoretical means via a custom grid of MARCS synthetic spectra, and make recommendations for where MARCS struggles to reproduce cool dwarf fluxes. Under leave-one-out cross-validation, our \textit{Cannon} model is capable of recovering $T_{\rm eff}$, $\log g$, [Fe/H], and [Ti/Fe] with precisions of 1.4\%, $\pm0.04\,$dex, $\pm0.10\,$dex, and $\pm0.06\,$dex respectively, with the recovery of [Ti/Fe] pointing to the as-yet mostly untapped potential of exploiting the abundant---but complex---chemical information within optical spectra of cool stars. 

%{\color{red}This document contains \total{citnum}\ references.}
\end{abstract}

\begin{keywords}
methods: data analysis -- techniques: spectroscopic -- stars: fundamental parameters -- stars: low-mass
\end{keywords}

%%%%%%%%%%%%%%%%%%%%%%%%%%%%%%%%%%%%%%%%%%%%%%%%%%%%%%%%%%%%%%%%%%%%%%%%%%%%%%%%%%%%%%%%%%%%%%%%%%%%%
% Introduction
%%%%%%%%%%%%%%%%%%%%%%%%%%%%%%%%%%%%%%%%%%%%%%%%%%%%%%%%%%%%%%%%%%%%%%%%%%%%%%%%%%%%%%%%%%%%%%%%%%%%%
\section{Introduction}\label{sec:intro}

The Solar Neighbourhood---and indeed the Universe more broadly---is dominated by cool dwarf stars of spectral types K and M  \citep[e.g.][]{henry_solar_1994, chabrier_galactic_2003, henry_solar_2006, winters_solar_2015, henry_solar_2018}. While Milky Way stars in general are expected to host at least one planet on average \citep{cassan_one_2012}, cool dwarfs are actually more likely to host small planets as compared to more massive stars \citep{howard_planet_2012, dressing_occurrence_2015} with many yet undiscovered \citep{morton_radius_2014}. Enabled by the space-based Kepler \citep{borucki_kepler_2010}, K2 \citep{howell_k2_2014}, and TESS \citep{ricker_transiting_2015} missions, exoplanetary astrophysics now has a large and ever-growing set of such systems to study both individually in detail, as well as collectively in a demographic sense. 

When presented as such, it is easy to come to the conclusion that cool dwarfs and their planets are as well-understood as their prevalence might imply. In reality though, these stars are intrinsically faint---especially at optical wavelengths---and possess complex spectra blanketed by innumerable overlapping molecular absorption features. In the infrared (IR) this absorption is dominated by molecules like H$_2$O, CO, FeH, and OH; and in the optical from oxides like TiO, ZrO, and VO, as well as hydrides like MgH, CaH, AlH, and SiH. Such complexity renders the spectral energy distribution (SED) not just a strong function of temperature, as with Solar--type stars, but also chemistry, making it difficult to ascribe an accurate or unique set of stellar parameters to any given star. This intense molecular absorption makes `true' continuum normalisation impossible at optical wavelengths, and poses severe challenges for traditional spectroscopic analysis techniques. As a result, our understanding of the chemistry of cool dwarfs and their planets typically lags far behind those of Solar-type stars.

This atmospheric complexity and the large impact a single molecular species can have on an emergent spectrum means that the generation of model spectra that accurately match observations has been, and continues to be, a challenge. While model spectra at cool temperatures demonstrate reasonable performance in the near infrared \citep[NIR, e.g.][]{allard_model_1997, baraffe_evolutionary_1997, baraffe_evolutionary_1998, allard_models_2012}, there have long been issues in the optical \citep[e.g.][]{baraffe_evolutionary_1998, reyle_effective_2011, mann_spectro-thermometry_2013, rains_characterization_2021}. The core reason is likely incomplete line lists for dominant sources of opacity, where the impact of not accurately knowing transition wavelengths or line depths can be severe \citep[e.g.][]{plez_spherical_1992-1, masseron_ch_2014}---particularly for TiO \citep[e.g.][]{hoeijmakers_search_2015, mckemmish_exomol_2019} which dominates absorption in the optical. All this means that, to this day, it is far from simple to produce accurate cool dwarf temperatures, radii, and \textit{especially} metallicities en masse---let alone individual elemental abundances. 

Given these complexities, it is thus critical to have a set of cool dwarfs of known chemistry to use as benchmarks for testing models or building empirical relations. The widely considered gold standard are cool dwarfs in binary systems with a warmer companion of spectral type F/G/K from which the chemistry can more easily be determined. This relies on the assumption that both stars formed at the same time and thus have the same chemical composition. Thankfully such chemical homogeneity is now well established for F/G/K--F/G/K pairs \citep[e.g.][]{desidera_abundance_2004, simpson_galah_2019, hawkins_identical_2020, yong_c3po_2023}, and while there remain edge-cases of chemically inhomogeneous pairs \citep[e.g.][]{spina_chemical_2021}---possibly the result of planet engulfment---the level of chemical homogeneity is more than sufficient for the precision of the current state of the art in cool dwarf chemical analysis.

The extreme sensitivity of cool dwarf spectra to stellar chemistry remains present in broadband optical photometry, though this is less the case in the IR where $K$ band \textit{photometry} at $2.2\,\mu$m is a comparatively [Fe/H]--insensitive\footnote{See Section 7.5: `The Role of Metallicity' in \citet{mann_how_2019} for a detailed discussion of the [Fe/H] sensitivity of empirical cool dwarf $M_{K_S}$--$M_\star$ relations in conjunction with stellar evolutionary models.} probe of stellar mass ($M_\star$) for isolated main--sequence stars with $M_\star\lesssim0.7\,{\rm M}_\odot$. This is something that was initially predicted by theory (see e.g. \citealt{allard_model_1997}, \citealt{baraffe_evolutionary_1998}, and \citealt{chabrier_theory_2000} for summaries), and later confirmed observationally \citep[][]{delfosse_accurate_2000}, and allows for the development of photometric metallicity relations using an optical--NIR colour benchmarked on the aforementioned K/M--F/G/K benchmark systems \citep[e.g.][]{bonfils_metallicity_2005, johnson_metal_2009, schlaufman_physically-motivated_2010, neves_metallicity_2012, hejazi_optical-near_2015, dittmann_calibration_2016, rains_characterization_2021, duque-arribas_photometric_2023}. While purely photometric metallicity relations suffer from certain limitations, such as their sensitivity to unresolved binarity or young stars still contracting to the main sequence---they are widely applicable given the volume of data available from photometric surveys like \textit{2MASS} \citep{skrutskie_two_2006}, \textit{SkyMapper} \citep{keller_skymapper_2007}, SDSS \citep{york_sloan_2000}, Pan-STARRS \citep{chambers_pan-starrs1_2016}, and \textit{Gaia} \citep{gaia_collaboration_gaia_2016-1}.

Greater metallicity precision can be achieved by using low-resolution spectra and building empirical relations from [Fe/H]-sensitive spectral regions or indices, again benchmarked against K/M--F/G/K binary systems. Such low resolution spectra contain vastly more information than broadband photometry alone and are relatively observationally cheap to obtain, especially at redder wavelengths (e.g. the $YJHK$ bands) where these stars are brighter. The last ${\sim}$10 years has seen a number of studies develop such relations, which span a range of spectral resolutions and wavelengths \citep[e.g.][]{rojas-ayala_metal-rich_2010, rojas-ayala_metallicity_2012, terrien_h-band_2012, mann_full_2013, mann_spectro-thermometry_2013, newton_near-infrared_2014, mann_how_2015, kuznetsov_characterization_2019}, which importantly gives rise to a large \textit{secondary set} of fundamentally-calibrated cool dwarf benchmarks. This proves useful as the wide separation F/G/K--M/K binaries passing the quality cuts necessary to serve as benchmark systems are more rare---and thus also more distant on average---making the secondary set of benchmarks the brighter and more populous sample.

Other studies have opted to determine [Fe/H] from model fits to high-resolution spectra. Not only can this give access to unblended atomic lines not accessible for observations made at lower spectral resolution---especially in the (N)IR---but it also allows for more detailed testing of the models themselves\footnote{There is no single `threshold' resolution at which this becomes possible due to the strong wavelength dependence of blending, but at the low end studies like \citet{souto_detailed_2022} have found success using R${\sim}22\,500$ spectra from APOGEE.}. These studies span a similarly wide range of optical and IR wavelengths \citep[e.g.][]{woolf_metallicity_2005, bean_accurate_2006, bean_metallicities_2006, woolf_calibrating_2006, rajpurohit_high-resolution_2014, passegger_fundamental_2016, lindgren_metallicity_2017, veyette_physically_2017, souto_chemical_2017, passegger_carmenes_2018, marfil_carmenes_2021, cristofari_estimating_2022}, and have helped in pushing the boundaries of what we know about cool dwarfs and how best to model and analyse them. 

However, despite these advances in the determination of cool dwarf metallicities, it is at best an approximation to assume that their spectra can reliably be parameterised by only three atmospheric parameters in $T_{\rm eff}$, $\log g$, and [M/H] (or [Fe/H], its common proxy\footnote{We adopt [Fe/H], meaning the abundance of iron on the $\log$ 12 scale, as one of our fundamental stellar parameters for the remainder of this study.}). In reality, individual elemental abundances are able to dramatically change the shape of the observed `pseudocontinuum'---and thus the measured stellar properties---via their effect on various dominant molecular absorbers. As a specific example, \citet{veyette_physical_2016} demonstrated that independently changing carbon and oxygen abundances by just $\pm0.2\,$dex can result in an inferred metallicity ranging over a full order of magnitude ($>1\,$dex), with typical metallicity indicators---like those from low-resolution spectra previously discussed---showing a strong dependence on the C/O ratio. In cool atmospheres the carbon abundance affects how much oxygen gets locked up in CO, a low energy molecule that preferentially forms, with only the leftover oxygen able to go into other dominant opacity sources like H$_2$O and TiO.  

Understanding elemental abundances of cool dwarfs beyond just the bulk metallicity is thus a critically important task. This important work is well underway \citep[e.g.][]{tsuji_near-infrared_2014, tsuji_near-infrared_2015, tsuji_near-infrared_2016, tsuji_near-infrared_2016-1, veyette_physical_2016, veyette_physically_2017, souto_chemical_2017, souto_stellar_2018, ishikawa_elemental_2020, souto_stellar_2020, maldonado_hades_2020, ishikawa_elemental_2022, souto_detailed_2022, cristofari_estimating_2022-1}, but more research is needed to fundamentally calibrate the results using a larger set of more chemically diverse binary benchmarks, do this at the scale of large spectroscopic surveys containing thousands of stars, and to use this knowledge to improve upon the current generation of cool dwarf model spectra. 

Data-driven models present another method to tackling this problem. Provided they are trained on spectra from a set of benchmarks with precise fundamental or fundamentally-calibrated properties, such an approach becomes an effective way of teasing apart the complex chemistry of these stars. Absent the limitations that come with \textit{physical} models (e.g. incomplete molecular line lists), \textit{data-driven} models have the potential to turn what is traditionally considered a \textit{weakness} of cool dwarfs---strong and innumerable overlapping absorption features from multiple different atomic and molecular species---into a \textit{strength} given the sheer amount of information present---assuming of course this chemical information can be properly exploited. This is a particularly important problem to solve in preparation for upcoming massive spectroscopic surveys like 4MOST \citep{de_jong_4most_2019} and SDSS-V \citep{kollmeier_sdss-v:_2017}.

Data-driven models like the \textit{Cannon} \citep[][]{ness_cannon:_2015} have been successfully applied to F/G/K stars observed by spectroscopic surveys like GALAH, APOGEE, LAMOST, and SPOCS \citep[e.g.][]{buder_galah_2018-1, ho_cannon:_2016, casey_cannon_2016, casey_data-driven_2019, wheeler_abundances_2020, rice_stellar_2020, nandakumar_combined_2022}, often with the goal of inter-survey comparison or the computational speed of data-driven stellar property determination versus more traditional modelling. Other studies have extended this work to cool (and brown) dwarfs using a variety of modelling approaches \citep[][]{behmard_data-driven_2019, birky_temperatures_2020, galgano_fundamental_2020, li_stellar_2021, feeser_using_2022} for the recovery of properties like spectral type, $T_{\rm eff}$, $\log g$, [Fe/H], [M/H], $M_\star$, stellar radius ($R_\star$), or stellar luminosity ($L_\star$), with \citet{maldonado_hades_2020} even reporting the impressive recovery of 14 different chemical abundances with $\Delta{\rm[X/H]}\lesssim0.10\,$dex for their sample of K/M--F/G/K binaries. Finally, these models can also be used to explore complex parameter spaces and make new physical insights---for example wavelength regions sensitive to particular elemental abundances---something more challenging to do with traditional analysis methods.%   

Here we present a new implementation of the \textit{Cannon} trained on low-to-medium resolution (R${\sim}3\,000$--$7\,000$) optical spectra ($4\,000 < \lambda < 7\,000\,$\SI{}{\angstrom}) of cool dwarfs observed with the WiFeS instrument \citep{dopita_wide_2007} on the ANU 2.3 m Telescope at Siding Spring Observatory (NSW, Australia). Our four label model in $T_{\rm eff}$, $\log g$, [Fe/H], and [Ti/Fe] draws its accuracy from a relatively small, but hand-selected, set of 103 stellar benchmarks primarily composed of stars with interferometric $T_{\rm eff}$, [Fe/H] and [Ti/Fe] measurements from a wide binary companion of spectral type F/G/K, or [Fe/H] determined from binary-benchmarked empirical relations based on low-resolution NIR spectra. We use our \textit{Cannon} model in conjunction with a custom grid of MARCS model spectra \citep{gustafsson_grid_2008} to investigate the sensitivity of optical cool dwarf fluxes to variations in chemical abundances, as well as limitations in reproducing optical fluxes. Our data and stellar benchmark selection are described in Section \ref{sec:benchmarks}; our \textit{Cannon} model and its training, validation, and performance in Section \ref{sec:cannon}; an investigation into cool dwarf optical flux sensitivity to elemental abundance variations using MARCS spectra in Section \ref{sec:marcs_abundances}; a discussion of results, comparison to previous work, MARCS flux recovery assessment, and future prospects in Section \ref{sec:discussion}; and concluding remarks in Section \ref{sec:conclusion}.

%%%%%%%%%%%%%%%%%%%%%%%%%%%%%%%%%%%%%%%%%%%%%%%%%%%%%%%%%%%%%%%%%%%%%%%%%%%%%%%%%%%%%%%%%%%%%%%%%%%%%
% Spectroscopic Data
%%%%%%%%%%%%%%%%%%%%%%%%%%%%%%%%%%%%%%%%%%%%%%%%%%%%%%%%%%%%%%%%%%%%%%%%%%%%%%%%%%%%%%%%%%%%%%%%%%%%%
\section{Cool Dwarf Benchmark Sample}\label{sec:benchmarks}
\subsection{Spectroscopic Data}\label{sec:data}
\begin{figure}
    \centering
    \includegraphics[width=\columnwidth, trim=0cm 0.8cm 0cm 0.8cm]{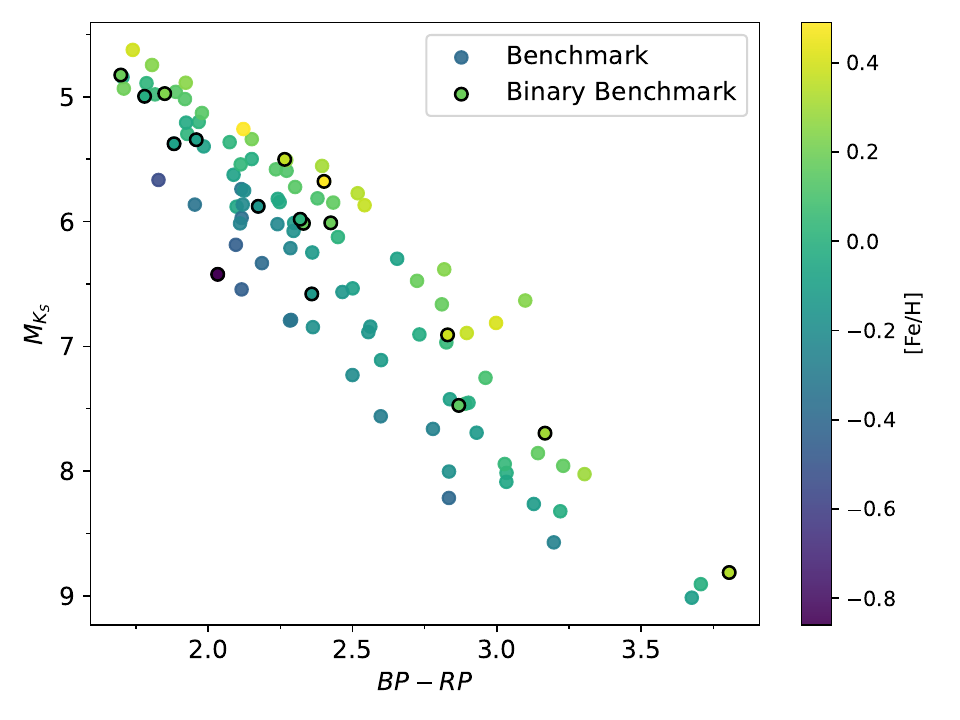}
    \caption{\textit{2MASS} $M_{K_S}$ versus \textit{Gaia} DR3 $(BP-RP)$ colour magnitude diagram for our 103 selected cool dwarf benchmarks, coloured according to their adopted [Fe/H]. The subsample of benchmarks with chemistry from an F/G/K binary companion are outlined.}
    \label{fig:cmd}
\end{figure}

Our data consists of low and medium resolution optical benchmark stellar spectra observed with the dual-camera WiFeS instrument \citep{dopita_wide_2007} on the ANU 2.3\,m Telescope as part of the spectroscopic surveys published as \citet{zerjal_spectroscopically_2021} and \citet{rains_characterization_2021}. All stars were observed with the B3000 and R7000 gratings using the RT480 beam splitter, yielding low-resolution blue spectra ($3\,500 \leq \lambda \leq 5\,700\,$\SI{}{\angstrom}, $\lambda/\Delta\lambda\sim3\,000$) and moderate resolution red spectra ($5\,400 \leq \lambda \leq 7\,000\,$\SI{}{\angstrom}, $\lambda/\Delta\lambda\sim7000$). This benchmark sample is described in Section \ref{sec:data:benchmarks}, and is plotted as a colour--magnitude diagram in Fig. \ref{fig:cmd}.

Our spectra were reduced using the standard PyWiFeS pipeline \citep{childress_pywifes:_2014} using the flux calibration approach of \citet{rains_characterization_2021}, but remain uncorrected for telluric absorption which we treat by simply masking the worst-affected wavelength regions. Radial velocities were determined by fitting against a template grid of MARCS synthetic spectra as described in \citet{zerjal_spectroscopically_2021}, and our spectra were subsequently shifted to the rest frame via linear interpolation using the \texttt{interp1d} function from \texttt{scipy}'s interpolate module in Python\footnote{All code for this project can be found at \url{https://github.com/adrains/plumage}, including a general-use script to run our fully-trained \textit{Cannon} model on non-WiFeS optical spectra for parameter determination.}.

%%%%%%%%%%%%%%%%%%%%%%%%%%%%%%%%%%%%%%%%%%%%%%%%%%%%%%%%%%%%%%%%%%%%%%%%%%%%%%%%%%%%%%%%%%%%%%%%%%%%%
\subsection{Benchmark Stellar Parameters}\label{sec:data:benchmarks}
\begin{table}
\centering
\caption{Summary of literature $T_{\rm eff}$, $\log g$, [Fe/H], and [Ti/Fe] sources referenced for our stellar benchmark sample, ordered from highest to lowest preference of label adoption. We list the median label uncertainty of each source and for our adopted set of labels as a whole, any inter-sample label systematics, benchmark stars \textit{with} labels from each sample, benchmark stars \textit{without} labels from each sample, and benchmark stars whose labels we \textit{adopt} from each sample. We use \citet{valenti_spectroscopic_2005} as our reference for computing [Fe/H] and [Ti/Fe] offsets which accounts for both reference abundance scale differences and other related systematics.}
\resizebox{\columnwidth}{!}{%
\begin{tabular}{ccccccc}
\hline
Label & Sample & Median $\sigma_{\rm label}$ & Offset & $N_{\rm with}$ & $N_{\rm without}$ & $N_{\rm adopted}$ \\
\hline
$T_{\rm eff}$ & All & 67\,K & - & 103 & 0 & 103 \\
& Interferometry & 25\,K & - & 17 & 86 & 17 \\
& Rains+21 & 67\,K & - & 103 & 0 & 86 \\
\hline
$\log g$ & All & 0.02\,dex & - & 103 & 0 & 103\\
& Rains+21 & 0.02\,dex & - & 103 & 0 & 103 \\
\hline
[Fe/H] & All & 0.08\,dex & - & 103 & 0 & 103\\
& Brewer+2016 & 0.01\,dex & +0.00\,dex & 7 & 96 & 7 \\
& Rice \& Brewer 2020 & 0.01\,dex & +0.00\,dex & 4 & 99 & 4 \\
& Valenti \& Fischer 2005 & 0.03\,dex & +0.00\,dex & 10 & 93 & 1 \\
& Montes+2018 & 0.04\,dex & +0.02\,dex & 15 & 88 & 4 \\
& Sousa+2008 & 0.03\,dex & - & - & - & 1 \\
& Mann+2015 & 0.08\,dex & +0.00\,dex & 75 & 28 & 69 \\
& Rojas-Ayala+2012 & 0.12\,dex & +0.01\,dex & 33 & 70 & 12 \\
& Other NIR & 0.08\,dex & - & - & - & 3 \\
& Photometric & 0.19\,dex & - & 98 & 5 & 2 \\
\hline
[Ti/Fe] & All & 0.04\,dex & - & 103 & 0 & 103 \\
& Brewer+2016 & 0.02\,dex & -0.02\,dex & 7 & 96 & 7 \\
& Rice \& Brewer 2020 & 0.04\,dex & +0.00\,dex & 4 & 99 & 4 \\
& Valenti \& Fischer 2005 & 0.05\,dex & +0.00\,dex & 10 & 93 & 1 \\
& Montes+2018 & 0.07\,dex & -0.03\,dex & 15 & 88 & 4 \\
& Adibekyan+2012 & 0.06\,dex & - & - & - & 1 \\
& This Work & 0.07\,dex & -0.03\,dex & 103 & 0 & 86 \\
\hline
\end{tabular}}
\label{tab:benchmark_sample_summary}
\end{table}

To train a data-driven model we require a set of stellar parameters to train on. Like previous studies, we adopt the three `core' stellar parameters of $T_{\rm eff}$, $\log g$, and [Fe/H], but distinguish ourselves by studying an additional chemical dimension in [Ti/Fe]. The primary motivation for selecting [Ti/Fe] as our abundance of choice to investigate is due to the strong expected signature of TiO on our optical spectra, something we expect to correlate with [Ti/Fe]\footnote{While not the point of our work here, from a Galactic Archaeology perspective [Ti/Fe] also traces [$\alpha$/Fe] at early times \citep{kobayashi_origin_2020}.}. Though we also expect strong optical signatures from C and O per \citet{veyette_physical_2016}, these elements have fewer absorption lines in the optical than Ti, and thus literature abundances sources from high--resolution spectroscopy are less prevalent. Finally, while we expect strong signatures from other oxides and hydrides on our spectra, we limit ourselves to two chemical dimensions in [Fe/H] and [Ti/Fe] for the purposes of this initial study with our relatively small sample size.

When selecting our benchmark sample of cool dwarfs, our objective was to use only those stars with fundamental---or fundamentally calibrated---stellar parameters\footnote{Where fundamental in this context refers to parameters derived as independently of models as possible, such as $T_{\rm eff}$ or $R_\star$ measured or benchmarked from interferometry rather than those derived purely from isochrone fitting based on physical stellar atmosphere and evolutionary models.}. A large factor motivating our decision to pursue a data-driven approach stems from the incomplete and physically inaccurate nature of current generation synthetic optical spectra used in more traditional analyses. Given our goal is to avoid, or even shed light on these limitations, and the accuracy of a data-driven model is only as good as its training sample, we must then be very selective.

Thus, our benchmark sample is composed of 103 cool dwarfs with stellar parameters from at least one of the following categories\footnote{While beyond the scope of our work here, a potential fourth category of benchmark or science target are stars in moving groups or open clusters. While nominally chemically homogeneous at the current precision of our \textit{Cannon} model (meaning that cool dwarf chemistry could be adopted from warmer cluster members), cluster chemical inhomogeneities have been observed when taking advantage of the extreme measurement precision offered by differential abundance analysis \citep[e.g.][]{liu_hyades_2016}---inhomogeneities which could plausibly be revealed using a data-driven model trained with an appropriate training sample.}:
\begin{enumerate}
    \item {[Fe/H]} or [Ti/Fe] from an F/G/K companion,
    \item {[Fe/H]} from empirical relations based on low resolution NIR spectra and calibrated to i),
    \item $T_{\rm eff}$ from interferometry,
\end{enumerate}
where 17 stars have [Fe/H] and [Ti/Fe] from a warmer binary companion, 93 have [Fe/H] from NIR empirical relations, and 17 stars have interferometric $T_{\rm eff}$. For those stars not in binary systems, we determine [Ti/Fe] based on chemodynamic trends present in \textit{Gaia} DR3 \citep{vallenari_gaia_2023} and GALAH DR3 \citep{buder_galah_2021} data. While small, we note that such a limited training sample has precedent in studies like \citet{behmard_data-driven_2019}, \citet{birky_temperatures_2020}, and \citet{maldonado_hades_2020}.

In addition to these parameter-specific requirements, we also impose two additional quality constraints on our sample. Firstly, to ensure we have a clean sample free from obvious unresolved binaries, we require benchmarks have a \textit{Gaia} DR3 Renormalised Unit Weight Error (RUWE) of $<1.4$, above which the single star astrometric fit is deemed poor by the \textit{Gaia} Consortium \citep[e.g.][]{lindegren_gaia_2021,belokurov_unresolved_2020}. Secondly, we reject binary benchmarks with inconsistent \textit{Gaia} DR3 kinematics between the primary and secondary components in order to better guarantee our gold standard chemical benchmarks are physically associated. 

The following sections describe our literature sources for each separate stellar parameter, as well as our adopted hierarchy between sources for those stars with more than a single source: Section \ref{sec:data:benchmarks:teff}: $T_{\rm eff}$, Section \ref{sec:data:benchmarks:logg}: $\log g$, Section \ref{sec:data:benchmarks:feh}: [Fe/H], Section \ref{sec:data:benchmarks:Ti_Fe_binary}: [Ti/Fe] from binaries, and Section \ref{sec:data:benchmarks:Ti_Fe_chemodynamic} [Ti/Fe] from empirical chemodynamic trends. Table \ref{tab:benchmark_sample_summary} serves as a summary of these subsections, with sources listed in order of preference when choosing which to adopt (included inter--sample systematics where measured).

\subsubsection{Stellar $T_{\rm eff}$}\label{sec:data:benchmarks:teff}
Interferometric temperature benchmarks form the cornerstone of our data-driven temperature scale, and thus we observed 17 stars from \citet{van_belle_directly_2009}, \citet{boyajian_stellar_2012-1}, \citet{von_braun_gj_2012}, \citet{von_braun_stellar_2014}, and \citet{rabus_discontinuity_2019} with a median literature $T_{\rm eff}$ uncertainty of $\pm 25\,$K. This left a decision on what temperatures to adopt for the remainder of our benchmark sample, specifically with how we handle systematics between different temperature scales or those stars without a previously reported value (mainly our binary benchmarks). As an example, the bulk of our NIR [Fe/H] benchmarks have $T_{\rm eff}$ from either \citet{rojas-ayala_metallicity_2012} or \citet{mann_how_2015}, with \citet{mann_how_2015} noting temperature systematics between the overlapping samples between the two studies at warmer $T_{\rm eff}$\footnote{Quoted as $28\pm14\,$K on average, but systematically higher at warmer $T_{\rm eff}$.}. Given this concern, we deemed the use of a single \textit{uniform} $T_{\rm eff}$ scale for our benchmark sample critical in order to have the best chance of investigating cool dwarf chemistry. To this end, for all non-interferometric benchmarks we adopt $T_{\rm eff}$ obtained via the fitting methodology of \citet[][]{rains_characterization_2021}---itself calibrated to our adopted interferometric scale. We describe this method below, and add our statistical uncertainties in quadrature with the median benchmark $T_{\rm eff}$ uncertainty for a final median uncertainty of $\pm 67\,$K.

\citet{rains_characterization_2021} undertook a benchmark-calibrated joint synthetic fit to spectroscopic and photometric data using flux calibrated WiFeS spectra and literature \textit{Gaia}/\textit{2MASS}/\textit{SkyMapper} photometry for their cool dwarf sample. As a necessity for model-based work with cool optical fluxes, they quantified model systematics by comparing synthetic MARCS spectra and photometry to observed spectra and integrated photometry from their benchmark sample of 136 cool dwarf benchmarks with $3,000 \lesssim T_{\rm eff} \lesssim 4,500\,$K. These systematics, parameterised as a function of \textit{Gaia} ($BP-RP$) colour, were used to correct synthetic photometry during fitting, with only the most reliable regions in the WiFeS R7000 spectral arm being included. $T_{\rm eff}$ was the principal output of their fit, with both $\log g$ and [Fe/H] fixed using empirical relations (the former from \citealt{mann_how_2015} and \citealt{mann_how_2019}, the latter developed in \citealt{rains_characterization_2021}) to avoid parameter degeneracies due to the complexity of cool dwarf fluxes. Reported temperatures were calibrated to a fundamental scale by correcting for temperature systematics observed between fits to the aforementioned benchmark sample, itself fundamentally calibrated to the interferometric $T_{\rm eff}$ scale.

\subsubsection{Stellar $\log g$}\label{sec:data:benchmarks:logg}
For stellar $\log g$, we also adopt the uniform values from \citet{rains_characterization_2021}. Due to model limitations and degeneracies, \citet{rains_characterization_2021} fixed $\log g$ when fitting for \teff and used a two step iterative process to determine the final gravity. Initial $M_\star$ and $R_\star$ values were obtained from the photometric $M_{K_S}$ band relations in \citet{mann_how_2019} and \citet{mann_how_2015} respectively, and were used to compute and fix $\log g$ for the initial fit. Following this initial fit, $R_\star$ was recalculated from the fitted $T_{\rm eff}$ and $f_{\rm bol}$ values via the Stefan Boltzmann relation, $\log g$ recalculated, and a final fit performed to give the adopted stellar $R_\star$ and $\log g$. It should be noted, however, that this process is almost entirely based on photometry, meaning that we are not sensitive to unresolved binarity or youth in the same way that spectroscopic techniques are. The median $\log g$ statistical uncertainty of our benchmark sample is $\pm 0.02\,$dex.

\subsubsection{Stellar [Fe/H]}\label{sec:data:benchmarks:feh}
The inability to recover [Fe/H] from optical cool dwarf spectra makes the selection of [Fe/H] benchmarks particularly crucial for any data-driven approach. The gold standard for cool dwarf metallicities continues to be those stars with a warmer companion of spectral type F/G/K from which traditional spectral analysis techniques like measuring equivalent widths or spectral synthesis are reliable. We adopt [Fe/H] from these stars where possible, sourced from \citet{valenti_spectroscopic_2005}, \citet{sousa_spectroscopic_2008}, \citet{brewer_spectral_2016}, \citet{montes_calibrating_2018}, and \citet{rice_stellar_2020}. These are our most precise [Fe/H] benchmarks, with $-0.86 < {\rm [Fe/H]} < 0.35$\footnote{Though clustered around Solar [Fe/H] $\pm{\sim}0.2\,$dex}, with a median literature uncertainty of $\pm 0.03\,$dex for the 17 such stars in our sample. 

We adopt \citet{valenti_spectroscopic_2005} as our reference when computing and correcting for [Fe/H] and [Ti/H] systematics (see the `offset' column in Table \ref{tab:benchmark_sample_summary}), though prioritise \citet{brewer_spectral_2016} and \citet{rice_stellar_2020}---both follow-up studies---for their higher precision. \citet{valenti_spectroscopic_2005} was chosen as our reference scale due to its large sample size and the fact that \citet{mann_prospecting_2013}---the original source for the [Fe/H] relations used in \citet{mann_how_2015}---also used it as their [Fe/H] reference point. Note that different sources in Table \ref{tab:benchmark_sample_summary} adopt different abundance reference points, either published reference abundance levels or their own solar-relative abundances, and this results in straightforward differences in [Fe/H] scales. However, there exists the possibility for other effective systematics present due to e.g. differences in the temperature scale, and these will not be apparent from a simple comparison between abundance scales---but are accounted for by our approach of identifying and removing systematics between samples. Finally, when computing offsets we do so using a larger crossmatched set of literature F/G/K--K/M binary benchmarks than we have spectra for here, with 256 primaries and 259 secondaries in total (noting that not all of these stars are in all publications).

Our largest sample of [Fe/H] comes from empirical relations built from [Fe/H] sensitive spectral regions in low-resolution NIR spectra based on these binary benchmarks. While there are many such relations in the literature, the bulk of our stars are drawn from just two of them. 69 of our stars have [Fe/H] from the work of \citet{mann_how_2015}, with $\sigma_{\rm [Fe/H]}=\pm0.08\,$dex. Another 12 are drawn from the work of \citet{rojas-ayala_metallicity_2012}, with $\sigma_{\rm [Fe/H]}=\pm0.18\,$dex. Only three stars have adopted [Fe/H] from other relations, with one star from \citet{terrien_near-infrared_2015} with $\sigma_{\rm [Fe/H]}=\pm0.07\,$dex, and another two from \citet{gaidos_trumpeting_2014} with $\sigma_{\rm [Fe/H]}=\pm0.1\,$dex.

Finally, for any star without [Fe/H], we adopt [Fe/H] from the photometric relation of \citet{rains_characterization_2021} with $\sigma_{\rm [Fe/H]}=\pm0.19\,$dex. This relation is only applicable to isolated single stars on the main sequence with reliable \textit{Gaia} parallaxes. In our case, only two stars---GJ 674 and GJ 832, both interferometric benchmarks---have a value from this relation. However, this ensures that all stars in our sample have an [Fe/H] value.

\subsubsection{Measured Stellar [Ti/Fe]}\label{sec:data:benchmarks:Ti_Fe_binary}

For those benchmark stars in F/G/K--K/M binary systems, we can adopt [Ti/Fe] abundances from the warmer primary where they it has previously been measured in the literature. Per Table \ref{tab:benchmark_sample_summary}, we adopt [Ti/Fe] for seven stars from \citet{brewer_spectral_2016}, four stars from \citet{rice_stellar_2020}, one star from \citet{valenti_spectroscopic_2005}, four stars from \citet{montes_calibrating_2018}, and one star from \citet{adibekyan_chemical_2012}. We note that \citet{brewer_spectral_2016} and \citet{rice_stellar_2020} are follow-up work to \citet{valenti_spectroscopic_2005}, and thus we preference them due to their higher [Ti/H] precision (while again adopting \citealt{valenti_spectroscopic_2005} as our adopted reference for [Ti/H] systematics), and for \citet{adibekyan_chemical_2012} we adopt the abundance derived from Ti I lines. Each of these works report abundances as [X/H], so we have calculated [Ti/Fe] using the adopted systematic-corrected [Fe/H] values (typically from the same literature source) and propagated the uncertainties accordingly, resulting in a median uncertainty of $\sigma_{\rm [Ti/Fe]}=\pm0.03\,$dex.

\subsubsection{Empirical Chemodynamic [Ti/Fe]}\label{sec:data:benchmarks:Ti_Fe_chemodynamic}

To assign [Ti/Fe] values for non-binary benchmarks, we make use of chemodynamic correlations in the Milky Way (MW) discs to ``map'' [Ti/Fe] values onto the benchmark stars. The chemical distinctness of the thick and thin discs in light elements (e.g. Mg, Ca, O, Si, Ti, or the $\alpha$-elements) across metallicities has become a well-accepted feature of our galaxy \citep[e.g. as observed early-on in high resolution studies of small samples and in the larger APOGEE survey,][]{nissen_two_2010, hayden_chemical_2015}. To map values of [Ti/Fe], we utilise the GALAH DR3 data release \citep{buder_galah_2021} to recover chemistry and the value added catalogue (VAC) of \citet{buder_galah_2022} for the stellar kinematics. Initial estimates of component membership (e.g. thick or thin disc) were made by comparing the energy (E) and $z$-component of the angular momentum (L$_{\mathrm{z}}$) of the benchmark stars to a subset of the GALAH DR3 sample. The GALAH subset was selected to only include dwarf stars with similar stellar parameters to the F/G/K primaries of our binary benchmark stars ($T_{\mathrm{eff}}>4500$\,K and $\mathrm{log}~g>3.0$) and to exclude stars with potentially unreliable chemistry (i.e. cool stars). The E, L$_{\mathrm{z}}$ values for the benchmarks were recovered assuming the \texttt{McMillan2017} 
\citep{mcmillan_mass_2017} approximation for the MW potential and assuming a solar radius of 8.21\,kpc and a circular velocity at the Sun of 233.1\,km\,s$^{-1}$. The local standard of rest (LSR) was selected to be in the same frame of reference as the GALAH VAC. That is, the Sun is set 25\,pc above the plane in keeping with \citet{juric_milky_2008} and has a total velocity of ($U, V, W$) = (11.1, 248.27, 7.25)\,km\,s$^{-1}$ in keeping with \citet{schonrich_local_2010}. 

\begin{figure}
    \centering
    \includegraphics[width=\columnwidth, trim=0cm 0.8cm 0cm 0.8cm]{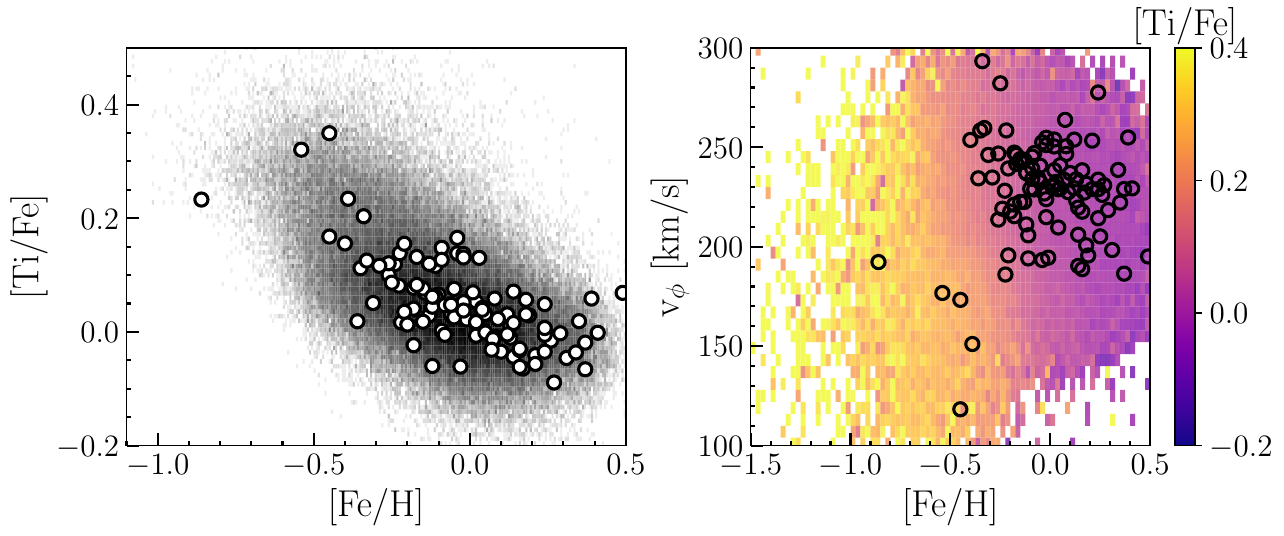}
    \caption{\textbf{Left:} The ``mapped'' values of [Ti/Fe] for the benchmarks (black circles) using their v$_{\phi}$, [Fe/H] values, and interpolation of the space presented in the right panel. The average values are calculated following 1\,000 draws sampling the errors associated with [Fe/H] and the input astrometric parameters from \textit{Gaia} DR3. The GALAH disc sample is plotted underneath. \textbf{Right:} The distribution of a subset of GALAH DR3 stars with large v$_{\phi}$ ($>100\,$km\,s$^{-1}$) selected to isolate the thick and thin MW discs. The subset is binned and coloured by the average [Ti/Fe] value of the bin (note the recovery of the [$\alpha$/Fe] bimodality). Benchmark stars are overplotted as open circles, where their v$_{\phi}$ values have been calculated under the LSR discussed in Section \ref{sec:data:benchmarks:Ti_Fe_chemodynamic}.}
    
    \label{fig:chemodynamic_fit}
\end{figure}

Unsurprisingly, all of the benchmark stars are found to be on nearly circular orbits, coincident with either the MW thick or thin disc in E, L$_{\mathrm{z}}$ space. In addition to calculating the E, L$_{\mathrm{z}}$ values for the benchmarks, we also calculated the v$_{\phi}$ values for the stars (the tangential velocity component in cylindrical coordinates) under the same assumption and orientation of the LSR. Following an exploration of various chemodynamic spaces, we found v$_{\phi}$ vs. [Fe/H] to isolate the [$\alpha$/Fe] bimodality \citep[associated with the thick and thin disks,][]{nissen_two_2010} the most cleanly. This is shown for a subset of GALAH stars in the left panel of Fig.~\ref{fig:chemodynamic_fit} where we have removed the bulk of the stellar halo by applying the cut, v$_{\phi}>100$\,km\,s$^{-1}$. The clean GALAH disc sample is binned into 100 bins in v$_{\phi}$, [Fe/H] and coloured by the the average value of [Ti/Fe] in each bin. The benchmark stars are overplotted in the right panel of Fig.~\ref{fig:chemodynamic_fit} to highlight their association with the thick (high [Ti/Fe], low v$_{\phi}$) and thin (low [Ti/Fe], high v$_{\phi}$) discs. 

To map a value of [Ti/Fe] to the benchmark stars, we performed a 2D interpolation of the clean GALAH disc sample in v$_{\phi}$, [Fe/H] space using the N--dimensional linear interpolator in \texttt{scipy} \citep{virtanen_scipy_2020}. To explore the uncertainty associated with the benchmark chemodynamics, we perform 1\,000 realisations sampling normal distributions in [Fe/H], and radial velocity (RV) and the multivariate distribution associated with the uncertainties in the astrometric parameters. We build the astrometric covariance matrix using the correlations and errors for the benchmarks within \textit{Gaia} DR3 \citep{lindegren_gaia_2021}. The v$_{\phi}$ values are then recalculated for each draw and the corresponding v$_{\phi}$, [Fe/H] values are used to infer a [Ti/Fe] for the star. The average recovered values of [Ti/Fe] for the benchmarks are shown in the left panel of Fig.~\ref{fig:chemodynamic_fit} as the open circles. They are overplotted on-top of the clean GALAH disc sample (shown as the black points). Note that the uncertainties are largest for the lowest metallicity benchmarks. This is likely a result of the metallicity distribution function of the thick disc being less well-sampled in our GALAH subset. This in itself it likely driven by our conservative cut in v$_{\phi}$ to remove the bulk of the MW halo (which presents a much more complex trend of light elements with metallicity).  

\begin{figure}
    \centering
    \includegraphics[width=\columnwidth, trim=0cm 0.8cm 0cm 0.8cm]{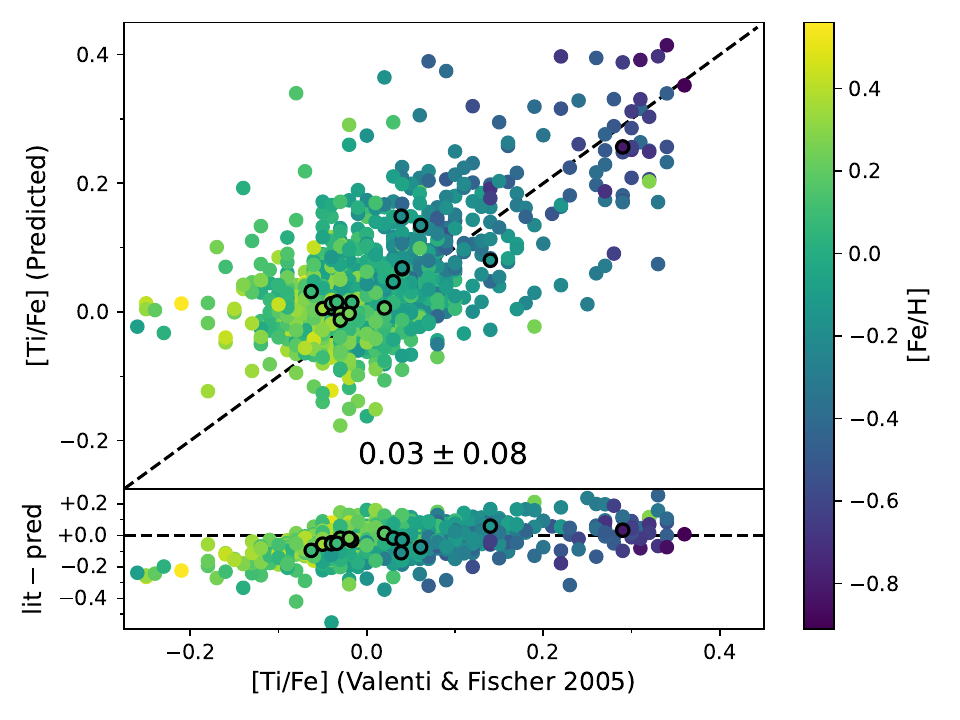}
    \caption{Literature [Ti/Fe] abundance for 1\,029 stars from \citet{valenti_spectroscopic_2005} versus [Ti/Fe] as predicted from \textit{Gaia} and GALAH chemodynamic trends as in Section \ref{sec:data:benchmarks:Ti_Fe_chemodynamic}, with the median and standard deviation of the residuals annotated. The black circled points correspond to the F/G/K primaries of our binary benchmarks. Note that we correct for the observed [Ti/Fe] systematic---that is we put our GALAH [Ti/Fe] values on the \citet{valenti_spectroscopic_2005} scale---and that the observed value of $\sigma_{\rm [Ti/Fe]}$ is comparable with the mapped [Ti/Fe] statistical uncertainties for our sample quoted in Table \ref{tab:benchmark_sample_summary} when taking into account [Fe/H] and kinematic uncertainties.}
    
    \label{fig:chemodynamic_recovery}
\end{figure}

To both validate our methodology and place GALAH and our benchmark stars on the same scale, we repeated the same exercise to predict [Ti/Fe] on the sample of dwarf stars from \citet{valenti_spectroscopic_2005}. Fig.~\ref{fig:chemodynamic_recovery} shows the comparison between our predicted values of [Ti/Fe] and those published in \citet{valenti_spectroscopic_2005} for stars with $\mathrm{[Fe/H]}\geq-1\,$dex. When considering the 1\,019 stars that meet our requirement in [Fe/H], we recover a median offset between the predicted and true values of [Ti/Fe] of $0.03\,$dex, with the residuals having a standard deviation of $0.08\,$dex. We correct for this offset in order to anchor our predicted [Ti/Fe] values to the \citet{valenti_spectroscopic_2005} scale, and take the uncertainty of our mapping to be $\sigma_{\rm [Ti/Fe]}=0.08\,$dex per the standard deviation.

%%%%%%%%%%%%%%%%%%%%%%%%%%%%%%%%%%%%%%%%%%%%%%%%%%%%%%%%%%%%%%%%%%%%%%%%%%%%%%%%%%%%%%%%%%%%%%%%%%%%%
% Methods
%%%%%%%%%%%%%%%%%%%%%%%%%%%%%%%%%%%%%%%%%%%%%%%%%%%%%%%%%%%%%%%%%%%%%%%%%%%%%%%%%%%%%%%%%%%%%%%%%%%%%
\section{The \textit{Cannon}}\label{sec:cannon}

Our adopted model is the \textit{Cannon}, first published in \citet{ness_cannon:_2015}. The \textit{Cannon} is a data-driven model trained upon a library of well-constrained benchmark stars and is able to learn a mapping between normalised rest frame stellar spectra and the corresponding set of stellar physical parameters. This mapping---essentially a form of dimensionality reduction between many pixels and few parameters---works by building a per-pixel model as a function of these parameters (also known as \textit{labels}), the simplest of which might be a single label model in terms of spectral type (as in \citealt{birky_temperatures_2020}), or a more complex---but more physically realistic---three parameter model in terms of e.g. $T_{\rm eff}$, $\log g$, and [Fe/H]. 

As with any data-driven or machine learning model, a given implementation of the \textit{Cannon} is only as accurate as its training sample of benchmark stars. When deployed in large spectroscopic surveys focusing primarily on warm stars (e.g. GALAH DR2, \citealt{buder_galah_2018-1}), the training sample can easily consist of thousands of stars all of which have a mostly complete, uniform, and reliable set of stellar labels. This is not possible for cool dwarfs, whose inherent faintness and spectroscopic complexity make it challenging to assemble a large and uniform set of benchmarks with a complete set of training labels. As an example, while the temperature of interferometric benchmarks and the chemistry of binary benchmarks are incredibly well-constrained, their other labels will be known to lower precision---or might even be outright missing. 

Given these challenges, for our work here we make use of two separate \textit{Cannon} models: a three label model in $T_{\rm eff}$, $\log g$, and [Fe/H]; and a four label model in $T_{\rm eff}$, $\log g$, [Fe/H], and [Ti/Fe]. We begin our methods section with a description of how we prepare our spectra in Section \ref{sec:cannon:preprocessing}, before giving an overview of our \textit{Cannon} models Section \ref{sec:cannon:model}, model training in Section \ref{sec:cannon:training}, and finally evaluating its performance in Section \ref{sec:cannon:results}.

%%%%%%%%%%%%%%%%%%%%%%%%%%%%%%%%%%%%%%%%%%%%%%%%%%%%%%%%%%%%%%%%%%%%%%%%%%%%%%%%%%%%%%%%%%%%%%%%%%%%%
\subsection{Spectra Normalisation}\label{sec:cannon:preprocessing}
Inherent in the use of the \textit{Cannon} is the assumption that the flux in each spectral pixel varies smoothly as a function of stellar labels, and that stars with identical labels will necessarily have near-identical fluxes. For this to be true, our spectra must be normalised and any pixels where this is not the case masked out and not modelled (e.g. wavelengths affected by stellar emission, telluric absorption, or detector artefacts). While it is possible to normalise optical spectra of warmer stars to the stellar continuum, this is not viable for cool dwarfs due to the intense molecular absorption present at such wavelengths. Fortunately, however, so long as the normalisation formalism is internally consistent, it is sufficient for input into the \textit{Cannon}. Our approach follows that initially implemented by \citet{ho_label_2017}, and later used by \citet{behmard_data-driven_2019} and \citet{galgano_fundamental_2020}, to normalise our spectra via a Gaussian smoothing process: 

\begin{equation}\label{eqn:flux}
    f(\lambda) = \frac{f_o}{\bar{f}}
\end{equation}
where $f$ is the Gaussian normalised flux associated with the rest-frame wavelength vector $\lambda$, $f_o$ is the observed flux calibrated WiFeS spectrum, and $\bar{f}$ is a Gaussian smoothing vector. Each term of this smoothing vector is defined as:
\begin{equation}
    \bar{f}(\lambda_n) = \frac{\sum_{i}{}\big[f_{o,i}\times\sigma_{o,i}^{-2}\times w_i(\lambda_n) \big]}{\sum_{i}{}\big[\sigma_{o,i}^{-2}\times w_i(\lambda_n) \big]}
\end{equation}
where $\bar{f}(\lambda_n)$ is the Gaussian smoothing term for rest-frame wavelength $\lambda_n$, $f_{o,i}$ is the observed flux at spectral pixel $i$, $\sigma_{o,i}$ is the observed flux uncertainty at spectral pixel $i$, and $w_i(\lambda_n)$ is the Gaussian weight for spectral pixel $i$ given $\lambda_n$. The Gaussian weight vector for spectral pixel $\lambda_n$ is computed as:
\begin{equation}
    w(\lambda_n) = e^{-\frac{(\lambda_n-\lambda)^2}{L^2}}
\end{equation}
 where $\lambda$ is our wavelength scale, and $L$ is the width of the Gaussian broadening in \SI{}{\angstrom}. We used $L=50\,$\SI{}{\angstrom}, noting that we find parameter recovery in cross-validation relatively insensitive for $25 < L < 100\,$\SI{}{\angstrom}.

In the end our data consists of 5\,024 spectral pixels with $4\,000 \leq \lambda \leq 7\,000$\SI{}{\angstrom}. We omit wavelengths with $\lambda < 4\,000\,$\SI{}{\angstrom} due to low SNR for our mid-M benchmarks, and mask out the hydrogen Balmer series and regions contaminated by telluric features. 

%%%%%%%%%%%%%%%%%%%%%%%%%%%%%%%%%%%%%%%%%%%%%%%%%%%%%%%%%%%%%%%%%%%%%%%%%%%%%%%%%%%%%%%%%%%%%%%%%%%%%
\subsection{\textit{Cannon} Model}\label{sec:cannon:model}

%%%%%%%%%%%%%%%%%%%%%%%%%%%%%%%%%%%%%%%

The traditional implementation of the \textit{Cannon} from \citet{ness_cannon:_2015} seeks to describe the stellar-parameter-dependent flux at a given spectral pixel with a model coefficient vector and an associated noise vector:
\begin{equation}
    f_{n,j} = \theta_j^T \cdot \ell_n + N_{n,j}
    %f_{n,\lambda} = \theta_{C,\lambda}^T \cdot \ell_{n,C} + N_\lambda
\end{equation}
where $f_{n,j}$ is the normalised model flux (from vector $f$ per Equation \ref{eqn:flux}) of star $n$ at spectral pixel $j$; $\theta_j$ is the model coefficient vector of length $N_{\rm coeff}$ describing spectral pixel $j$, $\ell_n$ is the label vector for star $n$ of length $N_{\rm coeff}$; and $N_j$ is a noise term for spectral pixel $j$, composed of the model intrinsic scatter $s_j$ and the observed flux uncertainty $\sigma_{n,j}$ added in quadrature as: 

\begin{equation}\label{eqn:noise}
    N_{n,j} = \sqrt{s_j^2 + \sigma_{n,j}^2}
\end{equation}
The full \textit{Cannon} model describing $\lambda$ spectral pixels thus has two unknown matrices which must be fit for: the coefficient vector $\theta$ of shape $\lambda \times N_{\rm coeff}$ and model scatter vector $s$ of length $\lambda$. To do so, we make use of our normalised observed flux and flux uncertainty vectors $f$ and $\sigma_f$ respectively with shapes $\lambda \times N_{\rm star}$, as well as the label vector $\ell$ of shape $N_{\rm star} \times N_{\rm coeff}$ constructed from the known stellar parameters of the training sample of $N_{\rm star}$ benchmark stars.

The \textit{Cannon} formalism is sufficiently generic that its model---that is the specifics of the coefficient and label vectors---can in principle be of any complexity and used to describe any number of labels, though typically a quadratic model in each label is considered sufficient (e.g. \citealt{ness_cannon:_2015}, \citealt{ho_label_2017}, \citealt{birky_temperatures_2020}). In the case of a three label quadratic model in $T_{\rm eff}$, $\log g$, and [Fe/H], this results in a 10 term $\ell_n$:
\begin{equation}
\begin{split}
    \ell_n  = \Big[& 1 + T_{\rm eff}^\prime + \log g^\prime + {\rm[Fe/H]}^\prime \\ & + \big(T_{\rm eff}^\prime \cdot \log g^\prime \big) + \big(T_{\rm eff}^\prime \cdot {\rm[Fe/H]}^\prime \big) + \big( \log g^\prime \cdot {\rm[Fe/H]}^\prime \big) \\ & + \big( T_{\rm eff}^\prime \big)^2  + \big( \log g^\prime \big)^2 + \big( {\rm[Fe/H]}^\prime \big)^2 \Big]
\end{split}
\end{equation}

or, alternatively, for a four term model in $T_{\rm eff}$, $\log g$, [Fe/H], and [Ti/Fe], a 15 term $\ell_n$:

\begin{equation}
\begin{split}
    \ell_n  = \Big[& 1 + T_{\rm eff}^\prime + \log g^\prime + {\rm[Fe/H]}^\prime + {\rm[Ti/Fe]}^\prime \\ & + \big(T_{\rm eff}^\prime \cdot \log g^\prime \big) + \big(T_{\rm eff}^\prime \cdot {\rm[Fe/H]}^\prime \big) + \big(T_{\rm eff}^\prime \cdot {\rm[Ti/Fe]}^\prime \big) \\ & + \big( \log g^\prime \cdot {\rm[Fe/H]}^\prime \big) + \big( \log g^\prime \cdot {\rm[Ti/Fe]}^\prime \big) + \big( {\rm[Fe/H]}^\prime \cdot {\rm[Ti/Fe]}^\prime \big)\\ & + \big( T_{\rm eff}^\prime \big)^2  + \big( \log g^\prime \big)^2 + \big( {\rm[Fe/H]}^\prime \big)^2 + \big( {\rm[Ti/Fe]}^\prime \big)^2 \Big]
\end{split}
\end{equation}

where $T_{\rm eff}^\prime$, $\log g^\prime$, ${\rm[Fe/H]}^\prime$, and ${\rm[Ti/Fe]}^\prime$ are the normalised stellar labels such that:
\begin{equation}
    \ell_{n,k}^\prime = \frac{\ell_{n,k} - \mu_{\ell_k}}{\sigma_{\ell_k}}
\end{equation}
where $\ell_{n,k}^\prime$ is the $k^{\rm th}$ normalised stellar label for star $n$, obtained from the stellar label $\ell_{n,k}$ and the mean ($\mu_{\ell_k}$) and standard deviations ($\sigma_{\ell_k}$) of the set of training labels $\ell_k$ such that the normalised labels each have zero-mean and unit-variance. 

Note that, for a quadratic model, the label vector $\ell_n$ contains three sets of terms: i) linear terms, including an offset term in the initial `1', ii) cross terms, and iii) quadratic terms. More generally, this allows the \textit{Cannon} to account for covariances between labels, in addition to the isolated contribution from each label. We discuss this in greater detail in Section \ref{sec:disussion:sensitivity}.
%%%%%%%%%%%%%%%%%%%%%%%%%%%%%%%%%%%%%%%

%%%%%%%%%%%%%%%%%%%%%%%%%%%%%%%%%%%%%%%%%%%%%%%%%%%%%%%%%%%%%%%%%%%%%%%%%%%%%%%%%%%%%%%%%%%%%%%%%%%%%
\subsection{Model Training}\label{sec:cannon:training}

\begin{figure*}%[hbt!]
    \centering
    \includegraphics[width=\textwidth, trim=0cm 1.25cm 0cm 0cm]{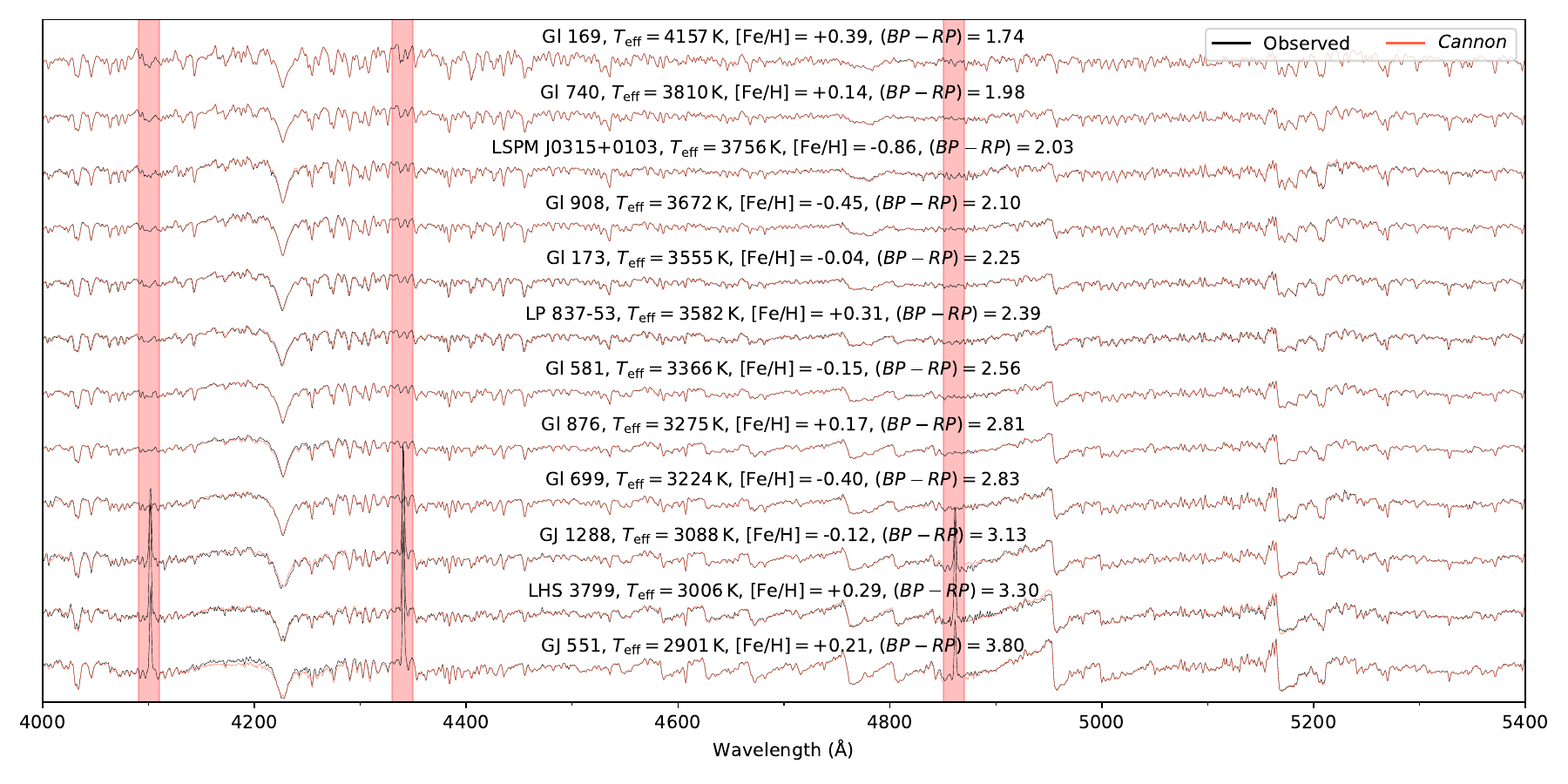}
    \caption{Spectra recovery for a representative set of benchmark stars with the WiFeS blue arm for $4\,000 < \lambda < 5\,400\,$\SI{}{\angstrom} at R${\sim}3\,000$, with observed spectra in black and \textit{Cannon} model spectra in red. We generate model spectra from our fully-trained three-label \textit{Cannon} model at the \textit{adopted} (rather than best-fit) benchmark labels. The vertical red bars correspond to H-$\beta$, H-$\gamma$, and H-$\delta$ from the hydrogen Balmer series which were masked out to avoid emission features. The stars are sorted by their \textit{Gaia} $(BP-RP)$ colour to show a smooth transition in spectral features across the parameter space considered.}
    \label{fig:blue_spectra_comp}
\end{figure*}

\begin{figure*}%[hbb!]
    \centering
    \includegraphics[width=\textwidth, trim=0cm 1.25cm 0cm 0cm]{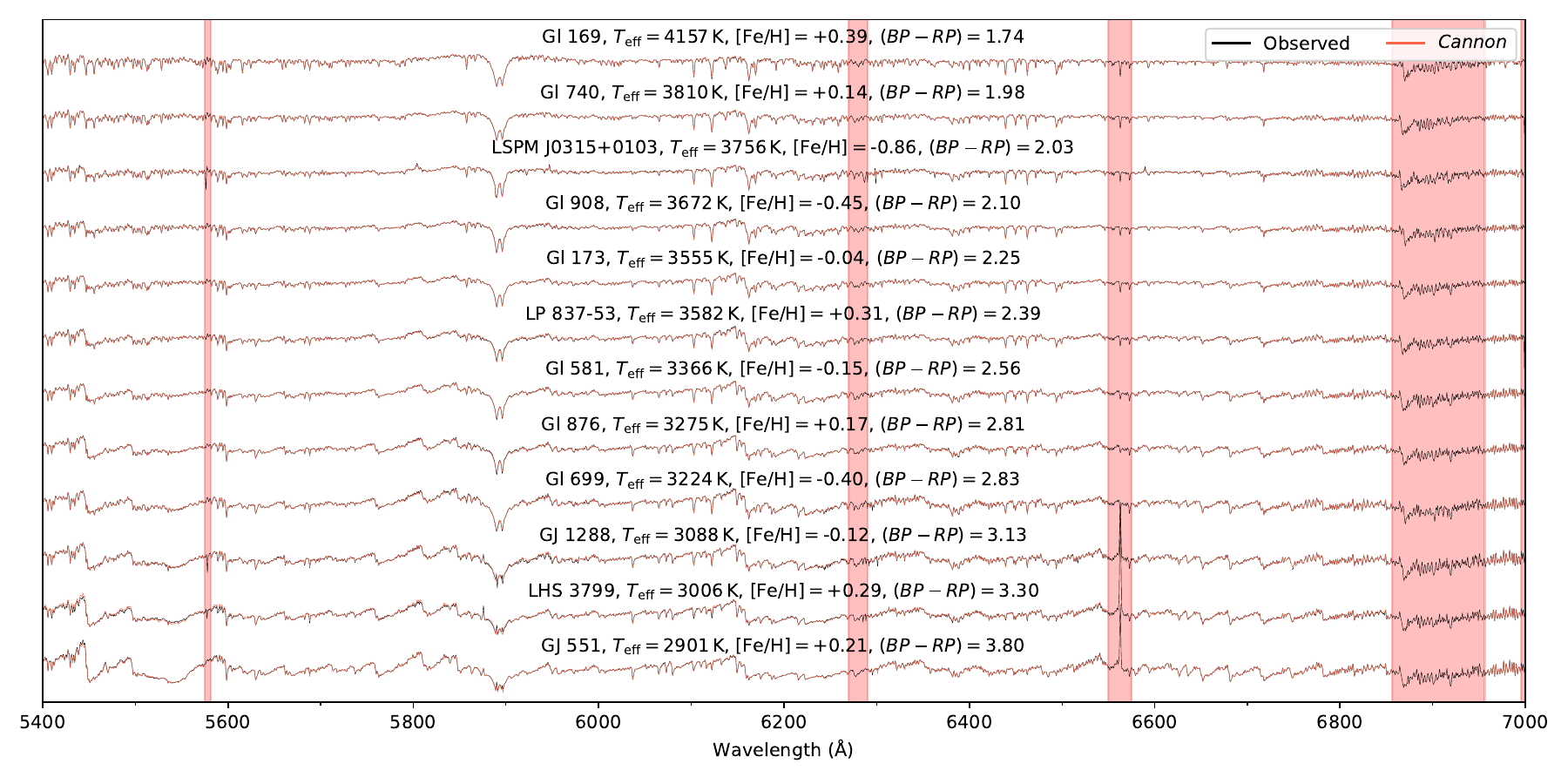}
    \caption{As Fig. \ref{fig:blue_spectra_comp}, but for WiFeS red arm spectra with $5\,400 < \lambda < 7\,000\,$\SI{}{\angstrom} at R${\sim}7\,000$. The vertical red bars (from left to right) correspond to a bad column on the WiFeS detector, atmospheric H$_2$O absorption, H-$\alpha$ from the hydrogen Balmer series, and O$_2$ telluric features, all of which were masked during modelling. }
    \label{fig:red_spectra_comp}
\end{figure*}

We implement our \textit{Cannon} model using \texttt{PyStan} v2.19.1.1 \citep{riddell_pystan_2021}, the Python wrapper for the probabilistic Stan programming language \citep{carpenter_stan_2017}. Training the \textit{Cannon} consists of optimising the model on a per-pixel basis for our two unknown vectors $\theta_\lambda$, our coefficient vector, and $s_\lambda$, the scatter per pixel using \texttt{PyStan}'s \texttt{optimizing} function. This is done via a log likelihood approach as follows: 
\begin{equation}
    \ln p\big(f_{n,\lambda}|\theta_\lambda^T,\ell_n, s_\lambda^2 \big) = - \frac{1}{2} \frac{\big[f_{n,\lambda} -  \theta_\lambda^T \cdot \ell_n\big]^2}{s_\lambda^2 + \sigma_{n,\lambda}^2} - \frac{1}{2} \ln\big(s_\lambda^2 + \sigma_{n,\lambda}^2 \big)
\end{equation}
where $\ln p\big(f_{n,\lambda}|\theta_\lambda^T,\ell_n, s_\lambda^2 \big)$ is the log likelihood.

We implemented a two-step training procedure to mitigate the impact of bad pixels on our model. Our initial model was trained and optimised on our benchmark set using only a global wavelength mask to exclude wavelength regions affected by telluric contamination or stellar emission. The resulting model was then used to predict fluxes for each benchmark star, with sigma clipping applied to exclude (via high inverse-variances during training) any pixel $6\sigma$ discrepant from the model fluxes. The final adopted model is then trained using a combination of the original global wavelength mask and the per-star bad pixel mask. Put another way, we do not directly adopt a per-star bad pixel mask as output from the WiFeS pipeline, but instead create one with reference to our initial \textit{Cannon} model.

Our three and four label models take of order ${\sim}1.5$ and ${\sim}2.5$ minutes to train respectively on an M1 Macbook Pro in serial for a single model without cross-validation. The spectral recovery for a representative benchmark sample with the blue and red arms of WiFeS respectively can be seen in Fig. \ref{fig:blue_spectra_comp} and \ref{fig:red_spectra_comp}, as generated from our fully--trained three label model\footnote{Where we use our three---rather than four---label model for ease of use when working with our existing MARCS grid and to avoid complexities that would arise when interpolating in [Ti/Fe].}. While spectral recovery struggles a little more for the bluest wavelengths of our coolest stars, we deem this primarily due to the lower SNR at these wavelengths and more sparse sampling of the parameter space at these temperatures. 

This recovery is particularly impressive when considering the deviations observed in \citet{rains_characterization_2021} between MARCS and Bt-Settl \citep{allard_model_2011} synthetic spectra versus the same flux normalised spectra we train our \textit{Cannon} model on here. In their Section 4.1, in particular fig. 4 and 5, they discuss $2-10$\,\% flux differences between synthetic and observed spectra for several optical bands---deviations large enough to be quite obvious to the eye. We discuss these differences more quantitatively in Section \ref{sec:discussion:marcs_comparison}

%%%%%%%%%%%%%%%%%%%%%%%%%%%%%%%%%%%%%%%%%%%%%%%%%%%%%%%%%%%%%%%%%%%%%%%%%%%%%%%%%%%%%%%%%%%%%%%%%%%%%
% Results
%%%%%%%%%%%%%%%%%%%%%%%%%%%%%%%%%%%%%%%%%%%%%%%%%%%%%%%%%%%%%%%%%%%%%%%%%%%%%%%%%%%%%%%%%%%%%%%%%%%%%
\subsection{Model Validation and Label Recovery Performance}\label{sec:cannon:results}
The accuracy of a data-driven or machine learning model can only be truly determined by testing the model on \textit{unseen} data---i.e. data the model was not trained upon. Ideally one would have a large enough sample to initially partition it into separate training and test sets without compromising on model accuracy. With only 103 benchmarks, however, this is not feasible for our work here, so we instead opt for a leave-one-out cross validation approach. Under this paradigm, we train $N$ different \textit{Cannon} models on $N$ different sets of $N-1$ benchmark stars, testing each model on the $N^{\rm th}$ benchmark left out of the model. Our final reported label recovery accuracy thus consists of the medians and standard deviations of the aggregate `left out' sample across all models. Labels are fit using the \texttt{curve\_fit} function from \texttt{scipy} given the coefficient and scatter arrays $\theta$ and $s_\lambda$ from a fully trained \textit{Cannon} model as described in Section \ref{sec:cannon:model}.

\begin{figure*}
    \centering
    \includegraphics[width=0.9\textwidth, trim=0cm 1.0cm 0cm 0cm]{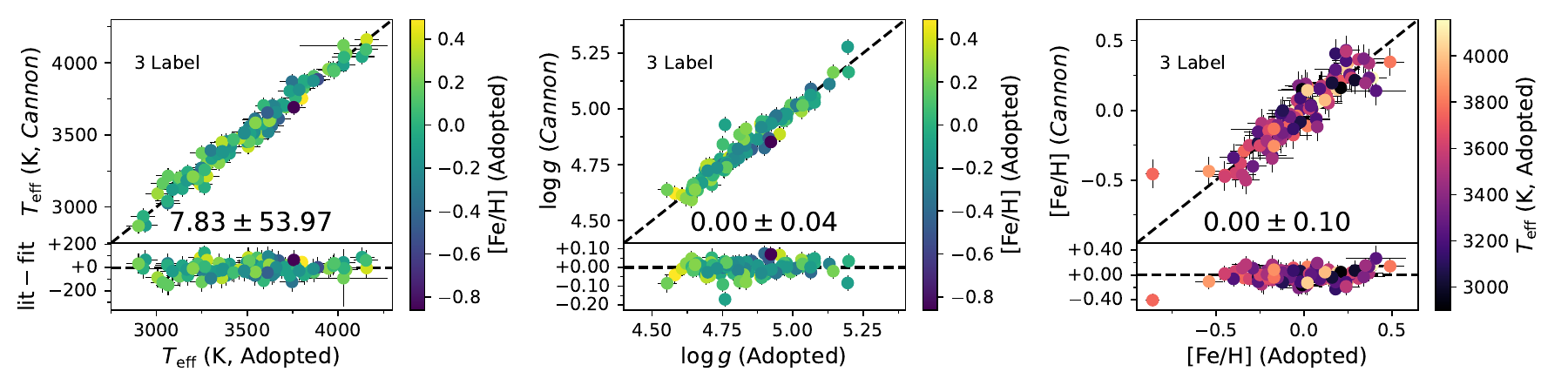}
    \includegraphics[width=0.9\textwidth, trim=0cm 1.0cm 0cm 0cm]{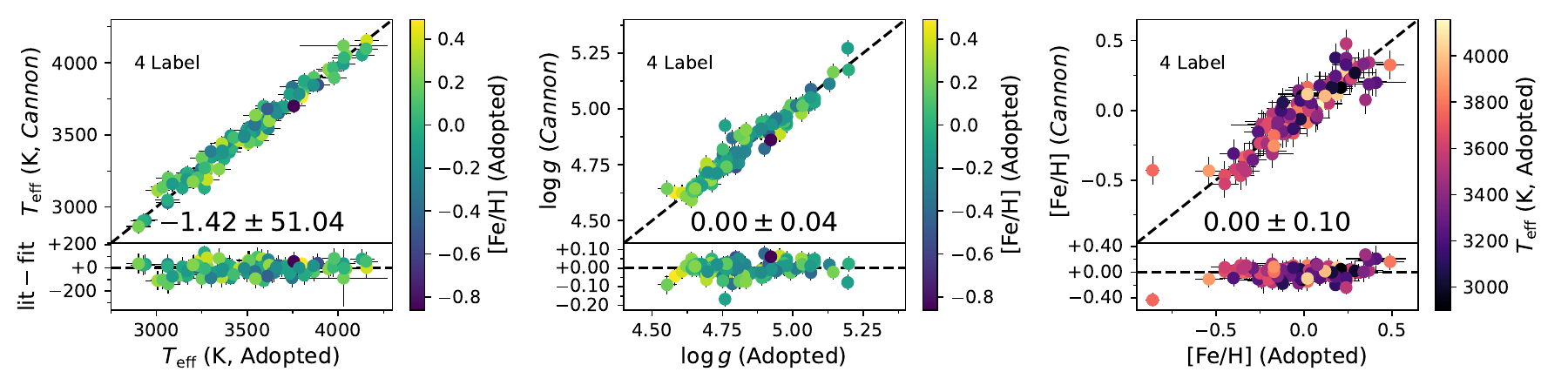}
    \caption{Leave-one-out cross validation performance for recovery of \textit{adopted} labels (per Table \ref{tab:benchmark_sample_summary}) for \textbf{Top:} 3 label (\teff, $\log g$, [Fe/H]) and \textbf{Bottom:} 4 label (\teff, $\log g$, [Fe/H], [Ti/Fe]) \textit{Cannon} models respectively. Each panel shows the median and standard deviation of the residuals computed between the adopted benchmark values and \textit{Cannon} predicted equivalents, where $\sigma_{\rm resid}$ is added in quadrature with the \textit{Cannon} statistical uncertainties to give our adopted label uncertainties.}
    \label{fig:cross_validation_bulk}
\end{figure*}

\begin{figure*}
    \centering
    \includegraphics[width=\textwidth, trim=0cm 1.0cm 0cm 0cm]{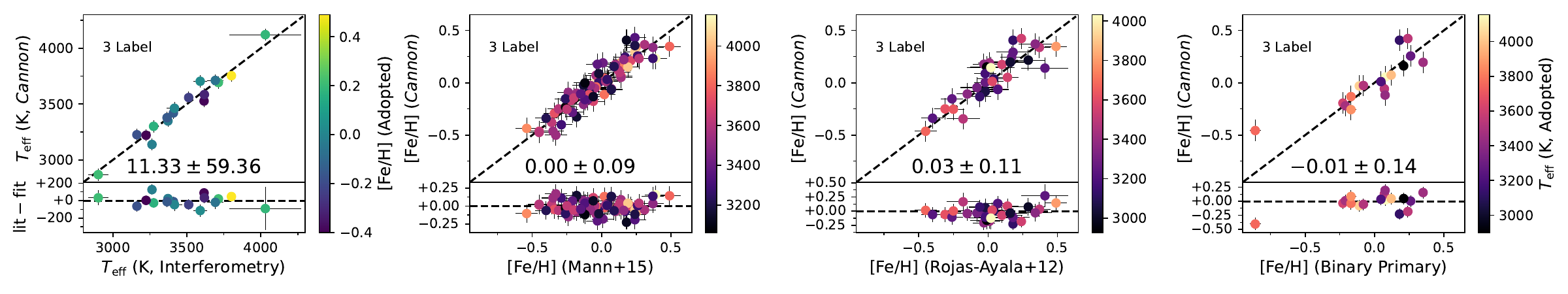}
    \includegraphics[width=\textwidth, trim=0cm 1.0cm 0cm 0cm]{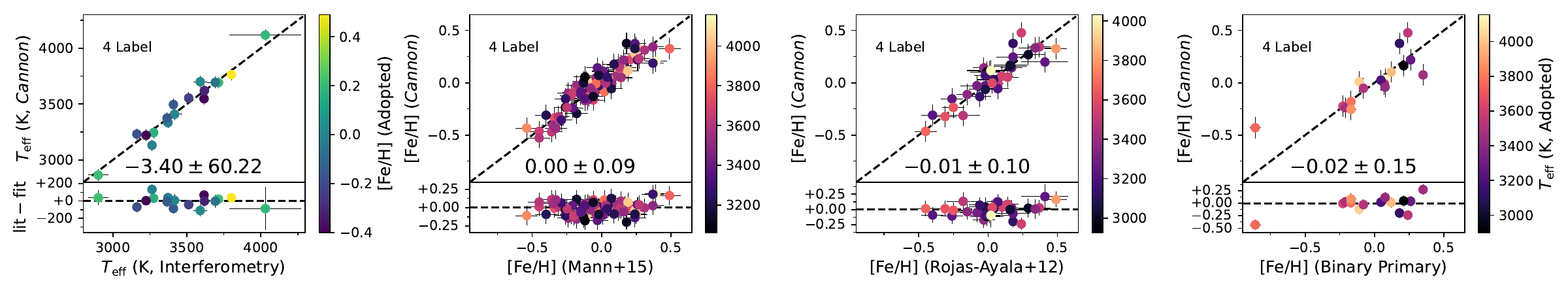}
    \caption{Leave-one-out cross validation performance for label recovery of literature parameter sources for \textbf{Top:} 3 label (\teff, $\log g$, [Fe/H]) and \textbf{Bottom:} 4 label (\teff, $\log g$, [Fe/H], [Ti/Fe]) \textit{Cannon} models respectively. Note that benchmark stars with [Fe/H] values published in multiple catalogues (e.g. both \citealt{rojas-ayala_metallicity_2012} and \citealt{mann_how_2015}) are plotted in multiple panels with their own colour--bars---not just the panel for the source we formally adopted. Each panel shows the median and standard deviation of the residuals computed between the adopted benchmark values and \textit{Cannon} predicted equivalents.}
    \label{fig:cross_validation_ps}
\end{figure*}

\begin{figure}
    \centering
    \includegraphics[width=0.75\columnwidth, trim=0cm 0.5cm 0cm 0.5cm]{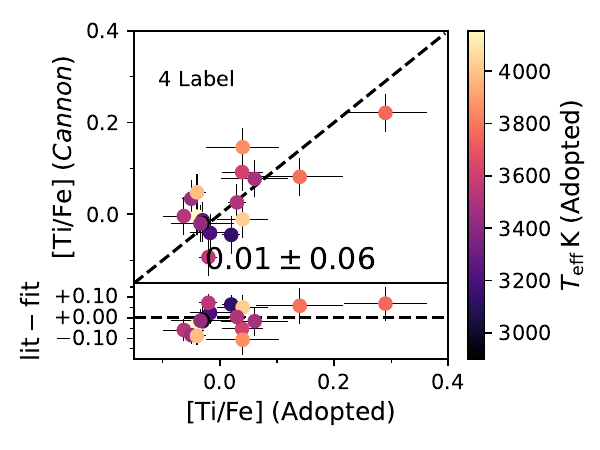}
    \caption{[Ti/Fe] leave-one-out cross validation performance for our binary benchmarks using our 4 label (\teff, $\log g$, [Fe/H], [Ti/Fe]) \textit{Cannon} model.}
    \label{fig:cross_validation_abundance}
\end{figure}

We plot our label recovery performance in leave-one-out cross validation for $T_{\rm eff}$, $\log g$, and [Fe/H] in Fig. \ref{fig:cross_validation_bulk} for both our three-- and four--label models. Similarly, Fig. \ref{fig:cross_validation_ps} illustrates our stellar parameter recovery as a function of label source: $T_{\rm eff}$ from interferometry, [Fe/H] from \citet{mann_how_2015}, [Fe/H] from \citet{rojas-ayala_metallicity_2012}, and [Fe/H] from F/G/K binary companions---again for our three and four parameter models respectively. Finally, Fig. \ref{fig:cross_validation_abundance} shows [Ti/Fe] recovery for our four--label model. Table \ref{tab:benchmark_parameters} lists the labels inferred from our fully--trained four label model in \teff, $\log g$, [Fe/H], and [Ti/Fe] for our benchmark, noting that while these values are from the model trained on all 103 stars, our uncertainties are derived from the leave-one-out cross validation performance standard deviations added in quadrature with the statistical uncertainties. For our four--label model, these statistical uncertainties are rather small with means 1.3\,K, 0.001\,dex, 0.004\,dex, and 0.001\,dex in $T_{\rm eff}$, $\log g$, [Fe/H], and [Ti/Fe] respectively, meaning that our reported errors are based primarily on how well we recover our adopted set of literature benchmark labels. We adopt (and correct for) $T_{\rm eff}$, $\log g$, [Fe/H], and [Ti/Fe] systematics and uncertainties of $-1\pm51\,$K, $0.00\pm0.04\,$dex and $0.00\pm0.10\,$dex, and $0.01\pm0.06\,$dex respectively. We note that these uncertainties purely refer to how well our recovered parameters \textit{reproduce} the adopted benchmark $T_{\rm eff}$, $\log g$, [Fe/H], and [Ti/Fe] scales---i.e. the quality of our label transfer using the \textit{Cannon}\footnote{A contributing factor for any systematics in label recovery is that the traditional \textit{Cannon} model does not internally consider label uncertainties. If our \textit{Cannon} model did properly model the uncertainties from benchmarks sourced from separate catalogues with varying label precisions we might expect any bias to be more solely a function of random noise.}. It is an altogether different task---and beyond the scope of our work here---to \textit{refine} these benchmark scales, with studies comparing e.g. physically realistic $T_{\rm eff}$ \citep{tayar_guide_2022} or abundance uncertainties (via differential analysis of Solar twins, e.g. \citealt{ramirez_solar_2014}) indicating that these uncertainties---including the adopted literature values for our benchmarks---are likely underestimates. 

Overall our three and four label models have similar label recovery performance (as distinct from \textit{spectral} recovery), with Fig. \ref{fig:cross_validation_bulk} showing the most significant difference between the two models being a reduced $T_{\rm eff}$ systematic ($+7.83\,$K to $-1.42\,$K) and scatter ($\pm53.97\,$K to $\pm51.04\,$K) for the four label model. Given that one of the main indicators of $T_{\rm eff}$ in cool dwarf atmospheres are the TiO bandheads---the characteristic `sawtooth' pattern in cool optical spectra and the traditional indicator of M dwarf spectral types---this improvement is consistent with expectations given the extra constraints provided by modelling [Ti/Fe]. By contrast, however, there are no similar improvements to $\log g$ and [Fe/H] recovery when using the four label model, and we hypothesise that there are three factors at play here. The first of which is that our three label model \textit{already} recovers $\log g$ and [Fe/H] at or nearly at the precision of the benchmark sample itself, meaning that further improvements would likely be marginal even in the best case. The second is that these parameters are less acutely sensitive to [Ti/Fe] or TiO absorption than $T_{\rm eff}$ is, or at least have constraints from unrelated spectral features. For example, other molecules like CaH are strongly sensitive to $\log g$ and have long been used as a discriminator between traditional stellar luminosity classes like subdwarf, dwarf, and giant \citep[e.g.][]{ohman_red_1936, jones_new_1973, mould_old_1978, kirkpatrick_standard_1991, mann_they_2012}. Thirdly is the effect of small number statistics, as our models are trained on only 103 benchmarks the influence of outliers (discussed more in subsequent paragraphs) is more significant when computing the scatter from the standard deviation of the residuals---an effect which might hide slight improvements in label recovery. When considering Fig. \ref{fig:cross_validation_ps}, which separates out label recovery for different literature sources, we are especially hesitant to draw firm conclusions about the differences between the two models given the smaller sample sizes---something especially acute for the binary sample---and lack of a \textit{Cannon} model which models label uncertainties as we can only expect label recovery at the level of the median $\sigma_{\rm [Fe/H]}$ of our benchmark sample. Nonetheless, our results are consistent with the expectation that a model constraining [Ti/Fe] would be better able to recover optical cool dwarf parameters given the significant influence of TiO on optical spectra.

NLTT 10349 (\textit{Gaia} DR3 3266980243936341248), our most metal poor star with [Fe/H]$=-0.86\pm0.04$, proves a consistent outlier in cross validation due to its uniqueness in our small benchmark sample, being roughly ${\sim}3.7\sigma$ from our sample's mean [Fe/H] (versus ${\sim}2.3\sigma$ from the mean value for the 2nd most metal poor star). Our model's inability to accurately recover its [Fe/H] in cross validation is a reflection on our sample size, rather than for instance a breakdown in the behaviour of cool dwarf spectra at low [Fe/H], leading to our conclusion that the \textit{Cannon} is unable to accurately extrapolate far beyond the label values of the benchmarks used to initially train it. 

Our $T_{\rm eff}$ recovery for both models is entirely consistent with the median $T_{\rm eff}$ precision of our sample ($\pm67\,$K vs our $\pm51\,$K here). There are similar systematics observed between the bulk and interferometric samples, which points to our adopted $T_{\rm eff}$ scale being correctly calibrated to a fundamental scale, despite the challenges noted in \citet{rains_characterization_2021} with their interferometric sample having saturated \textit{2MASS} photometry.

Our $\log g$ recovery is almost--consistent to within literature uncertainties ($\pm0.02\,$dex vs our $\pm0.04\,$dex here), noting that our gravities are mostly from photometric $M_{K_S}$--$M_\odot$ relations, and should be reliable for main sequence stars (but less accurate for unresolved binaries or stars still contracting to the main sequence). Assuming the \textit{Cannon} has successfully learnt how to identify gravity through gravity sensitive spectroscopic features, observed outliers in $\log g$ could be unresolved binaries as the \textit{Cannon} predicting higher gravities is consistent with underpredicted gravities from blended photometry. While cursory inspection of the spectra of these stars did not reveal any spectroscopic binaries, this hypothesis was validated once we made a RUWE cut upon the release of \textit{Gaia} DR3, which removed most of the targets (i.e. observed benchmarks which no longer pass the quality cuts to appear in our work here) with aberrant $\log g$ values, improving the precision of our $\log g$ recovery from an initial $\pm0.06\,$dex to the reported $\pm0.04\,$dex (which itself drops to $\pm0.03\,$dex were we to exclude the two most aberrant stars in the sample). 

Another alternative to unresolved binaries is that the \textit{Cannon} could be giving higher gravities to young stars still contracting to the main sequence as it has not been trained on such a sample. We note that \citet{behmard_data-driven_2019} included a $v \sin i$ dimension to their \textit{Cannon} model, but we are far less sensitive to rapid rotation with our $R{\sim}7\,000$ spectra versus their $R{\sim}60\,000$ resolution spectra. Either of these physical explanations would serve to increase the scatter on our primarily photometric $\log g$ values, but regardless of the physical origin we have flagged those remaining benchmarks with fitted $\log g$ aberrant by $>0.075$\,dex in Table \ref{tab:benchmark_parameters} using $\dagger$.

The primary motivation of our work here was to study the chemistry of our cool dwarf sample from \textit{optical} spectra, something \citet{rains_characterization_2021} was unable to accomplish previously using a model-based approach using the same spectra as we do here. Instead they developed a new photometric [Fe/H] relation precise to $\pm0.19\,$dex applicable to isolated main sequence stars. When compared this baseline, the ability of our \textit{Cannon} model to recover [Fe/H] to $\pm0.10$\,dex represents a significant improvement, even more so given our successful recovery of [Ti/Fe] to within benchmark uncertainties (corresponding to $\pm0.06$\,dex), something we discuss in Section \ref{sec:discussion}. Of note is that we achieve this [Ti/Fe] recovery with only 17/103 of our sample having measured [Ti/Fe] abundances, and the remaining 86/103 stars having [Ti/Fe] predictions informed by Galactic chemo-dynamic trends. This clearly demonstrates both a) the strength of the [Ti/Fe] signal in optical spectra, and b) the ability of the \textit{Cannon} to learn strong features, and points to greater precisions being possible with larger benchmark samples. When considering our 4 label model, we note that our [Fe/H] recovery is nearly consistent with the uncertainties of the \citet{mann_how_2015} sample (our largest single source of [Fe/H], $\pm0.08\,$dex vs our $\pm0.09\,$dex), and better than the quoted uncertainties for the \citet{rojas-ayala_metallicity_2012} sample (our second largest source of [Fe/H], $\pm0.12\,$dex vs our $\pm0.10\,$dex) suggesting that their uncertainties are overestimated.

%%%%%%%%%%%%%%%%%%%%%%%%%%%%%%%%%%%%%%%%%%%%%%%%%%%%%%%%%%%%%%%%%%%%%%%%%%%%%%%%%%%%%%%%%%%%%%%%%%%%%
% Influence of Stellar Abundances of Model Spectra
%%%%%%%%%%%%%%%%%%%%%%%%%%%%%%%%%%%%%%%%%%%%%%%%%%%%%%%%%%%%%%%%%%%%%%%%%%%%%%%%%%%%%%%%%%%%%%%%%%%%%
\section{Cool Dwarf Physical Model Spectra \& [X/F\lowercase{e}]}\label{sec:marcs_abundances}
%trim=0cm 1.25cm 0cm 0cm

\begin{figure*}
    \begin{turn}{-90}
    \begin{minipage}{\textheight}
    \centering
    \includegraphics[height=0.65\textheight]{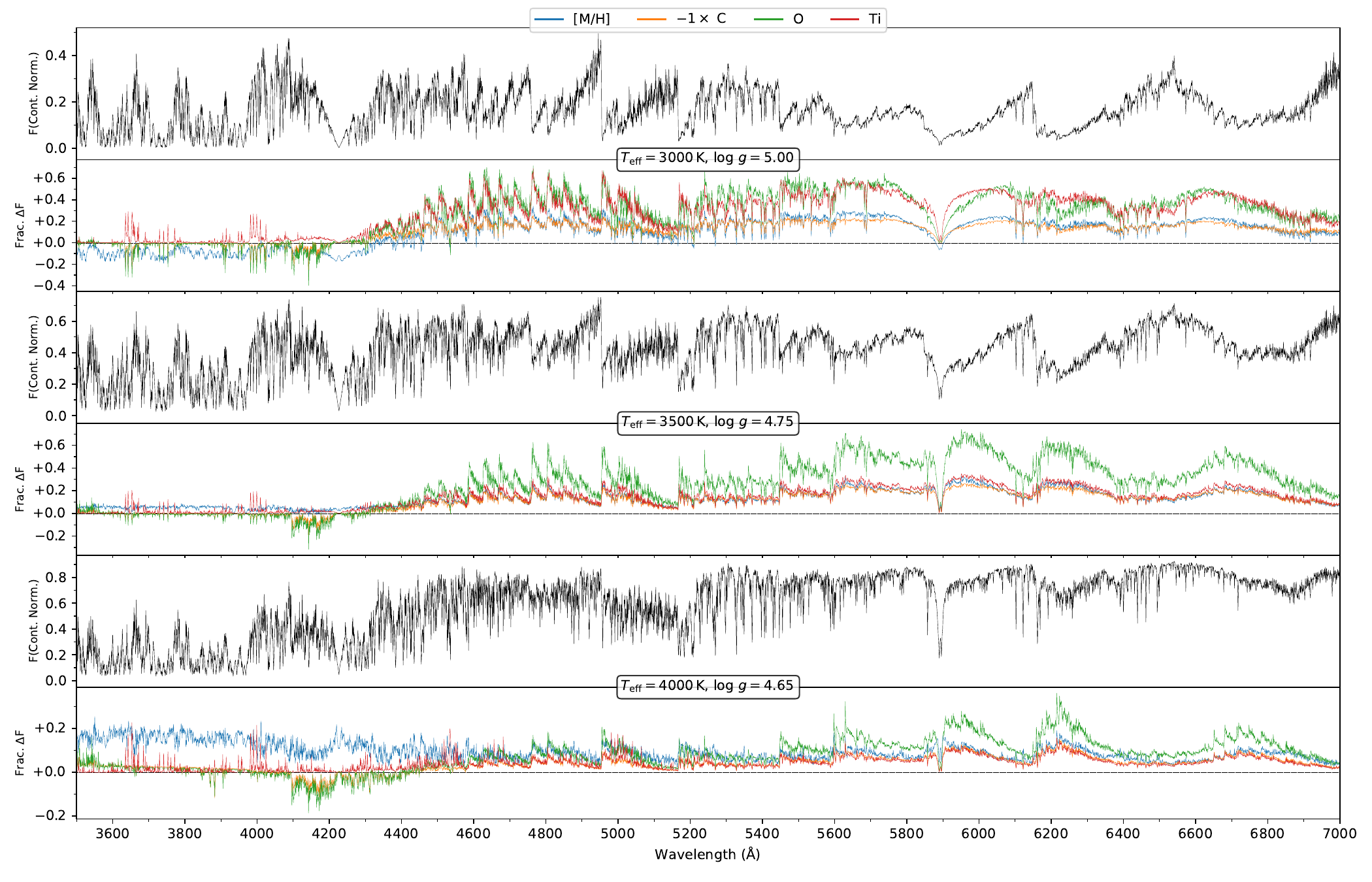}
    \caption{Sensitivity of cool dwarf MARCS model fluxes to variations in abundances for elements present in dominant molecular absorbers (C, O, and Ti) over the WiFeS wavelength range, as demonstrated with three different sets of solar [M/H] spectra (from top to bottom) with $T_{\rm eff}$ and $\log g$: $3\,000\,$K and $5.0$; $3\,500\,$K and $4.75$; $4\,000\,$K and $4.65$. The upper panel of each plot is the continuum normalised synthetic MARCS model spectra (model continuum level at 1.0) at Solar [M/H] and a Solar abundance pattern. The bottom panel shows the fractional change in continuum normalised flux when changing a single abundance from $+0.1$ to $-0.1\,$dex of the Solar value, as well as the bulk metallicity perturbed by the same amount for comparison. Note that the flux change from C has the opposite sign to O and Ti and as such has been multiplied by $-1$ for better comparison, and that the y-axis scale is different for each panel.}
    \label{fig:marcs_sensitivity_dominant_absorbers}
    \end{minipage}
    \end{turn}
\end{figure*}

\begin{figure*}
    \begin{turn}{-90}
    \begin{minipage}{\textheight}
    \centering
    \includegraphics[height=0.65\textheight]{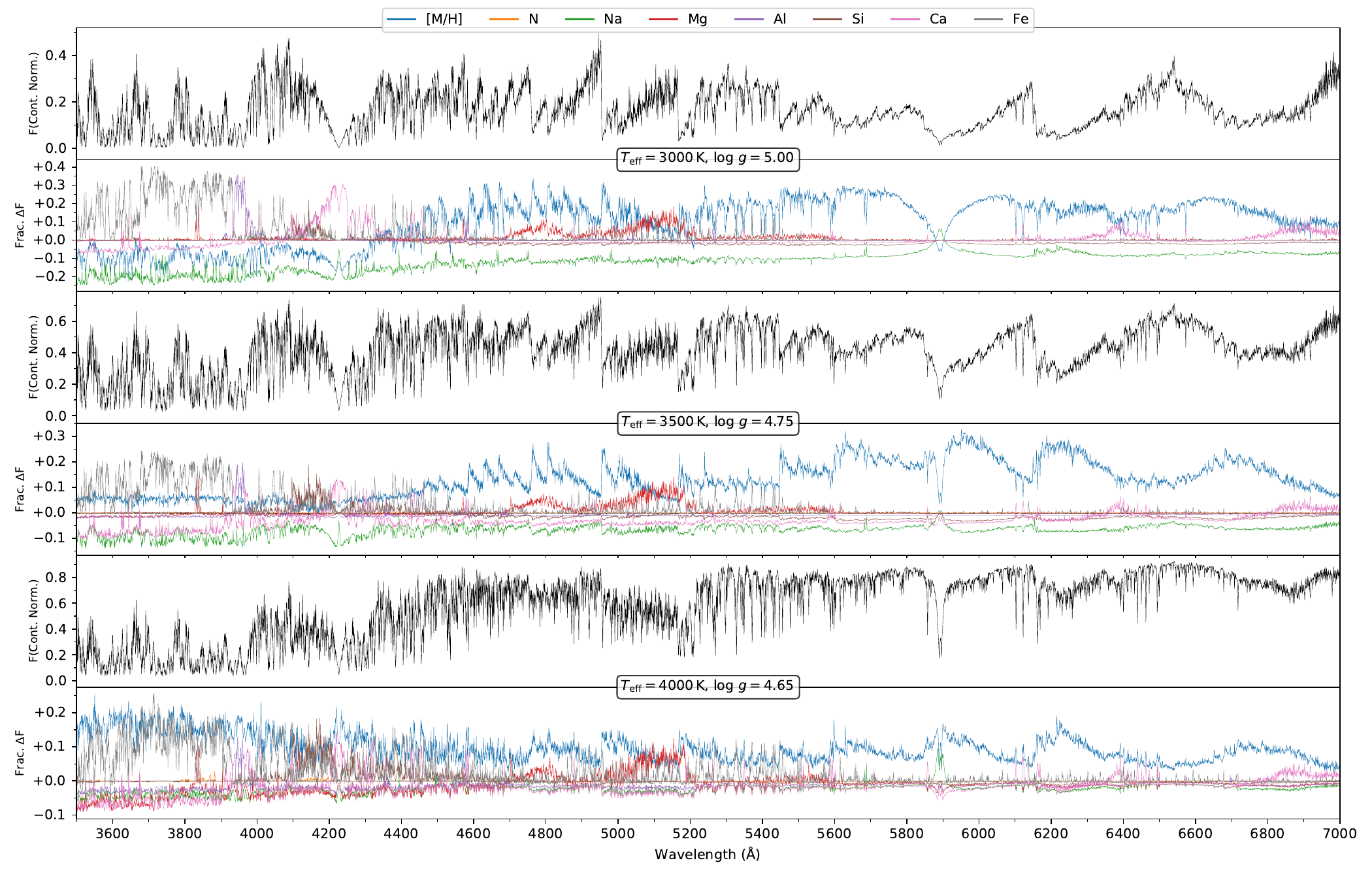}
    \caption{As Fig. \ref{fig:marcs_sensitivity_dominant_absorbers}, but for elements present in non-dominant molecular absorbers (N, Na, Mg, Al, Si, Ca, and Fe).}
    \label{fig:marcs_sensitivity_minor_absorbers}
    \end{minipage}
    \end{turn}
\end{figure*}

To complement our discussion of data-data-driven flux recovery and cool dwarf chemistry, we now turn to a grid of physical models for comparison. More specifically, in this section we set out to study the impact of [X/Fe] on model cool dwarf continuum normalised spectra for the wavelength range covered by our WiFeS spectra.

\subsection{Sensitivity of MARCS Pseudocontinuum to [X/Fe]}

To better understand the physics of cool dwarf atmospheres, as well as guide our interpretation of the results and performance of our \textit{Cannon} model, here we conduct a pilot investigation into the influence of atomic abundances on cool dwarf spectra using a bespoke grid of MARCS spectra. As in \citet{nordlander_lowest_2019}, our 1D LTE MARCS grid\footnote{While 3D models are in principle possible, \citet{ludwig_energy_2006} demonstrates that for the parameter space considered here 3D structures are very similar to their 1D equivalents---staying close to radiative equilibrium in the optically thin regions due to the convective velocities being too small to drive substantial deviations away from hydrostatic equilibrium. Given this---in addition to the substantial computation requirements of 3D models---we limit ourselves to standard 1D models for our work here.} was computed using the TURBOSPECTRUM code (v15.1; \citealt{alvarez_near-infrared_1998, plez_turbospectrum_2012}) and MARCS model atmospheres \citep{gustafsson_grid_2008}. The spectra were computed with a sampling resolution of $1\,$km$\,$s$^{-1}$, corresponding to a resolving power of $R{\sim}300\,000$, with a microturbulent velocity of $1\,$km$\,$s$^{-1}$. We adopt the solar chemical composition and isotopic ratios from \citet{asplund_chemical_2009}, except for an [$\alpha$/Fe] enhancement that varies linearly from $\text[\alpha / \text{Fe}] = 0$ when $\rm [Fe/H] \ge 0$ to $\text[\alpha/\text{Fe}] = +0.4$ when $\rm [Fe/H] \le -1$. We use a selection of atomic lines from VALD3 \citep{ryabchikova_major_2015} together with roughly 15 million molecular lines representing 18 different molecules, the most important of which for this work are CaH (Plez, priv. comm.), MgH \citep{kurucz_kurucz_1995,skory_new_2003}, and TiO \citep[with updates via VALD3]{plez_new_1998}. Finally, we note that while this suite of models has issues reproducing cool dwarf optical fluxes, likely due to missing opacities or incomplete line lists (see \citealt{rains_characterization_2021} for more information), the input physics should still be more than sufficient for qualitative analysis.

We say `bespoke grid' because of the limited parameter space covered: the grid has three $T_{\rm eff}$ values ($3\,000\,$K, $3\,500\,$K, $4\,000\,$K), two $\log g$ values (4.5, 5.0), and solar values of [Fe/H] and [$\alpha$/Fe] (0.0). While this might seem limiting, the strength of this grid comes from the added abundance dimensions, with each of C, N, O, Na, Mg, Al, Si, Ca, Ti, and Fe being able to be individually perturbed\footnote{Where the chemistry change is modelled by TURBOSPECTRUM while the underlying MARCS atmosphere remains the same.} by $\pm0.1\,$dex from the solar value. Thus, while the modelled star is broadly solar in composition, one can inspect the influence of small variations in abundance in isolation on optical fluxes where the pseudocontinuum position is strongly dependent on molecular absorption. 

Fig. \ref{fig:marcs_sensitivity_dominant_absorbers} shows the result of perturbing the abundances of C, O, and Ti---the three most influential elemental abundances in cool atmospheres---as well as the bulk metallicity [M/H] from $-0.1\,$dex to $+0.1\,$dex for the wavelengths covered by our WiFeS spectra at three different $T_{\rm eff}$ values. C, O, Ti, and [M/H] all have a significant---and mostly similar---impact on pseudocontinuum placement, with this similarity indicating the degenerate nature of these parameters. Notably C has the opposite sign to O, Ti, and [M/H], which we suggest relates to CO formation. While CO does not absorb at optical wavelengths it is, however, `energetically favourable' \citep{veyette_physical_2016}---for a lower C abundance there is a greater `relative' abundance of O available to form other molecules like TiO or H$_2$O.  At $T_{\rm eff}=3\,000\,$K, wavelengths longer than ${\sim}4\,500\,$\SI{}{\angstrom} see fractional change in flux of ${\sim}20$\% for C and [M/H], but closer to ${\sim}$40\% for Ti and O---likely due to the dominance of TiO. This effect diminishes with increasing temperature, though the pseudocontinuum can still change at the ${\sim}5-10$\% level---more than enough to complicate the measurement of equivalent widths via traditional spectroscopic analysis techniques. The very bluest wavelengths (below ${\sim}4\,300\,$\SI{}{\angstrom}) show a reduced influence from C, O, and TiO, potentially indicating that spectra in these regions would be less subject to degeneracies when fitting for the bulk metallicity. While the region between ${\sim}4\,100$--$4\,200\,$\SI{}{\angstrom} (likely due to SiH in our synthetic spectra due to our choice of line list\footnote{Studies using different line lists attribute features in this region to MgH--A-X or the tail of TiO $\alpha$ \citep{pavlenko_molecular_2014}; or AlH  \citep{pavlenko_alh_2022}.}, see Fig. \ref{fig:marcs_sensitivity_minor_absorbers} ) shows reduced influence from Ti as compared to C, O, and [M/H], on the whole it otherwise appears very difficult to disentangle the contributions from each element and the bulk metallicity.

Fig. \ref{fig:marcs_sensitivity_minor_absorbers} is the same, but for N, Na, Mg, Al, Si, Ca, and Fe instead. While none of these reach the significance of C, O, and Ti, being generally much more limited in the wavelength regions they affect, everything except N has at least one spectral region with a flux change of ${\sim}5-10$\%. Of note is that, outside of a number of strong lines, the influence of Fe is mainly below ${\sim}4\,000$\SI{}{\angstrom}. This means that our \textit{Cannon} model---as well as other optical [Fe/H] relations---are likely not actually sensitive to Fe spectral features directly when trained upon [Fe/H], rather how the shape of the pseudo--continuum correlates with [Fe/H]. We attribute the surprisingly large influence Na has on flux (at the ${\sim}5$--$20$\% level) to it being an electron donor, rather than being directly involved in atomic or molecular absorption itself. 

Collectively, even qualitative analyses like this allow insight into the sensitivity of different wavelength regions to varying elemental abundances to guide spectral analysis. Optical wavelengths clearly contain a wealth of chemical information, but limitations with current models in the cool dwarf regime and the parameters varied in model grids make this difficult to exploit using traditional methods.

%%%%%%%%%%%%%%%%%%%%%%%%%%%%%%%%%%%%%%%%%%%%%%%%%%%%%%%%%%%%%%%%%%%%%%%%%%%%%%%%%%%%%%%%%%%%%%%%%%%%%
% Discussion
%%%%%%%%%%%%%%%%%%%%%%%%%%%%%%%%%%%%%%%%%%%%%%%%%%%%%%%%%%%%%%%%%%%%%%%%%%%%%%%%%%%%%%%%%%%%%%%%%%%%%
\section{Discussion}\label{sec:discussion}

%%%%%%%%%%%%%%%%%%%%%%%%%%%%%%%%%%%%%%%%%%%%%%%%%%%%%%%%%%%%%%%%%%%%%%%%%%%%%%%%%%%%%%%%%%%%%%%%%%%%%
\subsection{Model Scatter and Wavelength Label Sensitivity}\label{sec:disussion:sensitivity}

\begin{figure*}
    \centering
    \includegraphics[width=\textwidth, trim=0cm 1.25cm 0cm 0cm]{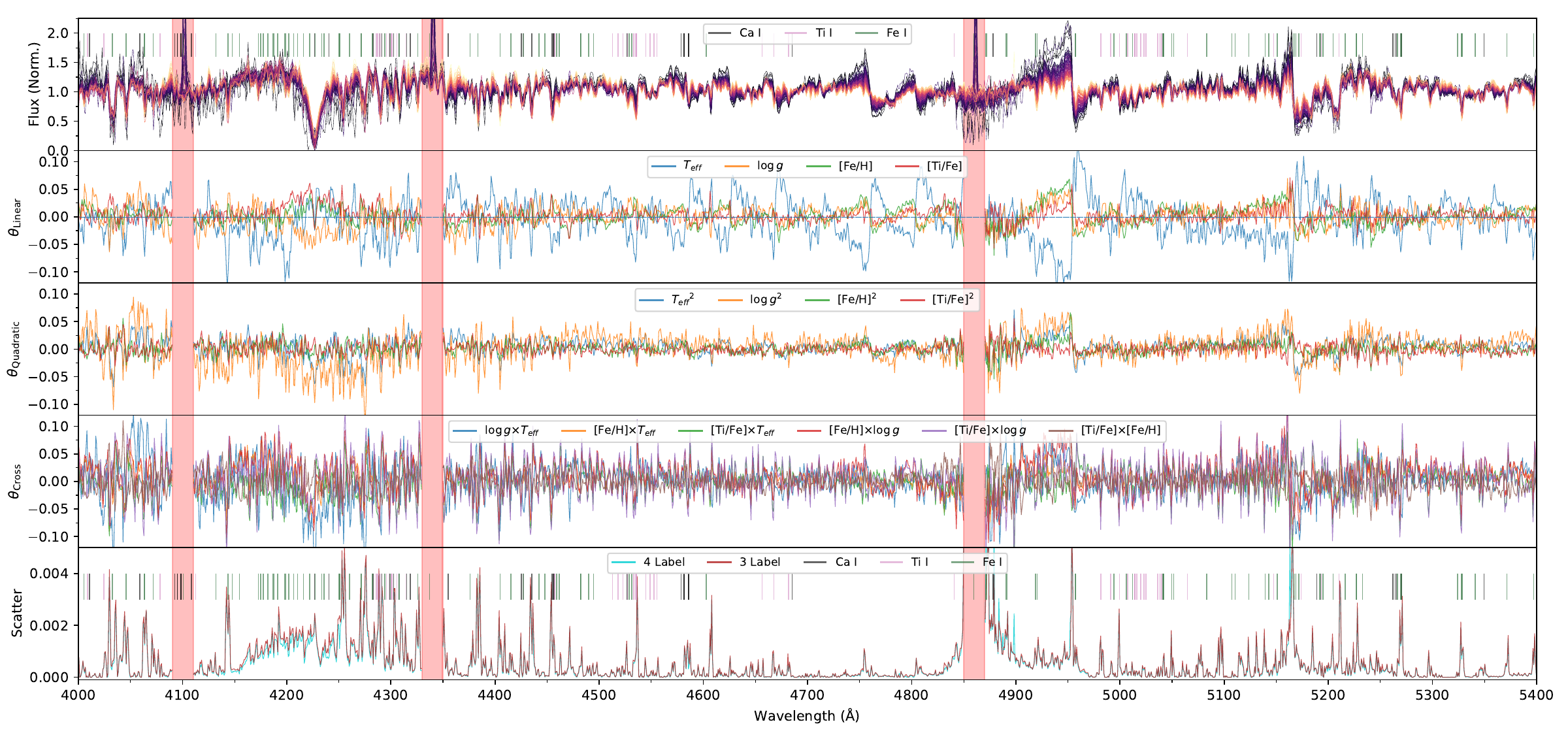}
    \caption{Four label \textit{Cannon} model pixel sensitivity to stellar labels and model scatter for WiFeS B3000 spectra with prominent atomic absorption features of \ion{Ca}{I}, \ion{Ti}{I}, and \ion{Fe}{I} labelled (EW$>400\,$m\SI{}{\angstrom}, calculated for $T_{\rm eff}=3
\,500\,$K, $\log g=5.0$, [Fe/H] = 0.0). \textbf{Panel 1:} Normalised B3000 spectra of stellar benchmarks used to train the model, where darker coloured spectra correspond to cooler stars. As before, the shaded red regions correspond to stellar emission, telluric absorption, or bad pixels. \textbf{Panel 2:} First order $\theta$ coefficients for each of the stellar labels ($T_{\rm eff}$, $\log g$, [Fe/H], and [Ti/Fe]). The further each coefficient is from 0, the more sensitive the flux in a given spectral pixel is to a specific label. \textbf{Panel 3:} As Panel 2, but for second order $\theta$ coefficients. \textbf{Panel 4:} As Panel 2 and 3, but for cross--term $\theta$ coefficients. \textbf{Panel 5:} \textit{Cannon} modelled scatter for each spectral pixel for both our 3 and 4 label \textit{Cannon} models, where higher values indicate that the adopted stellar labels are increasingly insufficient to completely parameterise the stellar flux. The same absorption features as before are overplotted.}
    \label{fig:param_sensitivity_b}
\end{figure*}

\begin{figure*}
    \centering
    \includegraphics[width=\textwidth, trim=0cm 1.25cm 0cm 0cm]{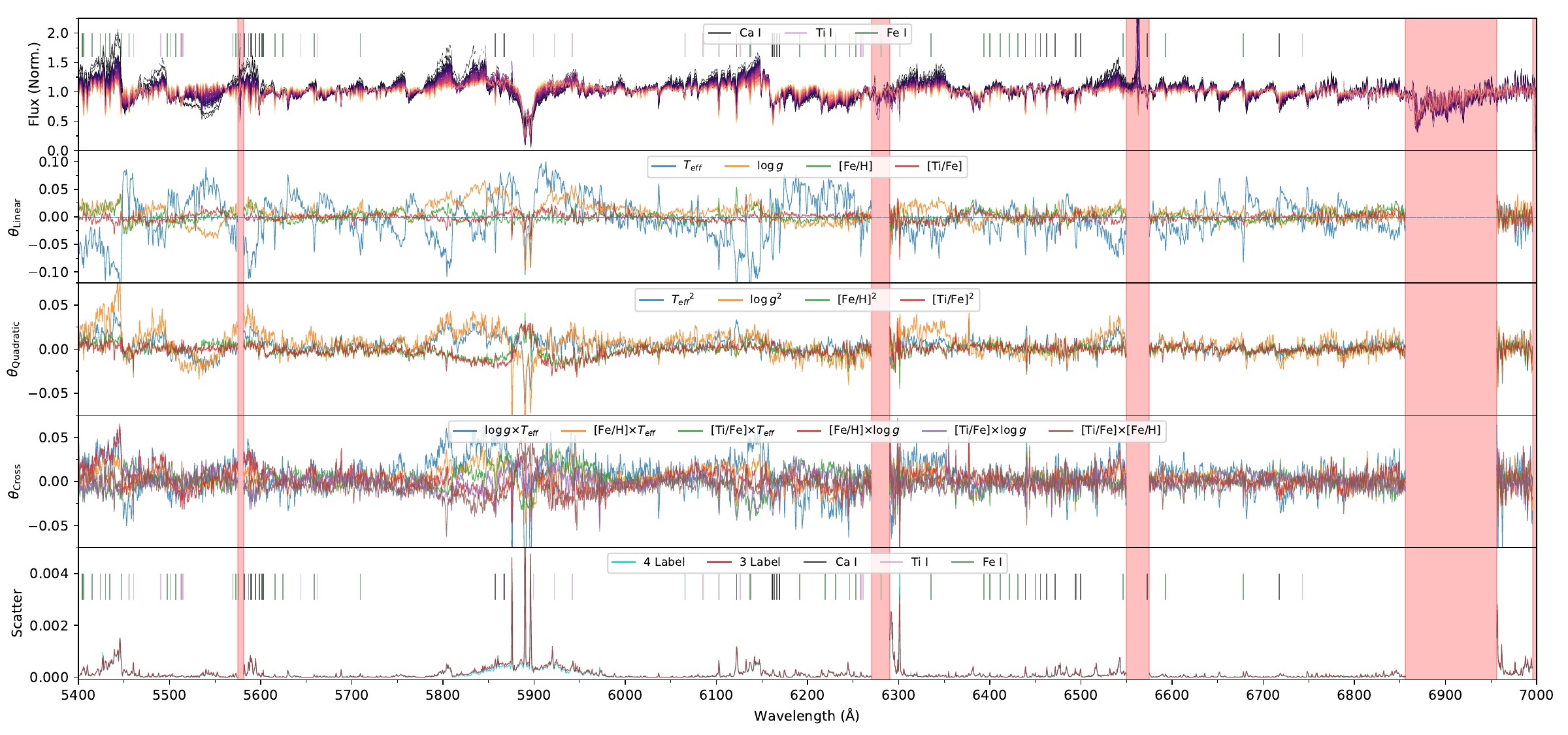}
    \caption{As Fig. \ref{fig:param_sensitivity_b}, but for WiFeS R7000 spectra.}
    
    \label{fig:param_sensitivity_r}
\end{figure*}

Now with some theoretical insight into how sensitive optical cool dwarf fluxes are to variations in abundance, we can begin to assess the performance of our \textit{Cannon} model in terms of how well it recovers these fluxes. The \textit{Cannon} attempts to parameterise the flux of each spectral pixel as a function of the adopted stellar labels, in our case $T_{\rm eff}$, $\log g$, [Fe/H], and [Ti/Fe]. Also inherent to the \textit{Cannon} model is a noise term (see Equation \ref{eqn:noise}) associated with each spectral pixel, which accounts for the fact that a) the training spectra are not known to arbitrarily high precision and thus have an associated flux uncertainty, and b) a scatter term to account for remaining flux variation unable to be parameterised as a function of the adopted labels with our adopted polynomial order. Broadly speaking, we would expect spectral pixels associated with strong atomic absorption features for atoms \textit{other} than Fe or Ti to be poorly modelled by the \textit{Cannon} as it does not have the constraints to properly learn these features. Similarly, since the entire optical region is affected by molecular absorption for cool stars, we in general expect a higher baseline model scatter for \textit{all} pixels, peaking where atomic or molecular features from elements unconstrained by our purely [Fe/H]--[Ti/Fe] chemical model dominate.

One important caveat to keep in mind before we begin assigning flux sensitivities or model scatter to being a function of stellar chemistry are \textit{observational} factors about our spectra or benchmark sample that could cause similar effects. The first of these is the SNR of our data, which is not uniform as a function of wavelength or $T_{\rm eff}$. In general, the warmer stars in our sample have higher SNR spectra from the WiFeS blue arm, with the coolest stars by comparison having much lower SNR in the blue---the very reason our \textit{Cannon} model begins at $4\,000\,$\SI{}{\angstrom} rather than covering the full extent allowed by the B3000 grating. The second effect is that of spectral resolution, with B3000 pixels having lower resolving power than their R7000 grating counterparts on the red arm. This could result in adjacent spectral features, which might be resolved at R${\sim}7\,000$, being blended at R${\sim}3\,000$ and being accordingly more complex for the \textit{Cannon} to parameterise as compared to the same information spread across two separate pixels. Finally, given the relatively small training sample, there exists the possibility of edge effects near the edge of the parameter space where the model is more poorly constrained---something which could also increase model scatter.

With that introduction and caveating out of the way, Fig. \ref{fig:param_sensitivity_b} and \ref{fig:param_sensitivity_r} present benchmark spectra, pixel sensitivity, and model scatter for the B3000 and R7000 wavelength regions respectively, and Table \ref{tab:theta_std_comp} offers a counterpart summary for each term in the label vector $\theta_N$. From top-to-bottom, each plot shows the normalised benchmark stellar spectra (labelled with prominent atomic features); spectral pixel sensitivity to $T_{\rm eff}$, $\log g$, [Fe/H], and [Ti/Fe] in the form of the linear \textit{Cannon} coefficients; quadratic \textit{Cannon} coefficients; cross-term \textit{Cannon} coefficients; and the \textit{Cannon} model scatter for each spectral pixel for our 3 and 4 label models (also labelled with prominent atomic features). For the purposes of comparison, note that the y-axis scale for the \textit{Cannon} scatter panels are the same for both plots. 

The model scatter in the blue is on average higher than in the red, which is consistent with both a higher contribution from unmodelled atomic absorption at bluer wavelengths, and the aforementioned lower spectral resolution of our B3000 spectra. We see a reduction in the median scatter value across all pixels by ${\sim}7.2\,$\% when switching from a three label to a four label model, which gives some hint as to the improved explanatory power of adding [Ti/Fe] as a label when attempting to reproduce observed spectra.

Table \ref{tab:theta_std_comp} lists the standard deviation $\sigma_{\theta_N}$ for each coefficient in our coefficient vector $\theta_N$---linear, quadratic, and cross-term in each label---for both WiFeS wavelength ranges, and for each of our two \textit{Cannon} models. Larger values of $\sigma_{\theta_N}$ correspond to the terms that are important for describing the largest-scale changes in the spectrum, for instance the linear and quadratic $T_{\rm eff}$ and $\log g$ terms for describing TiO bandheads. Values for all labels are uniformly larger in the blue versus the red, which again is consistent with the lower spectral resolution and higher atomic contribution in the blue. The cross-terms in particular give insight into the correlations between each label, and it will be interesting to see what would happen were we to add further [X/Fe] dimensions in follow-up work. Finally, of particular note is how important $\log g$ is as a label in linear, quadratic, and cross-terms---despite the fact it is often determined using photometry and was the label we put the least effort into sourcing literature values for. That said, for stars on the main sequence (i.e. the bulk of our sample bar the aforementioned outliers) $\log g$ is going to be well---or entirely---correlated with $T_{\rm eff}$ and [Fe/H] so it likely isn't truly an independent label.

There are numerous strong lines of Ca, Ti, Cr, and Fe in the spectrum (among other elements) which are clearly associated with many of the large `spikes' in model scatter. At this resolution we do not just expect a contribution from isolated strong lines, but also closely bunched multiplets which, whilst blended, still observably affect flux. As expected, there are many more atomic features towards bluer wavelengths---the exact regions where models are the least reliable, and cool dwarfs the faintest. Nonetheless, this points once more to the sheer amount of chemical information present at optical wavelengths for these stars, most of which is also inaccessible under our current \textit{Cannon} model formalism. 

Our present \textit{Cannon} model assumes that the flux of \textit{every} spectral pixel depends on $T_{\rm eff}$, $\log g$, [Fe/H], and [Ti/Fe]---the entire set of modelled stellar labels. While this is physically reasonable in the optical where much of the spectral information for these labels is contained in broad molecular features spanning a large range in wavelength, this assumption ceases to be valid when trying to model the abundances for elements whose primary influence is present only as discrete atomic lines. 
Thus, to better represent known physics it is more reasonable to implement a \textit{Cannon} model with regularisation whereby coefficients are actively encouraged to take on values of zero. \citet{casey_cannon_2016} demonstrated such an approach successful in modelling the full suite of APOGEE stellar parameters ($T_{\rm eff}$, $\log g$, and 15 abundances: C, N, O, Na, Mg, Al, Si, S, K, Ca, Ti, V, Mn, Fe, Ni), and we are hopeful that following their approach in the future would allow us to add additional [X/Fe] dimensions to our \textit{Cannon} model.

\begin{table}
\centering
\caption{Comparison of standard deviations computed for each $\theta$ term---offset, first order, second order, and cross term---for both our 3 and 4 label Cannon models. Larger values indicate a given coefficient (and thus the associated stellar parameter/s) is a more important term in modelling stellar fluxes.}
\begin{tabular}{cccc}
\hline
$\theta_{\rm 3 Label}$ & \multicolumn{3}{c}{$\sigma_{\theta_N}$}\\
& B3000 & R7000 & All \\
\hline
1 & 0.151 & 0.110 & 0.126 \\
$T_{\rm eff}$ & 0.036 & 0.035 & 0.035 \\
$\log g$ & 0.019 & 0.016 & 0.017 \\
${\rm[Fe/H]}$ & 0.013 & 0.008 & 0.010 \\
$T_{\rm eff}$$^2$ & 0.016 & 0.009 & 0.012 \\
$\log g$$\,\times\,$$T_{\rm eff}$ & 0.034 & 0.019 & 0.026 \\
${\rm[Fe/H]}$$\,\times\,$$T_{\rm eff}$ & 0.012 & 0.009 & 0.010 \\
$\log g$$^2$ & 0.028 & 0.014 & 0.020 \\
${\rm[Fe/H]}$$\,\times\,$$\log g$ & 0.016 & 0.010 & 0.012 \\
${\rm[Fe/H]}$$^2$ & 0.005 & 0.004 & 0.004 \\
\hline
\hline
$\theta_{\rm 4 Label}$ & \multicolumn{3}{c}{$\sigma_{\theta_N}$}\\
& B3000 & R7000 & All \\
\hline
1 & 0.151 & 0.110 & 0.126 \\
$T_{\rm eff}$ & 0.038 & 0.037 & 0.037 \\
$\log g$ & 0.019 & 0.017 & 0.018 \\
${\rm[Fe/H]}$ & 0.017 & 0.009 & 0.013 \\
${\rm[Ti/Fe]}$ & 0.014 & 0.007 & 0.010 \\
$T_{\rm eff}$$^2$ & 0.016 & 0.009 & 0.012 \\
$\log g$$\,\times\,$$T_{\rm eff}$ & 0.034 & 0.021 & 0.026 \\
${\rm[Fe/H]}$$\,\times\,$$T_{\rm eff}$ & 0.025 & 0.011 & 0.017 \\
${\rm[Ti/Fe]}$$\,\times\,$$T_{\rm eff}$ & 0.028 & 0.011 & 0.019 \\
$\log g$$^2$ & 0.028 & 0.015 & 0.020 \\
${\rm[Fe/H]}$$\,\times\,$$\log g$ & 0.032 & 0.012 & 0.021 \\
${\rm[Ti/Fe]}$$\,\times\,$$\log g$ & 0.035 & 0.011 & 0.022 \\
${\rm[Fe/H]}$$^2$ & 0.011 & 0.007 & 0.008 \\
${\rm[Ti/Fe]}$$\,\times\,$${\rm[Fe/H]}$ & 0.018 & 0.011 & 0.014 \\
${\rm[Ti/Fe]}$$^2$ & 0.010 & 0.007 & 0.008 \\
\hline
\end{tabular}
\label{tab:theta_std_comp}
\end{table}

Finally, with reference back to Fig. \ref{fig:marcs_sensitivity_dominant_absorbers} and \ref{fig:marcs_sensitivity_minor_absorbers}, it is possible to qualitatively compare model molecular absorption with increased \textit{Cannon} model scatter $s_\lambda$ (noting that we cannot expect a 1:1 match in light of known model spectra difficulties reproducing optical fluxes). Two regions of increased scatter seem to correspond with Ca--associated features (likely CaH) $4\,100-4\,300\,$\SI{}{\angstrom} (the region of largest scatter in the blue) and $6\,100-6\,150\,$\SI{}{\angstrom} (Ca I), though there are also relatively strong [Si/H] and [Fe/H] features in the former region. The Na doublet region ($5\,800-6\,000\,$\SI{}{\angstrom}) has the most scatter in the red, but it is difficult to ascribe this to Na alone. Curiously, $4\,700-5\,200\,$\SI{}{\angstrom} is the region where we expect Mg to have the largest impact, but the scatter is small compared to the aforementioned regions. While it is difficult to discuss this quantitatively with our existing model, it would be illustrative to revisit this with a future version of the \textit{Cannon} with regularisation and a broader set of chemical labels.

%%%%%%%%%%%%%%%%%%%%%%%%%%%%%%%%%%%%%%%%%%%%%%%%%%%%%%%%%%%%%%%%%%%%%%%%%%%%%%%%%%%%%%%%%%%%%%%%%%%%%
\subsection{Data-Driven vs Physical Model Flux Recovery}\label{sec:discussion:marcs_comparison}

\begin{figure*}%[hbb!]
    \begin{turn}{-90}
    \begin{minipage}{\textheight}
    \centering
    \includegraphics[height=0.7\textheight]{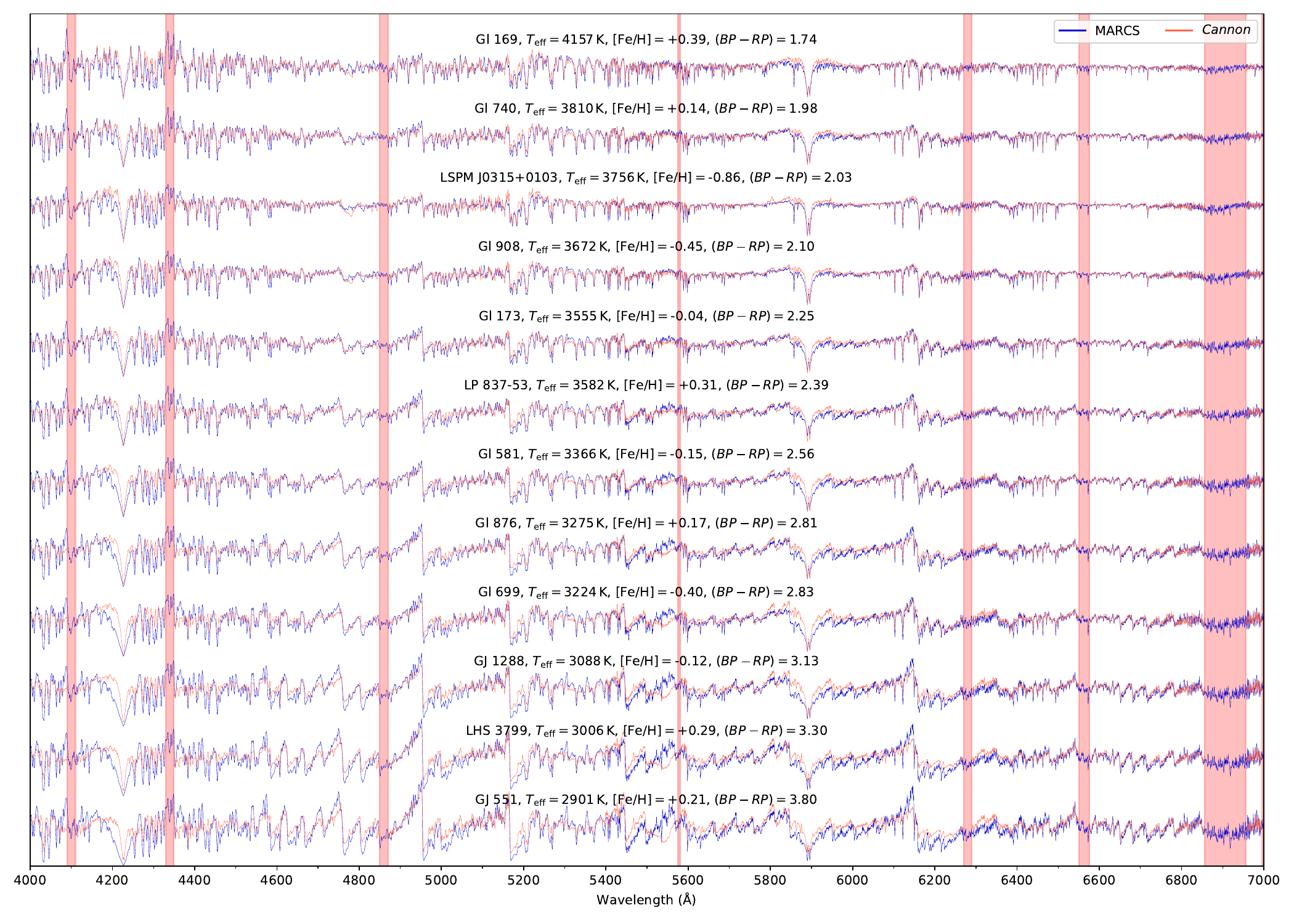}
    \caption{Comparison between our fully--trained 3 label \textit{Cannon} vs MARCS model spectra for a representative set of benchmark stars for $4\,000 < \lambda < 7\,000\,$\SI{}{\angstrom} at R${\sim}3\,000-7\,000$, with \textit{Cannon} model spectra in red and MARCS model spectra in blue. Both the MARCS and \textit{Cannon} spectra have been generated at our \textit{adopted} benchmark labels, rather than our fitted labels. The vertical red bars (from left to right) correspond to H-$\delta$, H-$\gamma$, H-$\beta$, a bad column on the WiFeS detector, atmospheric H$_2$O absorption, H-$\alpha$, and O$_2$ telluric features, all of which were masked out during modelling. The stars are sorted by their \textit{Gaia} $(BP-RP)$ colour to show a smooth transition in spectral features across the parameter space considered.}
    \label{fig:marcs_spectra_comp}
    \end{minipage}
    \end{turn}
\end{figure*}

\begin{figure*}%[hbb!]
    \centering
    \includegraphics[width=\textwidth, trim=0cm 0.8cm 0cm 0cm]{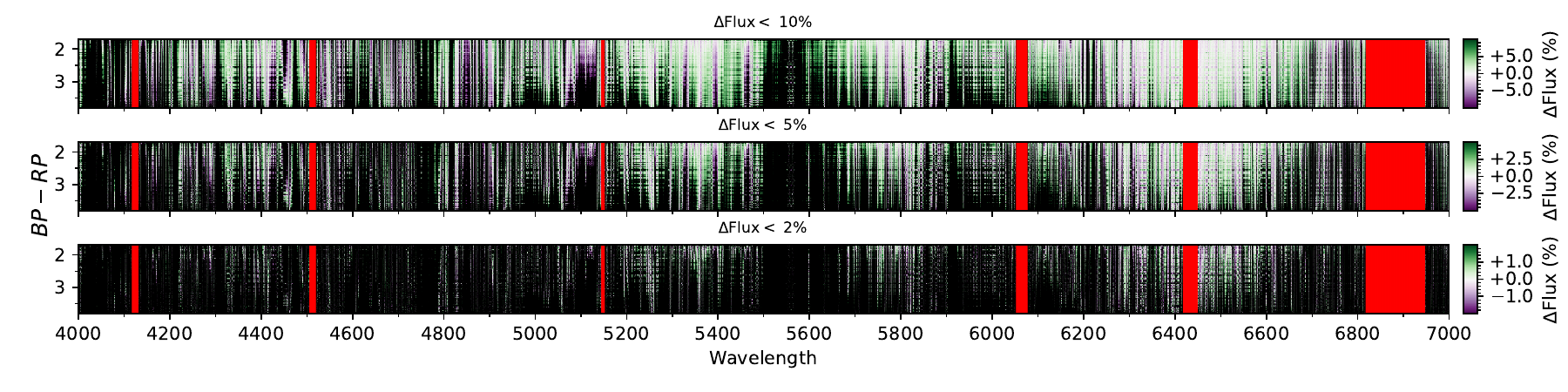}
    \caption{Percentage flux difference between our fully-trained 3 label \textit{Cannon} vs MARCS model spectra for all 103 cool dwarf benchmarks, again sorted by \textit{Gaia} DR3 $(BP-RP)$ colour. From top to bottom the panels show wavelengths where the percentage flux difference is $<$ 10\%, $<$ 5\%, and $<$ 2\% respectively. Each panel has its own colour bar, where white regions indicate a good match between the \textit{Cannon} and MARCS spectra ($\Delta$Flux $\approx0$), with the colour getting progressively darker green ($\Delta$Flux $>0$) or purple ($\Delta$Flux $< 0$) as the match worsens. Any wavelength region with a percentage flux difference beyond the 10\%, 5\%, or 2\% levels respectively is set to black, and the vertical red bars are again global masked telluric/emission/bad pixel regions. Note that these percentage differences only apply to our \textit{normalised} spectra, and not unnormalised flux-calibrated spectra, but they remain a useful metric for quantifying the accuracy of physical models like MARCS. These benchmarks span the parameter ranges $1.7 < (BP-RP) < 3.8$ and $4.55 < \log g < 5.20$.}
    \label{fig:marcs_delta_flux}
\end{figure*}

Given the good performance of our \textit{Cannon} model in recovering optical fluxes, it is illustrative to perform a comparison between data-driven and physical model spectra. For this comparison we treat our Cannon--produced spectra essentially as a self--consistent and interpolatable proxy for the observed spectra which, among other things, removes the influence of SNR or artefacts present in any single observed spectrum. Fig. \ref{fig:marcs_spectra_comp} shows a comparison between our 3 label \textit{Cannon} model in \teff, $\log g$, and [Fe/H] and the equivalent MARCS model spectra for the same set of benchmarks stars and adopted benchmark parameters presented in Fig. \ref{fig:blue_spectra_comp} and \ref{fig:red_spectra_comp} (again using the three--label model to avoid interpolating the MARCS grid in [Ti/Fe]). Alongside this is Fig. \ref{fig:marcs_delta_flux} which shows the percentage flux difference between the \textit{Cannon} and MARCS spectra for \textit{all} 103 benchmark stars (spanning the ranges $1.7 < (BP-RP) < 3.8$ and $4.55 < \log g < 5.20$). Our MARCs spectra were generated using the same grid used for \citet{rains_characterization_2021}, and normalised using the same normalisation formalism applied to our observed spectra as described in Section \ref{sec:cannon:preprocessing}.

Takeaways from this comparison are as expected from previous studies: performance is worse at bluer wavelengths and cooler temperatures. A few more detailed observations are as follow:
\begin{itemize}
    \item The best matching wavelength region is the few 100\,\SI{}{\angstrom} wide region surrounding H$\alpha$ (nominally ${\sim}6\,400$--$6\,800\,$\SI{}{\angstrom}, though even this gets worse towards mid-M spectral types).
    \item The wings of the Na D feature are consistently poorly reproduced by MARCS, even for the warmest stars in our sample.
    \item Discrepancies at blue wavelengths are obvious for even the warmest stars in our sample, and continue to get (sometimes dramatically) worse with cooler temperatures.
    \item While the \textit{positions} of TiO bandheads are well matched, their fluxes are not, likely posing problems for temperature or spectral type determination.
    \item Several broad absorption features (e.g. ${\sim}4\,400$--$4\,500\,$\SI{}{\angstrom}, and ${\sim}5\,500$--$5\,600\,$\SI{}{\angstrom}) are not reproduced in the MARCS spectra---indicative of missing opacity sources.
    \item We suspect the mismatch between $4\,000-4\,500\,$\SI{}{\angstrom} is caused by a spectral depression centred on the ${\sim}4\,227\,$\SI{}{\angstrom} neutral Ca resonance line that is poorly reproduced by MARCS. This feature---long known in the literature \citep[e.g.][]{lindblad_absorption_1935, vyssotsky_dwarf_1943}---was recently investigated in detail by \citet{jones_blue_2023}, who conclude the mismatch between models and observations is due to a `lack of appropriate treatment of line broadening for atomic calcium'. We direct interested readers to this paper and the references within for more information.
\end{itemize}
This isn't a perfect comparison, since we cannot control for the effect of elemental abundance variations which strongly affect optical fluxes. Where our MARCS models have a uniform scaled solar abundance pattern---critically uniform C, O, and Ti abundances---our \textit{Cannon} model is trained on stars from the Solar Neighbourhood which will instead show a spread in abundances. Further, our normalisation formalism is likely not robust to spectra as dramatically different as the \textit{Cannon} vs MARCS blue spectra are due to incomplete line lists or opacities in the latter. Nonetheless, even a qualitative comparison is illustrative the degree to which current generation model optical fluxes at low and medium resolution are a poor match to observations, and we advise caution when considering a purely model based approach when working with optical spectra of cool dwarfs.

%\end{landscape}

%%%%%%%%%%%%%%%%%%%%%%%%%%%%%%%%%%%%%%%%%%%%%%%%%%%%%%%%%%%%%%%%%%%%%%%%%%%%%%%%%%%%%%%%%%%%%%%%%%%%%
\subsection{Comparison with Previous Data-Driven Studies}\label{sec:discussion:comparison}

\begin{table*}%[!htb]
\scriptsize
%\centering
\caption{Comparison of previous data driven analyses on cool dwarfs.}
\label{tab:data_driven_comp}
\begin{tabular}{ccccccccccc}
\hline

Study & Method & Data & R & $\lambda$ & $T_{\rm eff}$ & $N_{\rm train}$ & Label Source & Labels & $\sigma_{T_{\rm eff}}$ & $\sigma_{\rm [Fe/H]}$ \\
 & & & & nm & K & & & & K & \\
\hline
\citet{behmard_data-driven_2019} & Cannon & CPS$^a$ &$60\,000$ & $499-610$ & $3\,000-5\,200$ & 141 & NIR$^b$, I$^c$ & $T_{\rm eff}$, $R_\star$, [Fe/H], $v\sin i$ & 68 & 0.08 \\

\citet{birky_temperatures_2020} & Cannon & APOGEE$^d$ & $22\,500$ & $1\,500-1\,700$ & $2\,860-4\,130$ & 87 & NIR & $T_{\rm eff}$, [Fe/H] & 77 & 0.09 \\

\citet{galgano_fundamental_2020} & Cannon & LAMOST$^e$ & $1\,800$ & $450-750$ & $3\,000-4\,100$ & 1\,388 & TCD$^f$ & $T_{\rm eff}$, $R_\star$, $M_\star$, $L_\star$ & 110 & -- \\

\citet{maldonado_hades_2020} & PCA$^g$ & HARPS$^h$ & $1\,000-2\,000^i$ & $534-691$ & $3\,100-4\,100$ & 19 & B$^j$ & [Fe/H], [X/H] $\times14^k$ & -- & 0.04 \\

\citet{li_stellar_2021} & SLAM$^l$ & LAMOST & $1\,800$ & $369-910$ & $2\,800-4\,500$ & 3\,785 & APOGEE & $T_{\rm eff}$, [M/H] & 48 & 0.13 \\

This Work & Cannon & WiFeS & $3\,000/7\,000$ & $400-700$ & $2\,900-4\,200$ & 103 & NIR, B, I & $T_{\rm eff}$, $\log g$, [Fe/H], [Ti/Fe]& 51 & 0.10 \\

%\citet{galgano_fundamental_2020} & \\
%\citet{li_stellar_2021} & \\
%This Work & $3000-7000$ & $4\,000 < \lambda < 7\,000$
\hline
\end{tabular}
\begin{minipage}{\linewidth}
\vspace{0.1cm}
\textbf{Notes:} $^a$California Planet Search (CPS, \citealt{howard_california_2010}); $^b$cool dwarfs with [Fe/H] from NIR empirical relations; $^c$cool dwarfs with interferometric $T_{\rm eff}$ and $R_\star$; $^d$Apache Point Galactic Evolution Experiment (APOGEE, \citealt{majewski_apache_2017}); $^e$the LAMOST Survey \citep{cui_large_2012}; $^f$TESS Input Catalogue ID \citep{muirhead_catalog_2018}; $^g$Principal Component Analysis; $^h$HARPS High-Accuracy Radial velocity Planetary Searcher (HARPS, \citealt{mayor_setting_2003}); $^i$Convolved with a Gaussian of width 12$\,$nm to smooth from HARPS native resolution of $R{\sim}115\,000$ down to $1\,000 \lesssim R \lesssim 2\,000$; $^j$[Fe/H] or [X/H] from cool dwarfs in a F/G/K--K/M binary system; $^k$Abundances for C, Na, Mg, Al, Si, Ca, Sc, Ti, V, Cr, Mn, Co, Ni, and Zn; $^l$Stellar LAbel Machine (SLAM, \citealt{zhang_deriving_2020}), which implements the Support Vector Machine (SVM) machine learning algorithm.\\
\end{minipage}
\end{table*}

Now we turn to putting our work in the broader context of other data-driven studies of cool dwarfs, with a broad overview shown in Table \ref{tab:data_driven_comp}. While the studies referenced span a range of data-driven algorithms, spectroscopic datasets, and modelling choices, we ultimately find it useful to evaluate them against two metrics that underpin much data-driven work in astronomy: label transfer and domain exploration.

The goal of label transfer is to quickly and precisely propagate a smaller set of potentially computationally expensive labels to some other, often larger, set of data. This might allow the transfer of elemental abundances from high-resolution spectroscopic surveys like APOGEE to their low-resolution counterparts like LAMOST \citep[e.g.][]{ho_cannon:_2016, wheeler_abundances_2020}, enable putting stellar parameters from distinct surveys on the same scale to enable cross-survey comparison (see e.g. comparing GALAH and APOGEE, \citealt{nandakumar_combined_2022}), or simply for computational efficiency due to the relative speed of data-driven methods versus more traditional modelling approaches (see e.g. GALAH DR2, \citealt{buder_galah_2018-1}). Of note is that label transfer---however precise---\textit{also} transfers any systematics or quirks from the original sample.

While all data-driven studies involve some level of label transfer, not all are---or need be---interested in domain exploration, where the goal is to better physically characterise a given parameter space and ideally learn something new or exploitable. A specific example of this is \citet{ness_spectroscopic_2016} using the \textit{Cannon} to recover spectroscopic red giant masses---historically an extremely challenging task---based on mass-dependent dredge-up signatures present in the CN absorption features of APOGEE spectra. In the case of cool dwarfs, domain exploration generally means attempting to overcome model limitations, account for $T_{\rm eff}$--chemistry degeneracies, and ideally gaining insight into the complex molecular physics governing their atmospheres. 

Collectively, the studies cited in Table \ref{tab:data_driven_comp}---\citet{behmard_data-driven_2019}, \citet{birky_temperatures_2020}, \citet{galgano_fundamental_2020}, \citet{maldonado_hades_2020} and \citet{li_stellar_2021}---demonstrate data-driven label transfer for cool dwarfs across a range of different wavelengths and spectroscopic resolutions. From these studies it is clear that the inability to continuum normalise cool dwarf spectra in any physically meaningful way does not appear to be a huge impediment to precise label transfer with data-driven models, which is fortunate for the potential of the method. That said, the accuracy\footnote{A note that use of the term `accuracy' is fraught when applied to chemical abundances---which are inherently model-derived---for anything other than the Sun's [X/H] $=0$ by definition, so we use it to refer to recovery of the original set of benchmark labels/accepted abundance scales, rather than something more physical like is appropriate in the case of parameters like $T_{\rm eff}$ or $\log g$.} of the transferred labels depend strongly upon the training sample used and the source of its labels, with each study also taking a different approach when it comes to domain exploration. In the next three subsections we put our work in context with these studies as we evaluate the state of the field for data-driven studies of cool dwarfs.

\subsubsection{Behmard et al. 2019 and Birky et al. 2020}

\citet{behmard_data-driven_2019} and \citet{birky_temperatures_2020} both use the \textit{Cannon} trained on small, but high-resolution, training sample of benchmark stars primarily drawn from \citet{mann_how_2015}. Their resulting [Fe/H] uncertainties prove consistent with the fundamentally-calibrated \citet{mann_how_2015} sample, while their $T_{\rm eff}$ values are marginally less precise. This is the opposite $T_{\rm eff}$ trend to what we observe with our results, where instead our values are consistent with literature uncertainties. We believe this discrepancy is a function of both spectral resolution and wavelength coverage, rather than inherent differences in our respective \textit{Cannon} models. Our broader wavelength coverage---enabled by our medium-resolution spectra---encompasses a greater number of highly-temperature sensitive molecular features like optical TiO bandheads, something less present in the shorter optical range covered by the HIRES spectra from \citet{behmard_data-driven_2019}, or the IR APOGEE spectra of \citet{birky_temperatures_2020}. On the other hand, their higher spectral resolution gives them access to many more unblended atomic features to use as [Fe/H] indicators, features which in the IR are also less affected by degeneracies imposed by molecular absorption. 

Using the California Planet Search (CPS) HIRES sample, \citet{behmard_data-driven_2019} trains their \textit{Cannon} model on a much wider temperature range ($3\,000 < T_{\rm eff} < 5\,200\,$K range than the other studies. This appears to have resulted in larger $T_{\rm eff}$ residuals at cooler temperatures, potentially indicating that a given \textit{Cannon} implementation struggles to model the molecule dominated spectra of cool stars at the same time as the more `regular' spectra of their solar-type counterparts. The implication being that, short of a more complex \textit{Cannon} model, it is likely best to restrict oneself to modelling the two paradigms separately---something our early prototyping with the broader range of stars from \citet{rains_characterization_2021} supports.

Our results at medium resolution and moderate SNR bear out the prediction from \citet{behmard_data-driven_2019} using convolved and degraded HIRES spectra that the \textit{Cannon} continues to function effectively across a range of spectral resolutions and SNRs. In an attempt to prevent overfitting, they also implemented a regularised \textit{Cannon} model which yielded no gain in label prediction. Their hypothesis, similar to our discussion in Section \ref{sec:disussion:sensitivity}, is that regularisation is unnecessary where each stellar label affects each spectral pixel, but will become more important for models extended to include elemental abundances where a given abundance might only affect a smaller set of spectral pixels.

\citet{birky_temperatures_2020} takes advantage of the \textit{Cannon} being an \textit{interpretable} machine learning model, and uses the first order model coefficients to identify [Fe/H] sensitive atomic or molecular features in their APOGEE spectra---the most significant of which they make publicly available. While we see similar sensitivity when interpreting our \textit{Cannon} implementation, we don't report a specific list of features as in our spectra they are both more likely to be blended and outweighed in significance compared to the ever-present molecular absorption. Additionally, while they take the time to analyse their fundamentally-calibrated cool dwarf [Fe/H] propagated to the broader APOGEE sample in terms of Galactic dynamic, we leave such work until we're able to derive more detailed abundances for our (much smaller) sample of stars.

Importantly, by putting their stars on a fundamentally-calibrated scale, \citet{birky_temperatures_2020} are able to reveal both $T_{\rm eff}$ and [Fe/H] systematics in the stellar parameters reported by APOGEE. While they do not undertake a detailed analysis, they attribute these discrepancies to incomplete line lists or opacities in the stellar models used by ASPCAP (APOGEE Stellar Parameters and Chemical Abundances Pipeline, \citealt{garcia_perez_aspcap_2016}), resulting in systematically biased fits as the pipeline attempts to optimise the continuum level in lieu of the missing features. This result is significant, as it hints at the danger of naively trusting cool dwarf parameters produced by generalist pipelines---something not appropriately considered by \citet{li_stellar_2021} when propagating APOGEE stellar labels to LAMOST spectra (see discussion in Section \ref{sec:galgano_li_discussion}).

\subsubsection{Maldonado et al. 2020}

Using a Principal Component Analysis (PCA) and Bayesian based approach in conjunction with F/G/K--K/M binary benchmarks \citet{maldonado_hades_2020} was able to recover a suite of 14 elemental abundances for their cool dwarf sample---the largest set of abundances to-date using a data-driven framework. Their method used HARPS spectra of 16 binary benchmarks convolved to a resolution of $R{\sim}1\,000-2\,000$, with the training sample selected to best match the metallicity distribution of nearby ($<70\,$pc) and kinematically similar F/G/K stars. While they do not undertake leave-one-out cross-validation as we do here, they validate their approach statistically by comparing abundances trends between K/M dwarfs and F/G/K stars, and methodologically by applying the technique to F/G/K stars to check for overfitting.

While the principal difference between our two approaches is of course the technique---PCA vs the \textit{Cannon}---there is also the difference in size between our two training samples. Although our binary benchmark sample is similar in size, 17 stars vs their 16 stars, we have a much larger sample of $T_{\rm eff}$ and secondary [Fe/H] benchmarks that serve to more robustly anchor our temperature and metallicity scales over a wider parameter space. \citet{maldonado_hades_2020} notes that only three stars in their training sample have [Fe/H] $<0.1\,$dex, and aimed to address this by matching the [Fe/H] distributions of their training sample with the Solar Neighbourhood F/G/K star sample. By comparison, six of our binary benchmarks---and roughly a third of our sample overall---have [Fe/H] below this threshold. Additionally, it is unclear to the extent that their approach is sensitive to $T_{\rm eff}-$chemistry degeneracies and how this affects their parameter recover, as this is something they do not discuss. Despite these limitations, they demonstrate clear recovery of their selected set of abundances---a remarkable achievement which points to the density of chemical information in optical spectra able to be retrieved when relying on a carefully considered set of benchmark stars.

Of their recovered abundances, it is interesting that both Ca and Mg show a comparatively large scatter as both these elements, and their derived molecules in CaH and MgH respectively, have a substantial number of absorption features in the optical. Hopefully follow-up work from both \citet{maldonado_hades_2020} with PCA, and our work with the \textit{Cannon}, can investigate whether this is astrophysical or sample-related in nature. It is, however, extremely encouraging to see that both C and Ti show much better recoveries given the extent both affect optical fluxes (per our discussion in Section \ref{sec:disussion:sensitivity}).

Finally, in terms of the merits between our respective models, it seems clear that PCA is both more effective and computationally less complex when it comes to dimensionality reduction versus our \textit{Cannon} model without regularisation. However, the \textit{Cannon}---being a generative model---remains the more interpretable approach, and we are better able to investigate label sensitivity as a function of wavelength and compare back to theoretical models. As such, it would be illustrative to apply both approaches to the same dataset to better compare, contrast, and capitalise on the respective advantages of each model in the pursuit of understanding the chemistry of cool dwarfs.

\subsubsection{Galgano et al. 2020 and Li et al. 2021}\label{sec:galgano_li_discussion}

As data-driven works, \citet{galgano_fundamental_2020} and \citet{li_stellar_2021} operate primarily in the label transfer space, with both using large ($>$ 1\,000 star) training samples to allow for more statistically robust model training and testing. \citet{galgano_fundamental_2020} apply the \textit{Cannon} to LAMOST spectra, and draw their labels ($T_{\rm eff}$, $R_\star$, $M_\star$, $L_\star$) from the TESS Input Catalogue (TIC, \citealt{muirhead_catalog_2018}) which is based primarily on photometric empirical relations such as those from \citet{mann_how_2015}. \citet{li_stellar_2021} also makes use of LAMOST data, but instead use the Support Vector Regression-based SLAM algorithm (Stellar LAbel Machine, \citealt{zhang_deriving_2020}) with APOGEE \citep{majewski_apache_2017} $T_{\rm eff}$ and [M/H] values. They chemically validate their methodology by checking against stars in known open clusters, as well as those in M--M binary systems whose stars should be chemically consistent.

Of the data-driven studies discussed, \citet{galgano_fundamental_2020} is unique in that it does not consider chemistry as a stellar label, primarily due to limitations with their use of the TIC as a source of labels. However, for those stars with stellar parallaxes in the TIC, $M_\star$ and $R_\star$ are computed from $M_{K_S}$, which is relatively insensitive to metallicity (as discussed in Section \ref{sec:intro}). As such, these labels---analogous to our adopted $\log g$ values also from photometric relations---will not be massively affected by parameter degeneracies in the source catalogue, even in the absence of chemical constraints. $T_{\rm eff}$, however, will be, which is part of the reason for the large $T_{\rm eff}$ uncertainties in the TIC to begin with. The TIC, however, is not entirely devoid of spectroscopic metallicity information for its brighter targets and, with the addition of photometric metallicities, an upgraded \textit{Cannon} model able to take into account label uncertainties would almost certainly prove useful. Nonetheless, there is---as the authors note---utility in transferring labels from photometric empirical relations sensitive to reddening to a more distant sample of stars observed at low spectral resolution like LAMOST, especially as a means of providing an empirically calibrated reference point for future model-based analyses.

An interesting comparison comes about when looking at the SNR investigation undertaken by \citet{galgano_fundamental_2020}. Their required SNR threshold for label uncertainties to be minimised is extremely high (SNR $> 150$), which indicates that at low spectroscopic resolution much higher-SNR values are necessary in order to constrain stellar labels from blended spectroscopic features, as compared to our much lower-SNR---but higher resolution---spectra.

\citet{li_stellar_2021} trains two separate SLAM models on LAMOST spectra: one with APOGEE labels, and another with labels from BT-Settl model spectra \citep{allard_model_2011}. Since both of these label sources are model-based, their stellar labels are not empirically calibrated on benchmark stars in the same way as the previously discussed studies, but they are able to observe and compare systematics between these two sources, as well as with the results of \citet{birky_temperatures_2020} which were also based on APOGEE spectra. While further comparison with our work is difficult as they neither use the \textit{Cannon}, nor benchmark--based labels, we note the power of working from a catalogue as large as LAMOST which allows validation using cluster stars or M--M binary systems, as well as the successful deployment of the SLAM algorithm to cool dwarfs in the spirit of algorithmic diversity.

Low-resolution optical spectra show great promise for data-driven studies of cool dwarfs since they are cheap to obtain observationally and, as discussed in Section \ref{sec:marcs_abundances}, should show the chemical imprints of a number of molecular species over a broad range in wavelength. With both \citet{galgano_fundamental_2020} and \citet{li_stellar_2021} having laid the groundwork for future LAMOST studies, it would be good to see a data-driven implementation with a more bespoke training sample based on well-characterised stellar benchmarks. Low-resolution surveys should have a particular advantage when it comes to collating a diverse sample of binary benchmarks as these systems rapidly become quite faint and difficult to observe at high-resolution in the optical. As such, there is much untapped potential for the existing LAMOST dataset, and these spectra should not be overlooked when it comes to understanding the physics and chemistry of cool dwarfs.

%%%%%%%%%%%%%%%%%%%%%%%%%%%%%%%%%%%%%%%%%%%%%%%%%%%%%%%%%%%%%%%%%%%%%%%%%%%%%%%%%%%%%%%%%%%%%%%%%%%%%
% Conclusions
%%%%%%%%%%%%%%%%%%%%%%%%%%%%%%%%%%%%%%%%%%%%%%%%%%%%%%%%%%%%%%%%%%%%%%%%%%%%%%%%%%%%%%%%%%%%%%%%%%%%%
\section{Conclusions}\label{sec:conclusion}
In the work presented above, we have detailed the development of a new four-label data-driven spectroscopic model in stellar $T_{\rm eff}$, $\log g$, [Fe/H], and [Ti/Fe] trained on 103 cool dwarf benchmarks observed in the optical ($4\,000<\lambda<7\,000\,$\SI{}{\angstrom}, R${\sim}$3\,000--7\,000) with the WiFeS instrument on the ANU 2.3\,m Telescope utilising the widely--used \textit{Cannon} algorithm. Not only do we put our work in context with other data-driven studies on cool dwarfs, but we conduct an investigation into the sensitivity of optical wavelengths to atomic and molecular features informed by both data-driven and physical models, and provide insight into the reliability of fluxes from physical models at cool temperatures. The main conclusions from our work are as follows:

\begin{enumerate}
    \item Our new four-label \textit{Cannon} model is trained on 103 cool dwarf benchmarks, 17 of which have literature abundance measurements from a binary companion. Under cross-validation our model is capable of recovering $T_{\rm eff}$, $\log g$, [Fe/H], and [Ti/Fe] with precisions of 1.4\%, $\pm0.04\,$dex, $\pm0.10\,$dex, and $\pm0.06\,$dex respectively---a very encouraging result given the extreme $T_{\rm eff}$--chemistry degeneracy of optical spectra. 
    
    \item Using kinematics from \textit{Gaia} DR3 and chemistry from GALAH DR3 we demonstrate the ability to predict [Ti/Fe] for Milky Way disc stars by interpolating in the empirical $v_\phi$--[Fe/H] space to $\sigma_{\rm [Ti/Fe]}\pm0.08\,$dex precision. Given the little that is known chemically about cool dwarfs due to their complex spectra, this approach shows promise for coarsely determining the abundances of $\alpha$ process elements prior to proceeding with more in--depth analyses of difficult-to-interpret optical spectra.

    \item We find our data-driven approach far superior at recovering optical cool dwarf fluxes compared to theoretical models using modern grids of synthetic spectra (see the discrepancies noted in e.g. \citealt{reyle_effective_2011}, \citealt{mann_spectro-thermometry_2013}, \citealt{rains_characterization_2021}). This demonstrates that data-driven techniques will be essential to fully exploiting optical spectra of cool stars until the next generation of physical models are able to update the currently incomplete line-lists for dominant molecular absorbers (e.g. TiO, \citealt{mckemmish_exomol_2019}).
    
    \item Using a custom grid of MARCS model cool dwarf spectra we conduct an investigation into the sensitivity of optical fluxes to chemical abundance variations. Our grid has 10 different chemical abundance dimensions (C, N, O, Na, Mg, Al, Si, Ca, Ti, and Fe) able to be individually perturbed by $\pm0.1\,$dex from Solar composition. Critically, this allows the inspection of the bulk effects of abundance variations on molecular absorption features. Our results indicate that a change in C, O, or Ti abundances affects the position of the pseudocontinuum to a similar or greater level than changing the bulk metallicity by the same amount (ranging from a $10-40\%$ change in flux), in concordance with prior work by \citet{veyette_physical_2016} on C and O abundances. 
    
    \item While not reaching the level of significance as C, O, or Ti, our grid also shows a number of spectral regions with a large ($10-20\%$) flux sensitivity to Na, Mg, Al, Si, Ca, and Fe, most likely arising from various molecular hydrides or oxides.
    
    \item Using the aforementioned model grid and a list of strong atomic features present in cool dwarf atmospheres, we interpret the modelled scatter of our \textit{Cannon} model---the pixel variation unable to be parameterised by our adopted four label quadratic model---in physical terms. We find the model scatter correlates with numerous strong lines of Ca, Ti, Cr, and Fe (among others), as well as regions we associate with molecular features like Ca or Si. 

    \item We perform a direct comparison between \textit{Cannon} and MARCS model spectra for wavelengths uncontaminated by strong telluric absorption within the spectral regions $4\,000 \leq \lambda \leq 5\,400\,$\SI{}{\angstrom} ($\lambda/\Delta\lambda\sim3\,000$) and $5\,400 \leq \lambda \leq 7\,000\,$\SI{}{\angstrom} ($\lambda/\Delta\lambda\sim7\,000$), with MARCS spectra showing large departures from the \textit{Cannon} fluxes that get worse at bluer wavelengths or cooler temperatures. We find only the few 100\,\SI{}{\angstrom} wide region surrounding H$\alpha$ (nominally ${\sim}6\,400-6\,800\,$\SI{}{\angstrom}) to be consistently reliable for the parameter space we consider, and warn anyone undertaking a model-based approach on optical cool dwarf spectra using current-generation models to proceed with caution.

\end{enumerate}
This study builds upon previous empirical, benchmark, and data-driven research on cool dwarfs and could not exist without such foundational work dedicated to understanding the most common kinds of stars in the Universe. While we have not yet resolved the $T_{\rm eff}$--chemistry degeneracy that has historically limited our understanding of such stars, we are given cause for cautious optimism. The sheer breadth of optical wavelengths that are sensitive to variations in chemical abundance in cool dwarfs are much greater than for Solar-type stars where most chemical information comes from isolated lines. This hints at cool stars being a powerful and as-yet-untapped method for studying the chemistry of our Galaxy and the demographics of planets---if only this information could be properly unlocked. This bodes well for cool dwarf focused work in current or upcoming optical surveys like GALAH, 4MOST, or SDSS-V, especially when combined with \textit{Gaia} DR3 for continuing to refine and broaden our benchmark sample. 

\section*{Acknowledgements}

We acknowledge and celebrate the traditional custodians of the land on which the ANU$\,$2.3 m Telescope stands, the Gamilaraay people, as well as the traditional custodians of the land on which the Australian National University is based, the Ngunnawal and Ngambri peoples, and pay our respects to elders past and present. 

We thank Michael Bessell and Ulrike Heiter for their helpful comments and discussions. We likewise thank the referee, Valentina D'Orazi, for her insightful suggestions that improved the quality and clarity of the manuscript.

A.D.R. acknowledges support by the Knut and Alice Wallenberg Foundation (grant 2018.0192). B.R-A acknowledges funding support from FONDECYT Iniciación grant 11181295 and ANID Basal project FB210003. Funding for M{\v Z} was provided by the European Union (ERC Advanced Grant, SUBSTELLAR, project number 101054354). This work was supported by computational resources provided by the Australian Government through the National Computational Infrastructure (NCI) under the National Computational Merit Allocation Scheme and the ANU Merit Allocation Scheme (project y89).

This work has made use of data from the European Space Agency (ESA) mission {\it Gaia} (\url{https://www.cosmos.esa.int/gaia}), processed by the {\it Gaia} Data Processing and Analysis Consortium (DPAC, \url{https://www.cosmos.esa.int/web/gaia/dpac/consortium}). Funding for the DPAC has been provided by national institutions, in particular the institutions participating in the {\it Gaia} Multilateral Agreement. This publication makes use of data products from the Two Micron All Sky Survey, which is a joint project of the University of Massachusetts and the Infrared Processing and Analysis Center/California Institute of Technology, funded by the National Aeronautics and Space Administration and the National Science Foundation. 

Software: \texttt{Astropy} \citep{astropy_collaboration_astropy:_2013}, \texttt{iPython} \citep{perez_ipython:_2007}, \texttt{Matplotlib} \citep{hunter_matplotlib:_2007}, \texttt{NumPy} \citep{harris_array_2020}, \texttt{Pandas} \citep{mckinney_data_2010, reback_pandas-devpandas_2020}, \texttt{SciPy} \citep{jones_scipy:_2016, virtanen_scipy_2020}. 

%%%%%%%%%%%%%%%%%%%%%%%%%%%%%%%%%%%%%%%%%%%%%%%%%%
\section*{Data Availability}

All adopted input benchmark parameters and fitted results are available in the article and in its online supplementary material, and stellar spectra will be shared on reasonable request to the corresponding author. All other data used is publicly available.



\bibliographystyle{mnras}
\bibliography{references} % if your bibtex file is called example.bib

%\begin{thebibliography}{99}
%\end{thebibliography}

%%%%%%%%%%%%%%%%%%%%%%%%%%%%%%%%%%%%%%%%%%%%%%%%%%

%%%%%%%%%%%%%%%%% APPENDICES %%%%%%%%%%%%%%%%%%%%%

\appendix

\section{Benchmarks and Parameter Fits}

\begin{table*}
\centering
\caption{Benchmark Stars}
\resizebox{0.925\textwidth}{!}{%
\begin{tabular}{cccccc|ccccc|cccc}
\hline
\multicolumn{6}{c}{} & \multicolumn{5}{c}{Adopted Parameters} & \multicolumn{4}{c}{Fitted Parameters} \\
Star & Gaia DR3 & $BP-RP$ & $BP$ & SNR$_{\rm B}$ & SNR$_{\rm R}$ & $T_{\rm eff}$ & $\log g$ & [Fe/H] & [Ti/Fe] & References & $T_{\rm eff}$ & $\log g$ & [Fe/H] & [Ti/Fe] \\
 &  &  &  &  &  & (K) & (dex) & (dex) & (dex) &  & (K) & (dex) & (dex) & (dex) \\
\hline
Gl 282B & 3057712188691831936 & 1.70 & 9.15 & 92 & 274 & $4154\pm67$ & $4.66\pm0.02$ & $+0.08\pm0.01$ & $-0.04\pm0.02$ & R21,R21,B16,B16 & $4172\pm51$ & $4.66\pm0.04$ & $+0.06\pm0.10$ & $-0.05\pm0.06$ \\
LSPM J2309+1425 & 2815543034682035840 & 1.70 & 10.46 & 44 & 136 & $4132\pm67$ & $4.66\pm0.02$ & $-0.05\pm0.08$ & $+0.01\pm0.07$ & R21,R21,M15,TW & $4074\pm51$ & $4.66\pm0.04$ & $+0.00\pm0.10$ & $-0.02\pm0.06$ \\
LSPM J1725+0206 & 4375233191015944192 & 1.71 & 7.75 & 65 & 197 & $4030\pm242$ & $4.55\pm0.03$ & $+0.19\pm0.08$ & $-0.02\pm0.05$ & vB09,R21,M15,TW & $4111\pm51$ & $4.63\pm0.04$ $\dagger$ & $+0.13\pm0.10$ & $-0.02\pm0.06$ \\
Gl 169 & 145421309108301184 & 1.74 & 8.54 & 93 & 282 & $4157\pm67$ & $4.61\pm0.02$ & $+0.39\pm0.08$ & $+0.02\pm0.08$ & R21,R21,M15,TW & $4150\pm51$ & $4.60\pm0.04$ & $+0.40\pm0.10$ & $+0.01\pm0.06$ \\
Gl 56.3 B & 2533723464155501952 & 1.78 & 10.93 & 46 & 144 & $4030\pm67$ & $4.68\pm0.02$ & $-0.11\pm0.02$ & $+0.04\pm0.04$ & R21,R21,M18,M18 & $3987\pm51$ & $4.66\pm0.04$ & $-0.02\pm0.10$ & $-0.01\pm0.06$ \\
HIP 12961 & 5077642283022422656 & 1.79 & 10.47 & 42 & 130 & $4033\pm67$ & $4.64\pm0.02$ & $+0.02\pm0.17$ & $+0.01\pm0.07$ & R21,R21,RA12,TW & $4034\pm51$ & $4.64\pm0.04$ & $+0.11\pm0.10$ & $-0.02\pm0.06$ \\
Gl 488 & 3689602277083844480 & 1.81 & 8.70 & 105 & 321 & $4044\pm67$ & $4.62\pm0.02$ & $+0.24\pm0.08$ & $-0.02\pm0.09$ & R21,R21,M15,TW & $4056\pm51$ & $4.61\pm0.04$ & $+0.27\pm0.10$ & $-0.02\pm0.06$ \\
Gl 208 & 3339921875389105152 & 1.82 & 9.08 & 108 & 318 & $3974\pm67$ & $4.66\pm0.02$ & $+0.05\pm0.08$ & $-0.02\pm0.07$ & R21,R21,M15,TW & $3978\pm51$ & $4.64\pm0.04$ & $+0.05\pm0.10$ & $-0.03\pm0.06$ \\
Gl 525 & 1244644727396803584 & 1.83 & 10.00 & 46 & 152 & $3888\pm67$ & $4.80\pm0.02$ & $-0.54\pm0.08$ & $+0.29\pm0.04$ & R21,R21,M15,TW & $3885\pm51$ & $4.78\pm0.04$ & $-0.47\pm0.10$ & $+0.25\pm0.06$ \\
HD 11964 B & 2462426800883156480 & 1.85 & 11.43 & 49 & 148 & $3979\pm67$ & $4.65\pm0.02$ & $+0.12\pm0.01$ & $-0.04\pm0.02$ & R21,R21,B16,B16 & $3898\pm51$ & $4.63\pm0.04$ & $+0.12\pm0.10$ & $+0.03\pm0.06$ \\
LP 754-50 & 6879662263797806976 & 1.88 & 11.52 & 43 & 136 & $3864\pm67$ & $4.73\pm0.02$ & $-0.17\pm0.04$ & $+0.04\pm0.06$ & R21,R21,M18,M18 & $3847\pm51$ & $4.71\pm0.04$ & $-0.26\pm0.10$ & $+0.13\pm0.06$ \\
Gl 79 & 5134635708766250752 & 1.89 & 9.12 & 182 & 529 & $3958\pm67$ & $4.66\pm0.02$ & $+0.14\pm0.08$ & $-0.02\pm0.06$ & R21,R21,M15,TW & $3950\pm51$ & $4.65\pm0.04$ & $+0.17\pm0.10$ & $-0.02\pm0.06$ \\
LSPM J2229+0139 & 2702655587447223168 & 1.92 & 10.75 & 74 & 219 & $3866\pm67$ & $4.64\pm0.02$ & $+0.13\pm0.08$ & $-0.01\pm0.06$ & R21,R21,M15,TW & $3897\pm51$ & $4.65\pm0.04$ & $+0.13\pm0.10$ & $-0.02\pm0.06$ \\
GJ 3108 & 301785537751949824 & 1.92 & 10.61 & 65 & 211 & $3948\pm67$ & $4.64\pm0.02$ & $+0.25\pm0.08$ & $-0.03\pm0.09$ & R21,R21,M15,TW & $3962\pm51$ & $4.60\pm0.04$ & $+0.33\pm0.10$ & $-0.03\pm0.06$ \\
Gl 471 & 3902785006045478656 & 1.92 & 9.91 & 106 & 322 & $3813\pm67$ & $4.68\pm0.02$ & $-0.04\pm0.08$ & $+0.02\pm0.06$ & R21,R21,M15,TW & $3867\pm51$ & $4.68\pm0.04$ & $-0.02\pm0.10$ & $-0.01\pm0.06$ \\
Gl 875 & 2611163717366876544 & 1.93 & 10.09 & 45 & 139 & $3828\pm67$ & $4.70\pm0.02$ & $+0.02\pm0.08$ & $-0.01\pm0.06$ & R21,R21,M15,TW & $3859\pm51$ & $4.70\pm0.04$ & $-0.08\pm0.10$ & $+0.00\pm0.06$ \\
Gl 239 & 3359074685047632640 & 1.95 & 9.86 & 91 & 282 & $3746\pm67$ & $4.79\pm0.02$ & $-0.34\pm0.08$ & $+0.15\pm0.06$ & R21,R21,M15,TW & $3817\pm51$ & $4.80\pm0.04$ & $-0.35\pm0.10$ & $+0.16\pm0.06$ \\
LP 873-74 & 6830027143525634432 & 1.96 & 11.66 & 43 & 131 & $3786\pm67$ & $4.70\pm0.02$ & $-0.17\pm0.03$ & $+0.14\pm0.08$ & R21,R21,M18,M18 & $3800\pm51$ & $4.67\pm0.04$ & $-0.05\pm0.10$ & $+0.10\pm0.06$ \\
Gl 846 & 2683023811628007296 & 1.97 & 9.39 & 159 & 462 & $3802\pm67$ & $4.67\pm0.02$ & $+0.02\pm0.08$ & $+0.00\pm0.08$ & R21,R21,M15,TW & $3813\pm51$ & $4.66\pm0.04$ & $+0.04\pm0.10$ & $+0.03\pm0.06$ \\
Gl 740 & 4282578724832056576 & 1.98 & 9.45 & 82 & 234 & $3810\pm67$ & $4.66\pm0.02$ & $+0.14\pm0.08$ & $-0.01\pm0.06$ & R21,R21,M15,TW & $3805\pm51$ & $4.66\pm0.04$ & $+0.06\pm0.10$ & $+0.01\pm0.06$ \\
Gl 678.1A & 4389844948935164544 & 1.98 & 9.53 & 47 & 148 & $3734\pm67$ & $4.69\pm0.02$ & $-0.09\pm0.08$ & $+0.02\pm0.07$ & R21,R21,M15,TW & $3781\pm51$ & $4.72\pm0.04$ & $-0.13\pm0.10$ & $+0.03\pm0.06$ \\
LSPM J0315+0103 & 3266980243936341248 & 2.03 & 14.89 & 28 & 93 & $3756\pm67$ & $4.92\pm0.02$ & $-0.86\pm0.04$ & $+0.29\pm0.07$ & R21,R21,M18,M18 & $3757\pm51$ & $4.91\pm0.04$ & $-0.86\pm0.10$ & $+0.27\pm0.06$ \\
GJ 229 & 2940856402123426176 & 2.07 & 8.37 & 117 & 330 & $3766\pm67$ & $4.71\pm0.03$ & $+0.02\pm0.11$ & $-0.00\pm0.07$ & R21,R21,G14,TW & $3759\pm51$ & $4.67\pm0.04$ & $+0.16\pm0.10$ & $-0.03\pm0.06$ \\
Gl 514 & 3738099879558957952 & 2.09 & 9.27 & 79 & 240 & $3707\pm67$ & $4.75\pm0.02$ & $-0.09\pm0.08$ & $+0.03\pm0.09$ & R21,R21,M15,TW & $3697\pm51$ & $4.73\pm0.04$ & $-0.10\pm0.10$ & $+0.02\pm0.06$ \\
Gl 908 & 2739689239311660672 & 2.10 & 9.22 & 73 & 213 & $3672\pm67$ & $4.85\pm0.02$ & $-0.45\pm0.08$ & $+0.21\pm0.08$ & R21,R21,M15,TW & $3657\pm51$ & $4.88\pm0.04$ & $-0.48\pm0.10$ & $+0.22\pm0.06$ \\
GJ 887 & 6553614253923452800 & 2.10 & 7.59 & 117 & 335 & $3692\pm57$ & $4.74\pm0.04$ & $-0.06\pm0.08$ & $+0.01\pm0.08$ & R19,R21,M15,TW & $3696\pm51$ & $4.76\pm0.04$ & $-0.16\pm0.10$ & $+0.03\pm0.06$ \\
Gl 686 & 4550763526539955712 & 2.11 & 9.82 & 38 & 123 & $3638\pm67$ & $4.81\pm0.02$ & $-0.25\pm0.08$ & $+0.06\pm0.05$ & R21,R21,M15,TW & $3680\pm51$ & $4.81\pm0.04$ & $-0.25\pm0.10$ & $+0.05\pm0.06$ \\
Gl 649 & 4572835535272627712 & 2.11 & 9.90 & 41 & 136 & $3590\pm45$ & $4.72\pm0.02$ & $+0.03\pm0.08$ & $-0.00\pm0.07$ & vB14,R21,M15,TW & $3696\pm51$ & $4.70\pm0.04$ & $+0.00\pm0.10$ & $-0.01\pm0.06$ \\
Gl 526 & 3741297293732404352 & 2.11 & 8.68 & 80 & 235 & $3618\pm31$ & $4.75\pm0.02$ & $-0.31\pm0.08$ & $+0.11\pm0.10$ & B12,R21,M15,TW & $3620\pm51$ & $4.76\pm0.04$ & $-0.34\pm0.10$ & $+0.13\pm0.06$ \\
Gl 821 & 6885776098199761024 & 2.12 & 11.11 & 41 & 131 & $3611\pm67$ & $4.91\pm0.02$ & $-0.45\pm0.08$ & $+0.23\pm0.05$ & R21,R21,M15,TW & $3675\pm51$ & $4.91\pm0.04$ & $-0.51\pm0.10$ & $+0.24\pm0.06$ \\
Gl 87 & 2515037264041041536 & 2.12 & 10.28 & 110 & 309 & $3635\pm67$ & $4.79\pm0.02$ & $-0.36\pm0.08$ & $+0.11\pm0.11$ & R21,R21,M15,TW & $3599\pm51$ & $4.78\pm0.04$ & $-0.42\pm0.10$ & $+0.12\pm0.06$ \\
Gl 701 & 4177731838628465408 & 2.12 & 9.60 & 43 & 132 & $3649\pm67$ & $4.79\pm0.02$ & $-0.22\pm0.08$ & $+0.05\pm0.07$ & R21,R21,M15,TW & $3685\pm51$ & $4.76\pm0.04$ & $-0.15\pm0.10$ & $+0.02\pm0.06$ \\
Gl 205 & 3209938366665770752 & 2.12 & 8.19 & 66 & 209 & $3801\pm9$ & $4.58\pm0.04$ & $+0.49\pm0.08$ & $-0.05\pm0.09$ & B12,R21,M15,TW & $3796\pm51$ & $4.59\pm0.04$ & $+0.45\pm0.10$ & $-0.06\pm0.06$ \\
GJ 3098 & 5042734468172061440 & 2.12 & 11.39 & 49 & 149 & $3660\pm67$ & $4.77\pm0.02$ & $-0.20\pm0.08$ & $+0.03\pm0.06$ & R21,R21,M15,TW & $3652\pm51$ & $4.75\pm0.04$ & $-0.14\pm0.10$ & $+0.01\pm0.06$ \\
Gl 390 & 3767878708888402816 & 2.15 & 10.38 & 53 & 163 & $3647\pm67$ & $4.70\pm0.02$ & $-0.02\pm0.08$ & $+0.00\pm0.07$ & R21,R21,M15,TW & $3664\pm51$ & $4.70\pm0.04$ & $+0.03\pm0.10$ & $-0.02\pm0.06$ \\
Gl 880 & 2828928008202069376 & 2.15 & 8.90 & 158 & 458 & $3713\pm11$ & $4.67\pm0.02$ & $+0.21\pm0.08$ & $-0.02\pm0.07$ & B12,R21,M15,TW & $3697\pm51$ & $4.66\pm0.04$ & $+0.23\pm0.10$ & $-0.03\pm0.06$ \\
PM I17052-0505 & 4364480521350598144 & 2.17 & 10.29 & 39 & 130 & $3615\pm67$ & $4.77\pm0.02$ & $-0.23\pm0.02$ & $+0.04\pm0.04$ & R21,R21,RB20,RB20 & $3547\pm51$ & $4.76\pm0.04$ & $-0.30\pm0.10$ & $+0.08\pm0.06$ \\
GJ 1 & 2306965202564744064 & 2.19 & 8.80 & 180 & 487 & $3616\pm14$ & $4.90\pm0.02$ & $-0.39\pm0.08$ & $+0.21\pm0.05$ & R19,R21,G14,TW & $3582\pm51$ & $4.91\pm0.04$ & $-0.47\pm0.10$ & $+0.22\pm0.06$ \\
Gl 382 & 3828238392559860992 & 2.23 & 9.49 & 92 & 268 & $3627\pm67$ & $4.72\pm0.02$ & $+0.13\pm0.08$ & $-0.02\pm0.05$ & R21,R21,M15,TW & $3630\pm51$ & $4.70\pm0.04$ & $+0.16\pm0.10$ & $-0.06\pm0.06$ \\
GJ 832 & 6562924609150908416 & 2.24 & 8.90 & 180 & 492 & $3512\pm23$ & $4.81\pm0.02$ & $-0.23\pm0.19$ & $+0.09\pm0.15$ & R19,R21,R21,TW & $3549\pm51$ & $4.78\pm0.04$ & $-0.17\pm0.10$ & $+0.03\pm0.06$ \\
Gl 361 & 614543647497149056 & 2.24 & 10.61 & 36 & 118 & $3583\pm67$ & $4.75\pm0.02$ & $-0.05\pm0.08$ & $+0.01\pm0.07$ & R21,R21,M15,TW & $3573\pm51$ & $4.73\pm0.04$ & $-0.09\pm0.10$ & $+0.03\pm0.06$ \\
Gl 173 & 3184351876391975808 & 2.25 & 10.58 & 44 & 130 & $3555\pm67$ & $4.75\pm0.02$ & $-0.04\pm0.08$ & $+0.02\pm0.10$ & R21,R21,M15,TW & $3577\pm51$ & $4.74\pm0.04$ & $+0.04\pm0.10$ & $-0.04\pm0.06$ \\
HD 46375 B & 3131777319158582272 & 2.26 & 12.39 & 28 & 97 & $3545\pm67$ & $4.65\pm0.02$ & $+0.24\pm0.01$ & $-0.02\pm0.02$ & R21,R21,B16,B16 & $3556\pm51$ & $4.61\pm0.04$ & $+0.39\pm0.10$ & $-0.07\pm0.06$ \\
GJ 1009 & 2316867885320845696 & 2.27 & 11.37 & 35 & 109 & $3622\pm67$ & $4.71\pm0.02$ & $+0.27\pm0.08$ & $-0.05\pm0.06$ & R21,R21,M15,TW & $3604\pm51$ & $4.68\pm0.04$ & $+0.25\pm0.10$ & $-0.05\pm0.06$ \\
Gl 104 & 87921523897823872 & 2.27 & 10.90 & 65 & 197 & $3601\pm67$ & $4.72\pm0.02$ & $+0.12\pm0.08$ & $-0.02\pm0.06$ & R21,R21,M15,TW & $3608\pm51$ & $4.73\pm0.04$ & $+0.17\pm0.10$ & $-0.06\pm0.06$ \\
Gl 745A & 4519789321942643072 & 2.28 & 11.01 & 40 & 135 & $3543\pm67$ & $4.94\pm0.02$ & $-0.33\pm0.08$ & $+0.08\pm0.08$ & R21,R21,M15,TW & $3578\pm51$ & $4.96\pm0.04$ & $-0.42\pm0.10$ & $+0.10\pm0.06$ \\
Gl 180 & 2979590513145784192 & 2.28 & 11.14 & 43 & 131 & $3522\pm67$ & $4.83\pm0.02$ & $-0.24\pm0.08$ & $+0.07\pm0.09$ & R21,R21,M15,TW & $3528\pm51$ & $4.81\pm0.04$ & $-0.14\pm0.10$ & $+0.02\pm0.06$ \\
Gl 745B & 4519789081415296128 & 2.29 & 11.01 & 40 & 134 & $3541\pm67$ & $4.94\pm0.02$ & $-0.35\pm0.08$ & $+0.10\pm0.09$ & R21,R21,M15,TW & $3570\pm51$ & $4.95\pm0.04$ & $-0.40\pm0.10$ & $+0.10\pm0.06$ \\
Gl 393 & 3855208897392952192 & 2.30 & 9.88 & 70 & 211 & $3540\pm67$ & $4.81\pm0.02$ & $-0.18\pm0.08$ & $+0.04\pm0.07$ & R21,R21,M15,TW & $3525\pm51$ & $4.76\pm0.04$ & $-0.05\pm0.10$ & $-0.02\pm0.06$ \\
GJ 2066 & 3089711447388931584 & 2.30 & 10.34 & 44 & 138 & $3534\pm67$ & $4.79\pm0.02$ & $-0.12\pm0.08$ & $+0.03\pm0.07$ & R21,R21,M15,TW & $3524\pm51$ & $4.78\pm0.04$ & $-0.18\pm0.10$ & $+0.07\pm0.06$ \\
Gl 176 & 3409711211681795584 & 2.30 & 10.21 & 78 & 233 & $3560\pm67$ & $4.73\pm0.02$ & $+0.14\pm0.08$ & $-0.03\pm0.06$ & R21,R21,M15,TW & $3548\pm51$ & $4.71\pm0.04$ & $+0.11\pm0.10$ & $-0.02\pm0.06$ \\
LSPM J1800+2933 & 4584638930036248192 & 2.32 & 12.84 & 31 & 108 & $3516\pm67$ & $4.76\pm0.02$ & $-0.08\pm0.01$ & $+0.03\pm0.02$ & R21,R21,B16,B16 & $3506\pm51$ & $4.77\pm0.04$ & $-0.06\pm0.10$ & $+0.02\pm0.06$ \\
Gl 250 B & 3101920046552857728 & 2.33 & 10.33 & 65 & 184 & $3533\pm67$ & $4.79\pm0.02$ & $+0.07\pm0.02$ & $-0.06\pm0.04$ & R21,R21,RB20,RB20 & $3530\pm51$ & $4.76\pm0.04$ & $+0.01\pm0.10$ & $-0.04\pm0.06$ \\
G 272-119 & 5142772953804963456 & 2.36 & 14.06 & 38 & 119 & $3483\pm67$ & $4.89\pm0.02$ & $-0.21\pm0.03$ & $+0.06\pm0.06$ & R21,R21,Sou08,A12 & $3488\pm51$ & $4.91\pm0.04$ & $-0.24\pm0.10$ & $+0.07\pm0.06$ \\
Gl 70 & 2560771312759450496 & 2.36 & 11.16 & 77 & 227 & $3497\pm67$ & $4.84\pm0.02$ & $-0.13\pm0.08$ & $+0.04\pm0.08$ & R21,R21,M15,TW & $3462\pm51$ & $4.82\pm0.04$ & $-0.21\pm0.10$ & $+0.04\pm0.06$ \\
LHS 3605 & 1838855376946302720 & 2.36 & 12.26 & 30 & 104 & $3484\pm67$ & $4.93\pm0.02$ & $-0.18\pm0.12$ & $+0.05\pm0.09$ & R21,R21,RA12,TW & $3524\pm51$ & $4.94\pm0.04$ & $-0.23\pm0.10$ & $+0.04\pm0.06$ \\
Gl 752 AB & 4293318823182081408 & 2.38 & 9.36 & 85 & 228 & $3555\pm67$ & $4.76\pm0.02$ & $+0.10\pm0.08$ & $-0.02\pm0.07$ & R21,R21,M15,TW & $3464\pm51$ & $4.74\pm0.04$ & $+0.07\pm0.10$ & $+0.02\pm0.06$ \\
LP 837-53 & 2910909931633597312 & 2.39 & 11.06 & 44 & 125 & $3582\pm67$ & $4.71\pm0.02$ & $+0.31\pm0.08$ & $-0.04\pm0.07$ & R21,R21,M15,TW & $3531\pm51$ & $4.72\pm0.04$ & $+0.37\pm0.10$ & $-0.09\pm0.06$ \\
LSPM J0546+0112 & 3219851121121855872 & 2.40 & 13.59 & 38 & 120 & $3456\pm67$ & $4.69\pm0.02$ & $+0.35\pm0.01$ & $-0.05\pm0.02$ & R21,R21,B16,B16 & $3474\pm51$ & $4.69\pm0.04$ & $+0.25\pm0.10$ & $-0.02\pm0.06$ \\
HD 24916 B & 3256334497479304832 & 2.42 & 11.73 & 41 & 123 & $3425\pm67$ & $4.75\pm0.02$ & $+0.08\pm0.02$ & $-0.03\pm0.04$ & R21,R21,RB20,RB20 & $3459\pm51$ & $4.76\pm0.04$ & $+0.01\pm0.10$ & $-0.05\pm0.06$ \\
Gl 811.1 & 6890353330746858368 & 2.43 & 11.72 & 42 & 124 & $3511\pm67$ & $4.76\pm0.02$ & $+0.16\pm0.08$ & $-0.02\pm0.06$ & R21,R21,M15,TW & $3455\pm51$ & $4.74\pm0.04$ & $+0.22\pm0.10$ & $-0.04\pm0.06$ \\
Gl 436 & 4017860992519744384 & 2.45 & 10.89 & 33 & 118 & $3416\pm54$ & $4.79\pm0.02$ & $+0.01\pm0.08$ & $-0.00\pm0.06$ & vB12,R21,M15,TW & $3404\pm51$ & $4.75\pm0.04$ & $+0.09\pm0.10$ & $-0.04\pm0.06$ \\
GJ 674 & 5951824121022278144 & 2.46 & 9.66 & 75 & 223 & $3409\pm25$ & $4.88\pm0.02$ & $-0.14\pm0.19$ & $+0.04\pm0.09$ & R19,R21,R21,TW & $3425\pm51$ & $4.87\pm0.04$ & $-0.23\pm0.10$ & $+0.07\pm0.06$ \\
LSPM J1804+1354 & 4498055584805914752 & 2.50 & 13.46 & 32 & 107 & $3385\pm67$ & $4.97\pm0.02$ & $-0.21\pm0.08$ & $+0.05\pm0.08$ & R21,R21,M15,TW & $3457\pm51$ & $4.97\pm0.04$ & $-0.22\pm0.10$ & $+0.03\pm0.06$ \\
Gl 109 & 114207651462714880 & 2.50 & 10.82 & 114 & 346 & $3402\pm67$ & $4.84\pm0.02$ & $-0.10\pm0.08$ & $+0.01\pm0.08$ & R21,R21,M15,TW & $3376\pm51$ & $4.87\pm0.04$ & $-0.17\pm0.10$ & $+0.03\pm0.06$ \\
HIP 79431 & 6245870673116518784 & 2.52 & 11.61 & 38 & 113 & $3360\pm67$ & $4.69\pm0.02$ & $+0.34\pm0.12$ & $-0.03\pm0.08$ & R21,R21,RA12,TW & $3411\pm51$ & $4.71\pm0.04$ & $+0.35\pm0.10$ & $-0.05\pm0.06$ \\
Gl 849 & 2627117287488522240 & 2.54 & 10.61 & 92 & 247 & $3506\pm67$ & $4.77\pm0.02$ & $+0.37\pm0.08$ & $-0.04\pm0.07$ & R21,R21,M15,TW & $3469\pm51$ & $4.79\pm0.04$ & $+0.37\pm0.10$ & $-0.06\pm0.06$ \\
LP 838-16 & 2911981886751531648 & 2.55 & 12.13 & 34 & 106 & $3362\pm67$ & $4.92\pm0.02$ & $-0.15\pm0.08$ & $+0.04\pm0.08$ & R21,R21,M15,TW & $3388\pm51$ & $4.93\pm0.04$ & $-0.09\pm0.10$ & $+0.01\pm0.06$ \\
Gl 581 & 6322070093095493504 & 2.56 & 10.81 & 42 & 133 & $3366\pm28$ & $4.90\pm0.02$ & $-0.15\pm0.08$ & $+0.04\pm0.07$ & R19,R21,M15,TW & $3372\pm51$ & $4.92\pm0.04$ & $-0.12\pm0.10$ & $+0.03\pm0.06$ \\
GJ 1236 & 4295619138933719808 & 2.60 & 12.68 & 30 & 96 & $3299\pm67$ & $5.01\pm0.02$ & $-0.29\pm0.08$ & $+0.09\pm0.08$ & R21,R21,M15,TW & $3380\pm51$ & $5.02\pm0.04$ & $-0.36\pm0.10$ & $+0.09\pm0.06$ \\
GJ 3325 & 2976850598890749824 & 2.60 & 11.98 & 28 & 95 & $3337\pm67$ & $4.96\pm0.02$ & $-0.12\pm0.08$ & $+0.03\pm0.07$ & R21,R21,M15,TW & $3377\pm51$ & $4.95\pm0.04$ & $-0.13\pm0.10$ & $+0.02\pm0.06$ \\
Gl 829 & 1784473016438653056 & 2.65 & 10.57 & 32 & 106 & $3303\pm67$ & $4.76\pm0.02$ & $-0.08\pm0.12$ & $+0.02\pm0.09$ & R21,R21,RA12,TW & $3342\pm51$ & $4.93\pm0.04$ $\dagger$ & $-0.04\pm0.10$ & $-0.00\pm0.06$ \\
Gl 179 & 3288082758293022848 & 2.72 & 12.23 & 27 & 88 & $3250\pm67$ & $4.81\pm0.02$ & $+0.18\pm0.12$ & $-0.02\pm0.05$ & R21,R21,RA12,TW & $3260\pm51$ & $4.76\pm0.04$ & $+0.16\pm0.10$ & $-0.01\pm0.06$ \\
Gl 628 & 4330690742322011520 & 2.73 & 10.32 & 62 & 195 & $3372\pm12$ & $4.91\pm0.02$ & $-0.03\pm0.08$ & $-0.00\pm0.07$ & R19,R21,M15,TW & $3351\pm51$ & $4.94\pm0.04$ & $+0.16\pm0.10$ & $-0.02\pm0.06$ \\
Gl 643 & 4339465360510320128 & 2.78 & 12.01 & 34 & 111 & $3247\pm67$ & $5.01\pm0.02$ & $-0.26\pm0.08$ & $+0.07\pm0.07$ & R21,R21,M15,TW & $3259\pm51$ & $5.01\pm0.04$ & $-0.18\pm0.10$ & $+0.05\pm0.06$ \\
Gl 876 & 2603090003484152064 & 2.81 & 10.45 & 100 & 281 & $3275\pm18$ & $4.89\pm0.02$ & $+0.17\pm0.08$ & $-0.02\pm0.06$ & R19,R21,M15,TW & $3244\pm51$ & $4.89\pm0.04$ & $+0.15\pm0.10$ & $-0.00\pm0.06$ \\
GJ 4333 & 2817999068780218368 & 2.82 & 11.94 & 32 & 103 & $3344\pm67$ & $4.83\pm0.02$ & $+0.24\pm0.08$ & $-0.02\pm0.06$ & R21,R21,M15,TW & $3255\pm51$ & $4.86\pm0.04$ & $+0.35\pm0.10$ & $-0.04\pm0.06$ \\
GJ 4065 & 4508377078422114944 & 2.82 & 12.75 & 26 & 82 & $3265\pm67$ & $4.89\pm0.02$ & $+0.04\pm0.08$ & $-0.01\pm0.07$ & R21,R21,M15,TW & $3244\pm51$ & $4.93\pm0.04$ & $+0.02\pm0.10$ & $-0.01\pm0.06$ \\
Gl 118.2 C & 116037376250217984 & 2.83 & 14.09 & 28 & 99 & $3282\pm67$ & $4.90\pm0.02$ & $+0.26\pm0.01$ & $-0.03\pm0.02$ & R21,R21,B16,B16 & $3293\pm51$ & $4.92\pm0.04$ & $+0.25\pm0.10$ & $-0.03\pm0.06$ \\
Gl 699 & 4472832130942575872 & 2.83 & 9.79 & 29 & 106 & $3224\pm10$ & $5.06\pm0.02$ & $-0.40\pm0.08$ & $+0.12\pm0.10$ & B12,R21,M15,TW & $3210\pm51$ & $5.09\pm0.04$ & $-0.38\pm0.10$ & $+0.10\pm0.06$ \\
Gl 729 & 4075141768785646848 & 2.83 & 10.73 & 32 & 110 & $3162\pm30$ & $5.06\pm0.02$ & $-0.18\pm0.08$ & $+0.04\pm0.08$ & R19,R21,M15,TW & $3222\pm51$ & $5.02\pm0.04$ & $-0.30\pm0.10$ & $+0.03\pm0.06$ \\
GJ 1207 & 4365609170043180416 & 2.84 & 12.52 & 31 & 105 & $3224\pm67$ & $4.98\pm0.02$ & $-0.09\pm0.08$ & $+0.03\pm0.08$ & R21,R21,M15,TW & $3207\pm51$ & $4.94\pm0.04$ & $-0.08\pm0.10$ & $-0.03\pm0.06$ \\
PM J19321-1119 & 4187836934609063552 & 2.87 & 14.17 & 31 & 102 & $3208\pm67$ & $4.98\pm0.02$ & $+0.05\pm0.02$ & $-0.02\pm0.04$ & R21,R21,RB20,RB20 & $3229\pm51$ & $4.94\pm0.04$ & $+0.05\pm0.10$ & $-0.06\pm0.06$ \\
GJ 3379 & 3316364602541746048 & 2.89 & 11.54 & 31 & 104 & $3206\pm67$ & $4.97\pm0.02$ & $+0.07\pm0.08$ & $-0.02\pm0.06$ & R21,R21,M15,TW & $3186\pm51$ & $4.93\pm0.04$ & $-0.02\pm0.10$ & $-0.04\pm0.06$ \\
GJ 4367 & 2435125446129139712 & 2.90 & 13.65 & 27 & 94 & $3278\pm67$ & $4.95\pm0.02$ & $+0.37\pm0.08$ & $-0.03\pm0.05$ & R21,R21,M15,TW & $3276\pm51$ & $4.93\pm0.04$ & $+0.27\pm0.10$ & $-0.03\pm0.06$ \\
LP 816-60 & 6882963066420161664 & 2.90 & 11.74 & 27 & 96 & $3173\pm67$ & $4.97\pm0.02$ & $-0.02\pm0.08$ & $+0.01\pm0.09$ & R21,R21,M15,TW & $3221\pm51$ & $4.96\pm0.04$ & $+0.05\pm0.10$ & $-0.00\pm0.06$ \\
LSPM J1845+1851 & 4512525016089353472 & 2.93 & 13.96 & 31 & 108 & $3167\pm67$ & $4.98\pm0.02$ & $-0.13\pm0.08$ & $+0.03\pm0.07$ & R21,R21,M15,TW & $3168\pm51$ & $4.96\pm0.04$ & $-0.05\pm0.10$ & $-0.03\pm0.06$ \\
GJ 3142 & 2507016253701863040 & 2.96 & 13.80 & 26 & 89 & $3227\pm67$ & $4.97\pm0.02$ & $+0.09\pm0.08$ & $-0.02\pm0.06$ & R21,R21,M15,TW & $3221\pm51$ & $4.94\pm0.04$ & $+0.19\pm0.10$ & $-0.05\pm0.06$ \\
Gl 285 & 3136952686035250688 & 3.00 & 11.42 & 25 & 87 & $3241\pm67$ & $4.88\pm0.02$ & $+0.41\pm0.17$ & $-0.06\pm0.11$ & R21,R21,RA12,TW & $3195\pm51$ & $4.90\pm0.04$ & $+0.25\pm0.10$ & $-0.09\pm0.06$ \\
LHS 3593 & 1750765357185849856 & 3.03 & 14.20 & 25 & 96 & $3121\pm67$ & $5.04\pm0.02$ & $+0.04\pm0.12$ & $+0.00\pm0.07$ & R21,R21,RA12,TW & $3184\pm51$ & $5.03\pm0.04$ & $+0.03\pm0.10$ & $+0.00\pm0.06$ \\
GJ 3323 & 3187115498866675456 & 3.03 & 12.41 & 23 & 85 & $3142\pm67$ & $5.08\pm0.02$ & $-0.06\pm0.08$ & $+0.01\pm0.07$ & R21,R21,M15,TW & $3144\pm51$ & $5.03\pm0.04$ & $-0.11\pm0.10$ & $+0.01\pm0.06$ \\
Gl 447 & 3796072592206250624 & 3.03 & 11.36 & 21 & 90 & $3264\pm24$ & $5.06\pm0.02$ & $-0.02\pm0.08$ & $+0.00\pm0.07$ & R19,R21,M15,TW & $3217\pm51$ & $5.04\pm0.04$ & $+0.08\pm0.10$ & $-0.02\pm0.06$ \\
GJ 1230A & 4535928365908540416 & 3.10 & 12.54 & 17 & 56 & $3201\pm67$ & $4.83\pm0.02$ & $+0.24\pm0.08$ & $-0.03\pm0.08$ & R21,R21,M15,TW & $3112\pm51$ & $4.82\pm0.04$ & $+0.37\pm0.10$ & $-0.06\pm0.06$ \\
GJ 1288 & 2868199402451064064 & 3.13 & 14.64 & 21 & 85 & $3088\pm67$ & $5.06\pm0.02$ & $-0.12\pm0.08$ & $+0.03\pm0.07$ & R21,R21,M15,TW & $3100\pm51$ & $5.05\pm0.04$ & $-0.03\pm0.10$ & $-0.02\pm0.06$ \\
LHS 495 & 1810616448010879488 & 3.14 & 13.69 & 27 & 98 & $3064\pm67$ & $4.99\pm0.02$ & $+0.16\pm0.12$ & $-0.02\pm0.07$ & R21,R21,RA12,TW & $3159\pm51$ & $5.01\pm0.04$ & $+0.12\pm0.10$ & $-0.00\pm0.06$ \\
Gl 777B & 2029432043779954432 & 3.17 & 14.65 & 11 & 52 & $3128\pm67$ & $4.98\pm0.02$ & $+0.18\pm0.01$ & $+0.02\pm0.02$ & R21,R21,B16,B16 & $3113\pm51$ & $5.01\pm0.04$ & $+0.28\pm0.10$ & $+0.00\pm0.06$ \\
PM J01125-1659 & 2358524597030794112 & 3.20 & 12.33 & 26 & 98 & $3062\pm67$ & $5.13\pm0.02$ & $-0.26\pm0.08$ & $+0.08\pm0.07$ & R21,R21,M15,TW & $3027\pm51$ & $5.12\pm0.04$ & $-0.27\pm0.10$ & $+0.06\pm0.06$ \\
GJ 1224 & 4145870293808914688 & 3.22 & 13.78 & 21 & 66 & $3061\pm67$ & $5.08\pm0.02$ & $-0.02\pm0.12$ & $+0.01\pm0.08$ & R21,R21,RA12,TW & $3046\pm51$ & $5.05\pm0.04$ & $-0.06\pm0.10$ & $-0.02\pm0.06$ \\
GJ 1214 & 4393265392168829056 & 3.23 & 14.91 & 22 & 76 & $3035\pm67$ & $5.03\pm0.02$ & $+0.16\pm0.12$ & $-0.02\pm0.07$ & R21,R21,RA12,TW & $3100\pm51$ & $5.01\pm0.04$ & $+0.14\pm0.10$ & $-0.01\pm0.06$ \\
LHS 3799 & 2595284016771502080 & 3.30 & 13.53 & 23 & 86 & $3006\pm67$ & $5.03\pm0.02$ & $+0.29\pm0.12$ & $-0.03\pm0.09$ & R21,R21,RA12,TW & $3090\pm51$ & $4.98\pm0.04$ & $+0.26\pm0.10$ & $-0.06\pm0.06$ \\
GJ 1002 & 2441630500517079808 & 3.67 & 14.07 & 18 & 82 & $2939\pm67$ & $5.19\pm0.02$ & $-0.11\pm0.07$ & $+0.03\pm0.06$ & R21,R21,T15,TW & $2929\pm51$ & $5.26\pm0.04$ & $-0.05\pm0.10$ & $-0.01\pm0.06$ \\
GJ 1286 & 2640434056928150400 & 3.71 & 14.95 & 9 & 64 & $2927\pm67$ & $5.20\pm0.02$ & $-0.01\pm0.12$ & $+0.02\pm0.07$ & R21,R21,RA12,TW & $2934\pm51$ & $5.19\pm0.04$ & $+0.11\pm0.10$ & $+0.01\pm0.06$ \\
GJ 551 & 5853498713190525696 & 3.80 & 11.37 & 49 & 187 & $2901\pm68$ & $5.14\pm0.02$ & $+0.21\pm0.03$ & $-0.03\pm0.05$ & R19,R21,VF05,VF05 & $2899\pm51$ & $5.16\pm0.04$ & $+0.22\pm0.10$ & $-0.04\pm0.06$ \\
\hline
\end{tabular}}
\label{tab:benchmark_parameters}
\begin{minipage}{\linewidth}
\vspace{0.1cm}
\textbf{References:}
R21: \citet{rains_characterization_2021}, B16: \citet{brewer_spectral_2016}, M15: \citet{mann_how_2015}, TW: This Work, vB09: \citet{van_belle_directly_2009}, M18: \citet{montes_calibrating_2018}, RA12: \citet{rojas-ayala_metallicity_2012}, G14: \citet{gaidos_trumpeting_2014}, R19: \citet{rabus_discontinuity_2019}, vB14: \citet{von_braun_stellar_2014}, B12: \citet{boyajian_stellar_2012-1}, RB20: \citet{rice_stellar_2020}, Sou08: \citet{sousa_spectroscopic_2008}, A12: \citet{adibekyan_chemical_2012}, vB12: \citet{von_braun_gj_2012}, T15: \citet{terrien_near-infrared_2015}, VF05: \citet{valenti_spectroscopic_2005}
\textbf{Notes:} $\dagger$: $|\Delta\log g|$ aberrant by $> 0.075\,$dex compared to literature value.
\end{minipage}
\end{table*}

%%%%%%%%%%%%%%%%%%%%%%%%%%%%%%%%%%%%%%%%%%%%%%%%%%

% Don't change these lines
\bsp	% typesetting comment
\label{lastpage}
\end{document}